\documentclass[10pt,a4paper,twoside]{article}

\usepackage{a4wide}

\usepackage[utf8]{inputenc} 
\usepackage[TS1,T1]{fontenc}

\usepackage{amssymb, amsmath,subeqnarray}
\usepackage{mathrsfs}
\usepackage{bbold}
\DeclareFontFamily{U}{bbold}{}
\DeclareFontShape{U}{bbold}{m}{n}{<-5.5> bbold5 <5.5-7.5> bbold7 <7.5-> bbold10}{}

\makeatletter

\@addtoreset{equation}{section}
\makeatother

\usepackage{graphicx}

\usepackage[english]{babel}

\newcommand{\be}{\begin{equation}}
\newcommand{\ee}{\end{equation}}
\newcommand{\bea}{\begin{eqnarray}}
\newcommand{\eea}{\end{eqnarray}}

\newcommand{\Aev}{\mathbb{A}}
\newcommand{\Aff}{A}

\newcommand{\betaeff}{{\beta_\star}}
\newcommand{\B}{\mathbb{B}}
\newcommand{\cc}{\widetilde{c}}
\newcommand{\ccstar}{c}
\newcommand{\cs}{\widetilde{s}}
\newcommand{\csstar}{s}
\newcommand{\C}{{\cal C}}
\newcommand{\Ch}{C}
\newcommand{\dexch}{d_\text{\mdseries exch}}

\newcommand{\Deltaexch}{\Delta_\text{\mdseries exch}}

\newcommand{\Deltastar}{\Delta}
\newcommand{\dint}{d_\text{\mdseries int}}
\newcommand{\En}{{\cal E}}
\newcommand{\Esp}[1]{\langle#1\rangle}
\newcommand{\Espcan}[1]{\langle#1\rangle_\text{\mdseries can}}
\newcommand{\Espeq}[1]{\langle#1\rangle_\text{\mdseries eq}}
\newcommand{\Espst}[1]{\langle#1\rangle_\text{\mdseries st}}
\newcommand{\fad}{\overline{f}}
\newcommand{\ft}{\widetilde{f}}
\newcommand{\F}{\mathbb{F}}
\newcommand{\Fth}{{\cal F}}
\newcommand{\gammaeff}{\gamma_\star}
\newcommand{\gapE}{\Delta e}
\newcommand{\Heat}{{\cal Q}}
\newcommand{\Heatav}{Q}
\newcommand{\Heatd}{{\cal Q}^\text{d}}
\newcommand{\Heatm}{q}

\newcommand{\iexp}{\text{\mdseries i}\,}
\newcommand{\Id}{\mathbb{I}}
\newcommand{\jad}{\overline{\jmath}}
\newcommand{\jinst}{j}
\newcommand{\Jcum}{{\cal J}}
\newcommand{\Jcumad}{\widetilde{\jmath}}

\newcommand{\kB}{k_{\scriptscriptstyle B}}
\newcommand{\Kc}{K_c}
\newcommand{\Ks}{K_s}
\newcommand{\lad}{\overline{\lambda}}
\newcommand{\M}{\mathbb{M}}
\newcommand{\nuad}{\overline{\nu}}
\newcommand{\Oobs}{{\cal O}}
\newcommand{\Pcb}{\mathbb{P}}
\newcommand{\Prob}{P}
\newcommand{\Probcan}{P_\text{\mdseries can}}
\newcommand{\Probeq}{P_\text{\mdseries eq}}

\newcommand{\Probst}{P_\text{\mdseries st}}
\newcommand{\Probdist}{\Pi}
\newcommand{\Sh}{S}

\newcommand{\SG}{S^{\scriptscriptstyle SG}}

\newcommand{\sighk}{\sigma_\text{\mdseries hk}}

\newcommand{\tad}{\tau}

\newcommand{\Trans}{\mathbb{W}}

\newcommand{\Uev}{\mathbb{U}}
\newcommand{\UevG}{\widehat{\mathbb{U}}}
\newcommand{\xcp}{x_c}
\newcommand{\xp}{x_+}

\title{\textbf{Thermal Contact. II. A Solvable Toy Model}}
\author{
F. Cornu\\ Laboratoire de Physique Théorique,  UMR 8627 du CNRS\\
Université Paris-Sud, Bât. 210
\\ F-91405 Orsay, France
 \\ \vspace{3mm}\\
 M. Bauer\\
 Institut de Physique Théorique de Saclay\footnote{CEA/DSM/IPhT, Unité de recherche associée au CNRS}, CEA Saclay
\\ F-91191 Gif-sur-Yvette Cedex, France
}

\date{May 2, 2015}

\begin{document}
\maketitle

\vspace{0.25cm}
\begin{itshape}

Most results contained in the  joined papers arXiv:1302.4538  and arXiv:1302.4540 put on cond-mat.stat-mech on February 19 2013  have now appeared in print as 

\vspace{0.15cm}
- [a]  \textnormal{Thermal Contact through a Diathermal Wall: A Solvable Toy Model},
\hfill\break\indent\indent
 F. Cornu and M. Bauer, J. Stat. Mech. (2013)  P10009

\vspace{0.15cm}
- [b] Part of \textnormal{Affinity and Fluctuations in a Mesoscopic Noria},
\hfill\break\indent\indent
 M. Bauer and F. Cornu,   J. Stat. Phys.   {\bf 155} (2014) 703, [arXiv:1402.2422]

\vspace{0.15cm}
- [c] \textnormal{ Local detailed balance : a microscopic derivation},
\hfill\break\indent\indent
M. Bauer and F. Cornu, J. Phys. A: Math. Theor. {\bf 48} (2015) 015008, [arXiv:1412.8179]

\vspace{0.25cm}

The present revised version takes into account the minor corrections made in the published articles but the bibliography is that of the first version (not updated). In reference [c] the denomination \textnormal{local detailed balance} has been used in place of \textnormal{modified detailed balance} in order to fit the  terminology that seems to be  most commonly used nowadays.

\end{itshape}
\vspace{0.5cm}

\begin{abstract}

  A diathermal wall between two heat baths at different  temperatures  can be mimicked by a layer of independent  spin pairs with some internal energy and where each
  spin $\sigma_a$ is flipped by  thermostat $a$ ($a=1,2$). The transition rates are determined from the
  modified detailed balance discussed in Ref.\cite{CornuBauerA}. Generalized
  heat capacities, excess heats, the housekeeping entropy flow and the thermal
  conductivity  in the steady state are calculated. The joint probability distribution of the heat
  cumulated exchanges at any time is computed explicitly. We obtain the large
  deviation function of heat transfer via a variety of approaches. In
  particular, by a saddle-point method performed accurately, we obtain the
  explicit expressions not only of the large deviation function, but also of the
  amplitude prefactor, in the long-time probability density for the heat
  current. The following physical properties are discussed : the effects of
  typical time scales of the mesoscopic dynamics which do not appear in
  equilibrium statistical averages and the limit of strict energy dissipation
  towards a thermostat when its temperature goes to zero.  We also derive some
  properties of the fluctuations in the two-spin system viewed as a thermal
  machine performing  cycles.

\vskip 0.25cm
{\bf PACS} ~: 05.70.Ln, 02.50.Ga, 05.60.Cd 
\vskip 0.25cm 
{\bf KEYWORDS}~: thermal contact; master equation; solvable model; excess heat;
thermal conductivity; large deviation function ; kinetic mean-field effect; strict dissipation ;
thermal cycles. 
 
 \vskip 0.25cm 
{\it Corresponding author ~:} 
CORNU Françoise,
E-mail: Francoise.Cornu@u-psud.fr

\end{abstract}

{\small{\tableofcontents}}

\section{Introduction}

\subsection{Issues at stake}

Though a rather specialized topic within non-equilibrium physics, the
problematic of thermal contact is in itself a vast subject, and one of foremost
theoretical and practical interest. A theoretic, microscopic, understanding of
non-equilibrium dynamics is still lacking today, even in the particular case of
heat exchanges. When compared with the most general non-equilibrium physics
world, certain specific settings have some advantages and this is the case of
thermal contact: when only quantities conserved by dynamics (energy for a
thermal contact) are exchanged between baths and an out-of-equilibrium system,
and different baths are in contact with different parts of this
out-of-equilibrium system, the fact that the baths remain at equilibrium all
along the experiment allows to keep a simple, thermodynamical interpretation of
various physical quantities (usually entropy variations) related to the
exchanges.

There are at least two important microscopic versions of thermal contact one
can have in mind. 

One is as an interface between two thermal baths, generically at different
temperatures. In our real (three-dimensional) world, the most natural model
geometry for the interface is a real or fictitious (two-dimensional) surface.
Each side of the surface consists of atoms of one bath. Heat flows from the high
temperature bath to the low temperature bath either via interactions among atoms
sitting on each side of an immaterial interface (a case relevant when the baths
consist of solid materials), or via interactions of the atoms with  a thin or
structureless material interface (a case relevant for instance when the baths
are gaseous, no matter is allowed to be exchanged, and the interface is a
diathermal wall). One can of course generalize to more than two baths, in contact via an
appropriate number of interfaces.

Another version of thermal contact is as an extended piece of material, with
specified physical properties, and such that two (or more) parts of its boundary
are in contact with thermal baths. A typical example is a bar of metal
whose extremities are maintained at different temperatures. Though the piece of
material is not in thermodynamic equilibrium stricto sensu, it is often of
practical value to assign a local temperature at each point of the bar, and the
famous Fourier law states that, for an isotropic material sustaining small
temperature gradients, the heat current is proportional to the temperature
gradient at each point. This law is of course approximate and phenomenological.
But finding a physically motivated model, even a crude one, for which the
relation between local temperature and heat current can be derived from first
principles, and compared with Fourier's law, is a major challenge in the field.
This is true even when the piece of material is a homogeneous bar and only its 
longitudinal dimension has to be taken into account, its section being
homogeneous with a good accuracy.

Our aims in this article can be considered as very modest, especially when
compared to the general issues raised by both versions of thermal contact
alluded above. But before describing in detail our model, let us put it
briefly in context and compare it to other, more or less similar approaches.

A significant trend in out-of-equilibrium statistical mechanics has been the
search for solvable models which could give some hints in the comprehension of
out-of-equilibrium phenomena in the absence of any theoretical framework which
would play the role of Gibbs' statistical ensemble theory for equilibrium
states.  Several kinds of models for heat conduction have been introduced. In
some of them the two heat baths are connected by a system with deterministic
dynamics, such as an anharmonic chain (Fermi-Pasta-Ulam model)
\cite{FermiPastaUlam1955,ChaosCampbell2005,ChaosProsen2005,NicolinSegal2010,Bernardin2011}
or a one-dimensional hard particle gas
\cite{GerschenfeldBrunetDerrida2010,GerschenfeldDerridaLebowitz2010}. 
In other models the system which ensures the energy transfer from one energy
reservoir to the other has a stochastic dynamics: it may be an Ising spin
system (see for instance among others \cite{LecomteRaczVanWijland2005,ZiaSchmittmann2007,LavrentovichZia2010,
  CorberiEtAl2011}) or a particle with Langevin stochastic dynamics
\cite{DerridaBrunet2005,Visco2006}.  In the latter case the heat exchanges are
described as the work performed by a force including a friction as well as a
random noise component. The latter interpretation of heat has been proposed and
investigated by Sekimoto \cite{Sekimoto1997} for ratchets models, and has been
used again in the interpretation of the Hatano-Sasa identity
\cite{HatanoSasa2001} as well as in the investigation of heat fluctuations in
Brownian transducers \cite{Gomez-MarinSancho2006}.

In the first part of the present work \cite{CornuBauerA}, referred to as
paper I in the sequel, we have investigated the generic statistical properties
for experimentally measurable quantities in the case of thermal contact models
with the following features: the system has a finite number of possible
configurations,  the heat exchanges are
described as changes in the populations of energy levels, and the configurations
evolve under a stochastic master equation with transition rates bound to obey
the modified detailed balance restated in \eqref{ergodicThermo}. We have shown
how the latter relation arises from the existence of an underlying ergodic
deterministic microscopic dynamics which conserves energy.

In the present article, we concentrate on the crudest possible version of a thermal
contact: two heat baths are connected via the smallest possible contact.  So in
the ``thermal contact at an interface'' image, we would replace a surface by a
point. And in the ``heat flow through a piece of material'' image, the bar we
consider has microscopic length, cutting any hope of understanding temperature
gradients.
We believe that the model has some interest with regard to
these interpretations despite its disarming simplicity. Especially for the "thermal contact at an interface", we could take a contact surface made of a collection of spin pairs, the two spins in a pair being coupled as above to allow for heat transfer, but the different pairs being independent. Of course in a more realistic thermal contact, some interactions between pairs, reflecting the two-dimensional geometry, would be present, but there is no obvious reason to believe that these interactions would change qualitatively the physics of heat transfer. For instance, within this model, the law of large numbers allows to quantify how fluctuations of the heat flow are suppressed when the size of the interface goes from microscopic to mesoscopic and macroscopic. We shall not embark on this study in the present paper, but we shall give in a moment a third view of thermal contact for which our model is relevant. 

Before that, we make yet another simplifying assumption on the
way heat is exchanged between the baths: heat bath $1$ (resp. $2$) can flip a dynamical variable $\sigma_1$ (resp. $\sigma_2$) which can take only two distinct values.
The energy $\En(\sigma_1,\sigma_2)$ changes when a contact dynamical
variable is flipped, and assuming an energy conserving dynamics, this means that
some energy comes from, or is given to, the heat bath responsible for the flip.
Without loss of generality, we may assume that the values taken by $\sigma_1$
and $\sigma_2$ belong to $\{-1,1\}$, and we shall use the name \textit{spins}
for $\sigma_1$ and $\sigma_2$. In general, the energy $\En(\sigma_1,\sigma_2)$
for the contact dynamical variables could take $4$ distinct values, but we even
concentrate on the case $\En(\sigma_1,\sigma_2)= \frac{1}{2}(1-\sigma_1\sigma_2)
\gapE$, where $\gapE>0$ is the energy gap. In the language of spins, this means
the absence of external magnetic fields. More abstractly, it implies a twofold
symmetry.

We shall concentrate on a description of the time evolution of interface states
$(\sigma_1,\sigma_2)$ by a Markov process, i.e. by a probabilistic description.
But other approaches are possible. For instance, as explained in paper I, the
motivation for our choices of transition rates comes from invoking an ergodicity
argument for a deterministic discrete time evolution of the compound ``heat bath
$1$ plus interface plus heat bath $2$''. 

As the toy model has only $4$ states, solving it can be reduced in some sense to
the diagonalization of a $4\times 4$ matrix, and the twofold symmetry of the
energy functional allows to reduce this task to the diagonalization of a pair of
$2\times 2$ matrices. However, this is not the end of the story, and this takes
us to the third interpretation of the model. 

The third view of a thermal contact mentioned above is not microscopic but
mesoscopic. We regard $\sigma_1$ and $\sigma_2$ as some relevant collective
variables and $\En(\sigma_1,\sigma_2)$ as an effective energy. Then the system
can be viewed, and analyzed, as a thermal machine. That is, our crude model
keeps track of one (and maybe only one) interesting feature: the system can make
cycles. Consider a sequence of flips in the interface, starting from the state
$(\sigma_1,\sigma_2)$:
\[\begin{array}{ccc} (\sigma_1,\sigma_2) & \rightarrow & (-\sigma_1,\sigma_2) \\
\uparrow  & & \downarrow \\
(\sigma_1,-\sigma_2) & \leftarrow & (-\sigma_1,-\sigma_2)
\end{array}\]
after which the interface has returned to its original state. Writing
$E_1$, $E_2$ for the initial energies in heat bath $1$ and $2$, the sequence
translates into  
\[\begin{array}{ccc} (E_1,E_2) & \rightarrow & (E_1-\sigma_1\sigma_2\gapE,E_2)\\
  & & \downarrow \\
(E_1-2\sigma_1\sigma_2\gapE,E_2+\sigma_1\sigma_2\gapE) & \leftarrow &
(E_1-\sigma_1\sigma_2\gapE,E_2+\sigma_1\sigma_2\gapE) \\
\downarrow & & \\
(E_1-2\sigma_1\sigma_2\gapE,E_2+2\sigma_1\sigma_2\gapE)
\end{array}\]
i.e. an amount of heat $2\sigma_1\sigma_2\gapE$ has been transferred from heat
bath $1$ to heat bath $2$. We use the term mesocopic (as opposed to macroscopic)
for two reasons: first the dynamics at the interface is not deterministic,
i.e. knowing $(\sigma_1,\sigma_2)$ at some time does not allow to know its value
in the future and second (this is somehow a consequence though) there may be
portions of time in which the net flow of heat is from the cold bath to the hot
bath.

When seen in this light, the model is already more interesting: the time
evolution of the heat bath energies is a random walk in continuous
time\footnote{And in two spatial dimensions. The sum of the two coordinates can
  only take a finite number of values, but the waiting times and some
  correlations prevent from concentrating only on one component.}, a subject
known to lead to a number of nontrivial mathematical problems, some of them
having a direct physical relevance. And indeed we shall concentrate mainly on
the physics, with the aim of performing detailed analytical computations.

Another interest of our specific solvable model is that it plays the role of a
pedagogical example where the general statements are made very explicit. 
For
instance, though the fluctuation relations entail a constraint upon large
deviation functions, they do not allow to determine them. The analytical
calculation of the large deviation functions may provide a deeper understanding
in the information which they contain.
We shall see that, within the model,
the computation of large deviation functions for the energy variations in the
baths can be remarkably simple or tricky, depending on the kind of techniques
one uses. 

Finally, let us note that in the absence of any general
framework for out-of-equilibrium statistical mechanics, the formul\ae\  obtained for
the solvable model can give a flavor of the physical effects. For instance, the
time scales of the microscopic dynamics, which do not show off in equilibrium
averages, play a role in out-of-equilibrium properties even at the macroscopic
level. Moreover the model can be considered in the limit where the temperature
of the cold thermostat vanishes ; then the strict dissipation of energy towards
the zero-temperature gives rise to specific phenomena.

\subsection{Contents of the paper}
\label{ContentsSec}

The results of the explicit analytical calculations for the solvable model where the system is reduced to two spins are the following.

In the case where the spin system involves only two spins, the transition rate
are determined by the modified detailed balance \eqref{ergodicThermo} up to the
typical inverse times $\nu_a$ of spin flips by each thermal bath $a$,
characterized by its temperature $T_a$. For an Ising interaction between the
two spins, the transition rates for the energy exchanges with one bath take a
form similar to that introduced by Glauber \cite{Glauber1963} in his
investigation of the time-dependent statistics of the Ising chain in contact
with a single thermal bath. Most of the time, our results will hold
whatever the values of $T_1$ and $T_2$ are. However, it is sometimes convenient
to know in which direction heat flows from one reservoir to the other on the average, and then
we shall always assume that $T_1 \leq T_2$. By symmetry, this induces no loss of
generality anyway: the results for $T_1 \geq T_2$ can be retrieved by permuting
$T_1$ with $T_2$ and $\nu_1$ with $\nu_2$.

The Non-Equilibrium Stationary State (NESS) of the model happens to have a very
specific property (subsection \ref{StationaryState}) : since the transition rates are invariant under the
simultaneous flips of both spins, the configuration probability distribution in
the NESS coincides with an equilibrium canonical distribution at some inverse
temperature $\betaeff$.

The linear and non-linear static responses are explicitly calculated (subsections \ref{secLinearStatic} and \ref{secNonLinearStatic}). The expressions for the generalized heat capacities   involve not only the temperatures of the energy reservoirs $a=1,2$ but also  the typical inverse time scales $\nu_a$'s of the heat
exchange dynamics with each reservoir. The $\nu_a$'s are also
called kinetic parameters in the following.  In the vicinity of equilibrium the
mean heat current is proportional to the difference $T_2-T_1$ between the bath
temperatures; then a linear thermal conductivity can be defined. When the
system is far from equilibrium the mean heat current is a bounded function of
the thermostat temperatures (saturation phenomenon); one can introduce a
non-linear thermal conductivity which vanishes in the limit where the relative
temperature difference goes to infinity.  The expression of the  housekeeping entropy flow is given, and the
excess mean heats, which are defined   in terms of the measurable averages of the
cumulative heats \cite{KomatsuETAL2011},  are explicitly calculated, for the static response protocol, from the  average heat amounts  received by the system from each bath during  a finite time $t$, and which are determined in subsection
 \ref{Excessheat}.

The joint probability distribution for received cumulative heats $\Heat_1$ and
$\Heat_2$ is determined at any finite time and for any initial distribution
probability through a generating function method (see section
\ref{JointProbsection}). Other distribution probabilities are then derived from
its expression \eqref{ProbDetailee}, and the explicit results are summarized in
subsection \ref{ExplicitProbabilities}. The results are given in terms of two
integrals in the complex plane. The system  obeys the finite-time
symmetry \eqref{OppositePQ1PQ2Probcan} enforced by the modified detailed balance for the ratio of the
probabilities to measure some given heat amounts $\Heat_1$ and $\Heat_2$ or
their opposite values when the system is initially prepared in an equilibrium
state. But it also satisfies another finite-time symmetry specific to the model
for the ratio of the same probabilities when the system has any initial
distribution probability.  
The latter fluctuation relation \eqref{OppositePQ1PQ2Prob0} is more subtle as it involves the initial probability distribution for the product of the spins (or equivalently for the energy of the spin pair).

The cumulants for the cumulative heat $\Heat_2$ are studied in section \ref{LongTimeCumulantsExp} from the  characteristic function of the probability density for  $\Heat_2$.  The relation between the characteristic function of a probability density  $\Probdist({\Heat};t)$ with the generating function for the probability function $\Prob(\Heat;t)$ when the variable $\Heat$ can take only discrete values is recalled in subsubsection \ref{CharFunction}. The explicit formul\ae\ for the first four cumulants per unit time in the infinite-time limit
are given in \eqref{longtimekappa1a4}. 
Even at equilibrium the cumulants are not those of a Gaussian.

The large deviation function for the cumulative heat current $\Heat_2/t$ is
calculated by three different methods (section \ref{LargeDeviationFunction}):
from the Gärtner-Ellis theorem (subsection \ref{GartnerEllisLD}), from a
saddle-point method (subsection \ref{DetailsCol}) and from Laplace's method on a
discrete sum (subsection \ref{LaplacesMethod}). The second and third methods
rely explicitly on the discrete nature of heat exchanges in the model and on an
ad hoc definition of large deviation functions discussed in paper I, but they allow to compute
subdominant contributions as well. The first method is straightforward, one just
has to check that the general applicability hypotheses (recalled in detail
below) are fulfilled and this is easy in our case. The third method is also
simple because it deals with a sum of nonnegative terms, so no compensation is
possible. The saddle point method however is remarkably tricky in our case, for
reasons that we shall detail below.  The expressions in terms of various
parameter sets are given in \eqref{fjadAB} and \eqref{fjadpppm}.  In order to
readily obtain the large deviation function in the case where the temperature of
the colder bath vanishes, its expressions for positive and negative currents are
explicitly distinguished in \eqref{fjneg}-\eqref{fjpos}.

The limit where the kinetic parameter of one thermostat becomes infinitely large
with respect to the kinetic parameter of the other thermostat is studied in
section \ref{DependenceTimeScales}.  In this limit the stationary distribution
of the spins is the equilibrium canonical probability at the temperature of the
``fast'' heat bath while the typical inverse time scale in the mean
instantaneous heat current is the kinetic parameter $\nu_\textrm{slow}$ of the
``slow'' heat bath. The probability distribution for the heat amount received
from the slow thermostat, $\Heat_\textrm{slow}$, at any finite time $t$ is that
of an asymmetric random walk with the inverse time scale $\nu_\textrm{slow}$. As
a consequence the probability distribution for $\Heat_\textrm{slow}$ obeys a
fluctuation relation at any finite time (see \eqref{FRHeat1}).  The probability
distributions of $\sigma_1\sigma_2$ and $\Heat_\textrm{slow}$ are independent
from each other, and this mean-field property is interpreted as a kinetic effect
in the considered limit. The very simple forms of the infinite-time cumulants per
unit time are given. The long-time distribution of the cumulative heat current is
exhibited : it vanishes exponentially fast over a time-scale given by the
inverse of the large-deviation function \eqref{fRW} with an amplitude which is
explicitly calculated.

In the limit where the temperature of the colder thermostat vanishes (section
\ref{T1zero}) the microreversibility is broken, but the system still reaches a
stationary state where all configurations have a non-vanishing weight, because
the Markov matrix is still irreducible. The large deviation function is
expressed in \eqref{fjnegposT1zero}. In the limit  where the kinetic parameter of one thermostat
 becomes infinitely large with respect to the kinetic parameter of the other thermostat, 
the probability distribution for the heat amount
$\pm\Heat _\text{slow}$, with sign $-$ ($+$) if the slow thermostat is the cold (hot) one,  becomes a Poisson process at any finite time $t$,
because the zero-temperature thermostat can only absorb energy (strict dissipation
towards the zero-temperature bath).  Again the very simple forms of the
infinite-time cumulants per unit time are given, as well as the large deviation
function \eqref{LDPoisson}.

The last section is devoted to a probabilistic study of the system seen as a
mesoscopic engineless thermal wheel, with an average heat flow from the hot reservoir to
the cold reservoir, but also fluctuations around the average which we try to
quantify. We compute the probability for the thermal machine to work backwards,
and the law of the fluctuations of the time it takes to the machine to do one
cycle. We present the argument for a system slightly more general than the
two-spin system, because the computations and their meaning are more transparent
this way, and then apply the formul\ae\ to the two-spin system.

\section{Model}

\label{sectionModel}

The physical system we deal with in this article is a toy model of thermal
contact, consisting of two heat baths, generically at different temperatures, put
indirectly in contact via a small subsystem made of two interacting Ising
spins $\sigma_1$ and $\sigma_2$. Each spin  $\sigma_a$, $a=1,2$ is   in
contact with a single bath denoted by $a$.
We aim at a statistical description, where the details of what happens in the
heat baths is not observed, but only the evolution of the two spins, i.e. of the configuration 
$\C\equiv (\sigma_1,\sigma_2)$.  We assume that this evolution is described by a
Markov process (in continuous time) with transition rate $(\C'\vert \Trans \vert
\C)$ from configuration $\C$ to configuration $\C'$.

As the system is out of equilibrium, the form of $\Trans$ is not a direct
consequence of known physical laws, and it is unclear whether a nature-given
preferred choice exists. So we start with a purely technical and down to earth
description of our choice for the transition rates, that we shall use for all
later explicit computations. The general principles
and steps that guided us to the modified detailed balance that the transition rates must obey  have been given in paper I. The main ideas are the following.

\subsection{Constraints upon transition rates arising from microscopic  discrete ergodic energy-conserving dynamics}

As usual, we view a heat bath as an ideal limit of some large but finite system.
So the system we describe is obtained via a limiting procedure from a large
system made of two large parts and a small one, which is reduced to
the two Ising spins $\sigma_1$ and $\sigma_2$, each one directly in contact
with one of the large parts.

We expect that in this limit many details become irrelevant, so we assume for
the sake of the argument that the degrees of freedom in the large parts are
discrete.

As in classical statistical mechanics, we take the viewpoint that the
statistical description of $\sigma_1,\sigma_2$ is an effective mesoscopic
description arising from a deterministic, energy-conserving dynamics for the
whole system. With
discrete variables, there is no general definition of time reversal
invariance, but we impose that the dynamics is ergodic. 

We also want the dynamics to reflect the fact that the two large parts interact
only indirectly: there is an interaction energy $\En(\sigma_1,\sigma_2)$ between
the two spins and the spin $\sigma_a$ is flipped thanks to energy exchanges with
the large part $a$ ($a=1,2$). Defining the operator $\F_a$ as the operator
flipping the spin $\sigma_a$ while leaving the other spin unchanged (e.g
$\F_1(\sigma_1,\sigma_2) =(-\sigma_1,\sigma_2)$), in this process the energy of
the large part $a$ is changed from $E_a$ to $E'_a$ according to the energy
conservation law \be
 \label{conserE}
E'_a-E_a=
\begin{cases}
-\left[\En(\C')-\En(\C)\right]   &\textrm{if}\quad  \C'=\F_a\C \\
0 &\textrm{otherwise},
\end{cases}
 \ee
while the energy of the other large part is unchanged. 

As shown in paper I, when the large parts are described at a statistical level
and in a transient regime where the large parts are described in the
thermodynamic limit, the transition rate $(\C'\vert \Trans \vert \C)$ from
configuration $\C$ to configuration $\C'$ obeys
three constraints: first the graph associated with the transition rates is
connected; second there is microscopic reversibility for any couple of
configurations $(\C,\C')$, 
\be
\label{MicroRevCond}
(\C'\vert \Trans \vert \C)\not=0 \qquad \Leftrightarrow \qquad (\C\vert \Trans
\vert \C')\not=0;
\ee 
third  the ratio of transition rates obeys the so-called modified detailed balance (MDB),
\be
\label{ergodicThermo}
\textrm{for $\C'=\F_a\C$}\quad
\frac{(\C'\vert \Trans \vert \C)}
{(\C\vert \Trans \vert \C')}=e^{-\beta_a\left[\En(\C')-\En(\C)\right]}.
\ee
We remind the reader that the latter relation is also referred to in the litterature as the ``generalized detailed balance''.

\subsection{Determination of transition rates}

The transition rates are non-zero only if the initial and final configurations $\C$ and $\C'$ differ only by the flip of one spin :
$(\C'\vert \Trans\vert \C)=0$
unless either $\C'=\F_1\C$ or $\C'=\F_2\C$.
Since $\sigma_a$ can take only the two values $+1$ and $-1$, the transition rate where $\sigma_a$ is flipped takes the generic form
\be
\label{TransStruct}
(\F_a\C\vert \Trans \vert \C)=\frac{\nu_a(\sigma_b)}{2}
\left[1-\sigma_a \Gamma_a(\sigma_b)\right].
\ee
The four parameters $\nu_a,\Gamma_a$, $a=1,2$, are a priori arbitrary except that the
$\nu$'s are $>0$ and the $\Gamma$'s are  of absolute value
$\leq 1$. 

Taking for simplicity an  interaction energy between the spins
 \be
 \label{energyExp}
 \En(\sigma_1,\sigma_2)=\frac{1-\sigma_1\sigma_2}{2}\gapE, \ee where $\gapE>0$
 is the energy gap between the two energy levels, one gets from the modified
 detailed balance in the form \eqref{ergodicThermo} that \be \textrm{for
   $\C'=\F_a\C$}\quad\frac{(\C'\vert \Trans \vert \C)} {(\C\vert \Trans \vert
   \C')}=e^{-\sigma_1\sigma_2\beta_a\gapE}, \ee a condition similar to the one
 obtained by Glauber \cite{Glauber1963} in the equilibrium case.  As
 $e^{2x}=\frac{1+\tanh x}{1-\tanh x}$ the generic form \eqref{TransStruct} of
 $(\F_a\C\vert \Trans \vert \C)$ has to satisfy \be (\F_a\C\vert \Trans\vert
 \C)=\frac{\nu_a(\sigma_b)}{2} \left[1-\sigma_1 \sigma_2 \gamma_a\right] \ee
 with \be \gamma_a\equiv \tanh\left(\beta_a\frac{\gapE}{2}\right).  \ee If
 $\beta_1$ and $\beta_2$ are finite, $0\leq \gamma_1<1$ and $0 \leq \gamma_2<1$,
 and the microscopic reversibility condition \eqref{MicroRevCond} is also
 satisfied. Without loss of generality we could, and will sometimes, assume that
 $T_1 \leq T_2$. Then $\gamma_1 \geq \gamma_2$.
 
For the sake of simplicity, in the following we assume that $\nu_a$ depends only
on the properties  of the thermostat and not on the value of $\sigma_b$. (This choice enforces the equality between the transition rate from  
$(\sigma_1,\sigma_2)$ and that from  $(-\sigma_1,-\sigma_2)$, which are two configurations with the same energy.)  Apart
from simplicity, we have no convincing argument that this should be THE nature-given
preferred choice. Anyway, we write
\be
\label{Transexp}
 (\F_a\C\vert \Trans\vert \C)=\frac{\nu_a}{2}
\left[1-\sigma_1 \sigma_2 \gamma_a\right].
\ee
This ends the argument explaining our choice of transition rates and gives a
physical interpretation of the parameters: $\gamma_a$ is formed with the
energy scale in the two-spin system and the temperature of  bath $a$, while
$\nu_a $ describes a rate at which bath $a$ attempts to flip spin
$\sigma_a$. 

We notice that, though the transition rate expressions have been derived from hypotheses implying the microscopic reversibility
\eqref{MicroRevCond}, these expressions still make sense if $\beta_2<\beta_1=+\infty$. (The limit $\beta_1\to +\infty$ where microscopic reversibility is broken is discussed in section \ref{T1zero}.)

Moreover, even if $\beta_1=+\infty$, the Markov matrix $\M$ defined by
\be
 \label{defMarokovM}
(\C'\vert \M\vert \C)=
\begin{cases}
(\C'\vert \Trans\vert \C)   &\textrm{if}\quad  \C'\not=\C \\
-\sum_{\C''}(\C''\vert \Trans\vert \C)  &\textrm{if}\quad  \C'=\C
\end{cases}
 \ee
is irreducible, namely
 any configuration $\C'$ can be reached by a succession of jumps with non-zero transition rates from any configuration $\C$.

\section{Non Equilibrium Stationary State (NESS) as a canonical distribution with an effective temperature}

\label{NESSsection}

\subsection{Stationary state distribution}
\label{StationaryState}

The master equation which rules the evolution of the probability $\Prob(\C;t)$ can be written in terms of the Markov matrix $\M$ defined in 
\eqref{defMarokovM} as
\be
\label{MarkovEquation}
\frac{d \Prob(\C;t)}{dt}=
\sum_{\C'}(\C\vert \M\vert \C') \Prob(\C';t).
\ee
In the basis where the probability $\Prob(\sigma_1,\sigma_2;t)$ is represented by the column vector
\be
\label{basis}
\vert\Prob(t))=\begin{pmatrix}
\Prob(++;t) \\
\Prob(--;t) \\
\Prob(+-;t) \\
\Prob(-+;t)
\end{pmatrix}
\ee
the matrix $\M$ takes the form
\be
\label{matrixM}
\M=\frac{\nu_1+\nu_2}{2}
\begin{pmatrix}
-1+\gammaeff & 0 & \nuad_2 (1+\gamma_2)& \nuad_1(1+\gamma_1) \\
0 & -1+\gammaeff &\nuad_1(1+\gamma_1)  & \nuad_2 (1+\gamma_2) \\
\nuad_2 (1-\gamma_2) & \nuad_1(1-\gamma_1) &-1 -\gammaeff  &0 \\
\nuad_1(1-\gamma_1) & \nuad_2 (1-\gamma_2) & 0 &-1-\gammaeff 
\end{pmatrix}.
\ee
In the latter equation we have introduced the dimensionless inverse time scales
\be
\label{defnuad}
\nuad_a=\frac{\nu_a}{\nu_1+\nu_2}
\quad\textrm{for $a=\{1,2\}$},
\ee
and we have set
\be
\label{gammaeff}
\gammaeff=\nuad_1 \gamma_1+\nuad_2 \gamma_2.
\ee

The Markov matrix $\M$ is irreducible (even if $\gamma_1=1$, namely $T_1\to0$):
for any pair of configurations $\C$ and $\C'$, there exists a succession of spin
flips, with non-zero transition rates, which allows to make the system evolve
from  $\C$ to $\C'$. Henceforth, according to the Perron-Frobenius theorem there
exists a single stationary state distribution $\Probst(\C)$ and it is nonzero for every configuration $\C$.

Moreover,  since the system is made of two discrete variables which can take only the values $\pm1$ and since the transition rates are invariant under the simultaneous flips
 $\sigma_1\rightarrow -\sigma_1$ and $\sigma_2\rightarrow -\sigma_2$, the stationary distribution $\Probst(\sigma_1,\sigma_2)$ takes the form 
$\Probst(\sigma_1,\sigma_2)=a+d\sigma_1\sigma_2$. Indeed, the generic form of 
$\Prob(\sigma_1,\sigma_2)$ reads 
$\Prob(\sigma_1,\sigma_2)=a+b \sigma_1+c\sigma_2 +d \sigma_1\sigma_2$. On the other hand,  the invariance of the transition rates under the simultaneous flips
 $\sigma_1\rightarrow -\sigma_1$ and $\sigma_2\rightarrow -\sigma_2$ entails that if $a+b \sigma_1+c\sigma_2 +d \sigma_1\sigma_2$  is a stationary solution,  $a-b \sigma_1-c\sigma_2 +d \sigma_1\sigma_2$ is also a stationary solution. But, since $\M$ is irreducible, there is only one stationary solution, so that $b=c=0$.
By solving explicitly the master equation \eqref{MarkovEquation} and using the normalization of a probability distribution, the stationary solution proves to be
\be
\label{structPst}
\Probst(\sigma_1,\sigma_2)=\frac{1}{4}[1+\gammaeff \sigma_1\sigma_2]
\ee
where $\gammaeff$ is defined in \eqref{gammaeff}.

The stationary distribution of the model has the following remarkable property: it coincides with some equilibrium distribution. More precisely, the stationary state distribution is equal to 
 the canonical state distribution at the effective inverse temperature $\betaeff$ 
 \be
 \label{Probcaneff}
\Probst(\sigma_1,\sigma_2)=\Probcan^\betaeff(\sigma_1,\sigma_2),
\ee
where $\betaeff$  is determined by the relation
\be
\gammaeff=\tanh\left(\betaeff\frac{\gapE}{2}\right),
\ee
and 
\be
\label{expProbcan}
\Probcan^{\beta}(\C)=\frac{e^{-\beta \En(\C)}}{Z(\beta)}, \ee where $Z(\beta)$
is the canonical partition function at the inverse temperature $\beta$,
$Z(\beta)=\sum_{\C} e^{-\beta \En(\C)}$.  We notice that the canonical form for
the state distribution implies that $\betaeff$ obeys the canonical ensemble
relation which is equivalent to the definition of the inverse temperature in the
microcanonical ensemble, namely 
\be 
\betaeff=\frac{\partial
  \SG\left[\Probst\right]}{\partial \Espst{\En}}, 
\ee 
where $\Espst{\En}\equiv
\sum_{\C} \En(\C) \Probst(\C)$ is the stationary mean value of the energy and
$\SG\left[\Probst\right]$ is the value of the dimensionless Shannon-Gibbs
entropy in the stationary state. The dimensionless Shannon-Gibbs entropy (where
the Boltzmann constant is set equal to 1) is defined from the configuration
probability distribution $\Prob(\C;t)$ as \be \SG\left[\Prob(t)\right]\equiv-
\sum_{\C}\Prob(\C;t)\ln\Prob(\C;t).  \ee Its evolution has been recalled in
paper I.

\subsection{Linear static response to a variation of some external parameter}
\label{secLinearStatic}

In the present section we consider the static linear response of some observable ${\cal O}$ to a change of some external parameter, namely the inverse temperature $\beta_a$  or the typical inverse time scale $\nu_a$ of   bath $a$, with $a=1,2$. 

In the  protocols for the study of static linear response, the system is prepared in some stationary state at time $t_0=0^-$ and the external parameters are instantaneously changed by infinitesimal amounts at time $t=0$. Then, in the infinite time limit,  the system reaches another stationary state corresponding to the new values of the external parameters.

\subsubsection{Relation with static correlations for a  ``canonical'' NESS}

Since the nonequilibrium stationary distribution  given by \eqref{structPst} involves only one parameter, namely $\betaeff$, 
the linear response coefficient $\partial \Espst{\Oobs}/\partial g_\text{ext}$ for the mean value of an observable $\Oobs$ in the stationary distribution when some external parameter $g_\text{ext}$ is varied is proportional to $\partial \Espst{\Oobs}/\partial \betaeff$, namely
$\partial \Espst{\Oobs}/\partial g_\text{ext}=\left(\partial\betaeff/\partial g_\text{ext}\right)\times
\left(\partial \Espst{\Oobs}/\partial \betaeff\right)$. 
Moreover, by virtue of \eqref{Probcaneff}, the stationary distribution  is the canonical distribution at the inverse temperature
$\betaeff$. Henceforth the  coefficient $\partial \Espst{\Oobs}/\partial \betaeff$ is merely  opposite to the correlation between $\Oobs$ and the energy $\En$ according to the canonical equilibrium identity 
\be
\frac{\partial \Espcan{\Oobs}^{\betaeff}}{\partial \betaeff}=-\left[\Espcan{\Oobs\En}^{\betaeff}-\Espcan{\Oobs}^{\betaeff}\Espcan{\En}^{\betaeff}\right],
\ee
where $\Espcan{\Oobs}^{\betaeff}$ denotes an average with respect to the canonical distribution $\Probcan^\betaeff$.
As a result, the  relation valid for responses to the variation of any external parameter in the nonequilibrium stationary state reads
\be
\frac{\partial \Espst{\Oobs}}{\partial g_\text{ext}}=-\frac{\partial\betaeff}{\partial g_\text{ext}}
\left[\Espst{\Oobs\En}-\Espst{\Oobs}\Espst{\En}\right].
\ee

\subsubsection{Dependance of the mean energy upon the time scales of the microscopic dynamics}

The main difference between the response of the mean energy in non-equilibrium and equilibrium states arises for the response to a variation of the time scales of the microscopic dynamics which rules the heat exchanges with the baths. When $\beta_1=\beta_2$ the equilibrium mean energy $\Espeq{\En}=\Espcan{\En}^{\beta_1}$ depends only on the thermodynamic temperature common to both baths. On the contrary, in the non-equilibrium case the stationary mean energy $\Espst{\En}$ does also depend on both inverse time scales $\nu_1$ and $\nu_2$. 
Indeed, since the stationary probability corresponds to the effective canonical distribution \eqref{Probcaneff}, the stationary mean energy reads
\be
\label{EspstEn}
\Espst{\En}=\left(1-\gammaeff\right)\frac{\gapE}{2}=\left(1-\nuad_1\gamma_1-\nuad_2\gamma_2\right)\frac{\gapE}{2}.
\ee 
Changing $\nu_a$  means changing the physical connection between  thermal bath $a$ and the spin system.
The linear response of the stationary energy associated with a variation of the inverse time scale $\nu_a$ is determined by the  coefficient
\be
\frac{\partial \Espst{\En}}{\partial \nu_a}=-\frac{\nu_b}{(\nu_1+\nu_2)^2}\, \gamma_a \frac{\gapE}{2}
\quad\textrm{for $\{a,b\}=\{1,2\}$}.
\ee

\subsubsection{Stationary mean energy and generalized heat capacities}
\label{HeatCapacity}

The heat capacity $C_\text{eq}$ is a measurable quantity defined as the ratio
\be  
C_\text{eq}(T)=\frac{\Esp{\delta\Heat}}{dT},
\ee
where $\Esp{\delta\Heat}$ is the mean heat amount received by the system in transformations which involve only heat transfers and make the system go from an equilibrium state at temperature $T$ to another equilibrium state at temperature $T+dT$, while all other thermodynamic parameters which determine the equilibrium state are kept constant.
( $\Esp{\delta \Heat}=\lim_{t\to+\infty}\Esp{\Heat}_t$ in the protocol mentioned in the introduction of the section.)
According to the energy conservation,  $\Esp{\delta\Heat}=\Espeq{\En}^{T+dT}-\Espeq{\En}^{T}$
and the heat capacity is related to a partial derivative of the equilibrium mean energy,
\be
\label{expCeq}
C_\text{eq}(T)=\frac{\partial \Espeq{\En}}{\partial T}
=-\beta^2\frac{\partial \Espeq{\En}}{\partial \beta}.
\ee

\par\medskip
When the system is in a stationary non-equilibrium state induced by thermal contact with two heat reservoirs at respective temperatures $T_1$ and $T_2$, we can introduce measurable heat capacities by similar definitions. When the temperature $T_1$ of  thermal bath $1$ is changed by $dT_1$, while the temperature $T_2$ of  thermal bath $2$ is kept fixed, and when the system evolves from a stationary state to another one only by heat transfers, then the generalized heat capacity $C_\text{st}^{[1]}$  is defined as
\be
C_\text{st}^{[1]}(T_1,T_2)=\frac{ \Esp{\delta\left(\Heat_1+\Heat_2\right)}}{d T_1}.
\ee
According to conservation energy, $\delta\Esp{\Heat_1+\Heat_2}=\Espst{\En}^{T_1+dT_1,T_2}-\Espst{\En}^{T_1,T_2}$
and the heat capacity is related to a partial derivative of the  stationary mean energy
\be
\label{defheatcapacity}
C_\text{st}^{[1]}(T_1,T_2)=\left.\frac{\partial \Espst{\En}^{T_1,T_2}}{\partial T_1}\right\vert_{T_2}.
\ee

In the present model the expression \eqref{EspstEn} of the stationary mean energy takes  the very specific  form
\be
\label{EspstEnDecomp}
\Espst{\En}=\nuad_1\Espeq{\En}^{T_1}+\nuad_2 \Espeq{\En}^{T_2}.
\ee
Indeed the relation $\nuad_1+\nuad_2=1$ and the expression of the equilibrium mean energy  at the  inverse temperature $\beta$,
\be
\label{EspeqEn}
\Espeq{\En}^{T}=\left(1-\gamma\right)\frac{\gapE}{2}\quad\textrm{when}\quad
\beta_1=\beta_2=\beta,
\ee
allow to rewrite the mean energy expression \eqref{EspstEn} in the non-equilibrium stationary state in the form \eqref{EspstEnDecomp}.
By virtue of the specific decomposition \eqref{EspstEnDecomp} of the mean energy,   the heat capacities $C_\text{st}^{[a]}(T_1,T_2)$'s  read
\be
\label{expCa}
C_\text{st}^{[a]}(T_1,T_2)=\nuad_a C_\text{eq}(T_a)\quad\textrm{with}\quad a=\{1,2\},
\ee
where, according to the relation \eqref{expCeq} and the expression \eqref{EspeqEn} of $\Espeq{\En}^T$,
\be
C_\text{eq}(T_a)= \left[1-\tanh^2\left(\frac{\beta_a\gapE}{2}\right)\right]\left(\frac{\beta_a\gapE}{2}\right)^2.
\ee

More generally, when  the temperatures $T_1$ and $T_2$ of both thermostats are varied independently
\be
\Espst{\En}^{T_1+dT_1, T_2+dT_2}-\Espst{\En}^{T_1, T_2}=
C_\text{st}^{[1]}(T_1,T_2)dT_1+C_\text{st}^{[2]}(T_1,T_2)dT_2.
\ee
If $T_1$ and $T_2$ are increased by the same infinitesimal quantity $dT$ the corresponding heat capacity, defined as
$C_\text{st}(T_1,T_2)\equiv \delta \Esp{\Heat_1+\Heat_2}/d T$ is equal to the sum $C_\text{st}^{[1]}(T_1,T_2)+C_\text{st}^{[2]}(T_1,T_2)$. For the present model
$
\label{expC}
C_\text{st}(T_1,T_2)=\nuad_1 C_\text{eq}(T_1)+\nuad_2 C_\text{eq}(T_2).
$
In the limit where $T_1=T_2=T$, by virtue of the relation $\nuad_1+\nuad_2=1$, we retrieve the equilibrium heat capacity $C_\text{eq}(T)$, as it should be.

\subsubsection{Stationary heat current and linear thermal conductivity}
\label{sectiongeneralizedconductivity}

The instantaneous heat current $\jinst_a(\C)$ received from  heat bath $a$  when the system jumps out of the configuration $\C$ has been defined in paper I as
\be
\label{defjinstHeat}
\jinst_a(\C)\equiv \jinst_{\delta q_a}(\C)\equiv
\sum_{\C'}(\C'\vert \Trans\vert \C) \,\delta\Heatm_a(\C'\leftarrow \C),
\ee
  where $\delta\Heatm_a(\C'\leftarrow \C)$ is the heat received from thermal bath $a$ when the system evolves from  configuration $\C$  to configuration $\C'=\F_a\C$, where $\F_a$ is the flip caused by thermal bath $a$, namely 
\be
\label{defdeltaq}
\begin{cases}
\delta\Heatm_a(\C'\leftarrow\, \C)=\left[\En(\C')-\En(\C)\right]  &\textrm{if}\quad  \C'=\F_a\C \\
 \delta\Heatm_a(\C'\leftarrow\, \C)=0 &\textrm{otherwise}.
  \end{cases}
\ee
 In the present model
  $\jinst_2(\sigma_1,\sigma_2)
 =\left[\En(\sigma_1,-\sigma_2)-\En(\sigma_1,\sigma_2)\right](\sigma_1,-\sigma_2\vert \Trans\vert \sigma_1,\sigma_2)=\nu_2\left[\sigma_1\sigma_2-\gamma_2\right](\gapE/2)$.

In the stationary state the mean energy is constant so that the mean currents received from both baths cancel, $\Espst{\jinst_1}+\Espst{\jinst_2}=0$. 
For the stationary state probability distribution \eqref{structPst}, one has $\Espst{\sigma_1\sigma_2}=\gammaeff=\nuad_1\gamma_1+\nuad_2\gamma_2$ and
\be
\label{jinstexp}
\Espst{\jinst_2}=\nuad_1\nuad_2\left(\gamma_1-\gamma_2\right)\frac{(\nu_1+\nu_2)\gapE}{2},
\ee
where $\gamma_1-\gamma_2$ may be rewritten as
\be
\gamma_1-\gamma_2=\tanh\left(\frac{(\beta_1-\beta_2)\gapE}{2}\right)
\left[1-\tanh\left(\frac{\beta_1\gapE}{2}\right)\tanh\left(\frac{\beta_2\gapE}{2}\right)\right].
\ee
When $T_1 \leq T_2$, $\Espst{\jinst_2}\geq 0$, as it should : the mean heat
current flows from the hot bath to the cold bath. 
Note that $\Espst{\jinst_2}$ is a bounded function of $T_1$ and $T_2$. Thus, 
 in the generic case $\Espst{\jinst_2}$ is not proportional  to the bath temperatures difference $T_2-T_1$. As for any system, the  linear dependence upon   $T_2-T_1$ (or $\beta_1-\beta_2$) appears  in the limit where $(\beta_1-\beta_2)\gapE \ll 1$. In the high temperature regime where both $\beta_1\gapE\ll 1$ and $\beta_2\gapE\ll 1$, the condition $(\beta_1-\beta_2)\gapE \ll 1$ is satisfied and $\Espst{\jinst_2}$ is proportional to $T_2-T_1$.

When $\beta_1=\beta_2$ the system is at equilibrium and $\Espeq{\jinst_2}=0$. Moreover, as shown in paper I, the partial derivatives of the current obey the generic symmetry 
\be
\label{oppositePD}
\left.\frac{\partial \Espst{\jinst_2}}{\partial \beta_1}\right\vert_{\beta_2}(\beta_1=\beta,\beta_2=\beta)
=-\left.\frac{\partial \Espst{\jinst_2}}{\partial \beta_2}\right\vert_{\beta_1}(\beta_1=\beta,\beta_2=\beta).
\ee
This property  can also be checked from  the expression \eqref{jinstexp} of $\Espst{\jinst_2}$. It
entails that
\be
\Espst{\jinst_2}\underset{(T_1,T_2)\to(T,T)}{\sim}
(T_2-T_1)\left.\frac{\partial \Espst{\jinst_2}}{\partial T_2}\right\vert_{T_1}(T,T).
\ee
In other words, when $T_1$ and $T_2$ independently  tend to the same value $T$,  at first order in the independent variables $T_1-T$ and $T_2-T$ the ratio $\Espst{\jinst_2}/(T_2-T_1)$ depends on $T$ but is independent of the ways $T_1-T$ and 
$T_2-T$ vanish.

As a consequence, for  a non-equilibrium stationary state near equilibrium, namely when the temperature difference between the thermostats is  such that $\left(\beta_1-\beta_2\right)\gapE\ll 1$, one can define the thermal conductivity as
\be
\label{defThermalCond}
\kappa_\text{th}\equiv
\lim_{(T_1,T_2)\to(T,T)}\frac{\Espst{\jinst_2}}{T_2-T_1}=\left.\frac{\partial \Espst{\jinst_2}}{\partial T_2}\right\vert_{T_1}(T,T).
\ee
From \eqref{jinstexp} we get the expression for the thermal conductivity,
\be
\label{expThermalCond}
\kappa_\text{th}=
\frac{\nu_1\nu_2}{\nu_1+\nu_2}
\left[1-\tanh^2\left(\frac{\beta\gapE}{2}\right)\right]\left(\frac{\beta\gapE}{2}\right)^2.
\ee
We remind the reader that the thermal conductivity, which is a positive transport coefficient, is
related to the kinetic coefficient (also called Onsager coefficient) introduced in phenomenological irreversible thermodynamics as
\be
\label{KineticCoef}
L\equiv\lim_{\Fth\to 0} \frac{\Espst{\jinst_2}}{\Fth},
\ee
where,  as recalled in paper I, the  thermodynamic force $\Fth$  can be defined from the 
 stationary entropy production rate, which is opposite to the exchange entropy flow, $\dint \SG/dt\vert_\text{st}=-\dexch S/dt\vert_\text{st}$,  through the relation  
\be
\label{dirrSst}
\left.\frac{\dint \SG}{dt}\right\vert_\text{st}=-\left.\frac{\dexch S}{dt}\right\vert_\text{st}=\Fth\Espst{\jinst_2}
\ee
when there is only one independent mean instantaneous current.
In the case of the thermal contact $\Fth=\beta_1-\beta_2$. Therefore
the relation between the kinetic coefficient and the thermal conductivity defined in \eqref{defThermalCond} reads 
\be
\label{relThermCondKinetic}
L=\frac{\kappa_\text{th}}{\beta^2}.
\ee

Now we compare the results about the  linear  static response  in  non-equilibrium stationary states which are either in the vicinity of equilibrium or far away from equilibrium.
When the system is far from equilibrium, namely when $\left(\beta_1-\beta_2\right)\gapE\gg 1$, \eqref{jinstexp} leads to
\be
\left.\frac{\partial \Espst{\jinst_2}}{\partial \beta_2}\right\vert_{\beta_1}=-\frac{\nu_1\nu_2}{\nu_1+\nu_2}
\left[1-\tanh\left(\frac{\beta_2\gapE}{2}\right)^2\right]\left(\frac{\gapE}{2}\right)^2
\ee
\be
\left.\frac{\partial \Espst{\jinst_2}}{\partial \beta_1}\right\vert_{\beta_2}=
\frac{\nu_1\nu_2}{\nu_1+\nu_2}
\left[1-\tanh\left(\frac{\beta_1\gapE}{2}\right)^2\right]\left(\frac{\gapE}{2}\right)^2.
\ee
The linear response coefficients $\left.\partial \Espst{\jinst_2}/\partial T_2\right\vert_{T_1}$ and $\left.\partial \Espst{\jinst_2}/\partial T_1\right\vert_{T_2}$ are no more opposite to each other. As a consequence, when $T_1$ and $T_2$ are varied independently, the corresponding variation of the stationary mean instantaneous current
  $\Espst{\jinst_2}^{[\beta_1,\beta_2]}$ 
at first order reads
\be
\label{difjinst}
\Espst{\jinst_2}^{[\beta'_1,\beta'_2]}-\Espst{\jinst_2}^{[\beta_1,\beta_2]}
\underset{(T'_1,T'_2)\to(T_1,T_2)}{\sim}
(T'_1-T_1)\left.\frac{\partial \Espst{\jinst_2}^{[\beta_1,\beta_2]}}{\partial T_1}\right\vert_{T_2}
+(T'_2-T_2)\left.\frac{\partial \Espst{\jinst_2}^{[\beta_1,\beta_2]}}{\partial T_2}\right\vert_{T_1}.
\ee
The latter variation depends not only on $T_1$, $T_2$ and the variation of the temperature difference 
$(T'_1-T'_2)-(T_1-T_2)$  but also on the way in which $T'_1$ and $T'_2$ are varied around the  given values $T_1$ and $T_2$. 

\subsection{Non-linear static response in the NESS}
\label{secNonLinearStatic}
\subsubsection{Non-linear thermal conductivity}

When the system is far from equilibrium, instead of introducing the linear response $\Espst{\jinst_2}^{[\beta'_1,\beta'_2]}-\Espst{\jinst_2}^{[\beta_1,\beta_2]}$ 
with $(\beta'_1-\beta_1)\gapE\ll 1$ and $(\beta'_2-\beta_2)\gapE\ll 1$ (and the associated linear response coefficients $\partial \Espst{\jinst_2}^{[\beta_1,\beta_2]}/\partial \beta_a$), one may rather consider
a non-linear  thermal conductivity defined as
\be
\label{defkappathnlin}
\kappa_\text{th}^\text{nlin}=\frac{\Espst{\jinst_2}^{[\beta_1,\beta_2]}}{T_2-T_1}.
\ee
From \eqref{jinstexp}
\be
\label{expkappathnlin}
\kappa_\text{th}^\text{nlin}=\frac{1}{T_1T_2}\frac{\nu_1\nu_2}{\nu_1+\nu_2}\frac{\tanh\left(\frac{(\beta_1-\beta_2)\gapE}{2}\right)}{\frac{(\beta_1-\beta_2)\gapE}{2}}
\left[1-\tanh\left(\frac{\beta_1\gapE}{2}\right)\tanh\left(\frac{\beta_2\gapE}{2}\right)\right]
\left(\frac{\gapE}{2}\right)^2.
\ee
According to the expression \eqref{jinstexp}, $\Espst{\jinst_2}$ is a bounded function of $T_1$ and $T_2$, so that $\kappa_\text{th}^\text{nlin}$ vanishes when $T_2-T_1$ becomes very large with respect to either $T_1$ or $T_2$.

We also notice that when both thermostats are at very high temperature, namely 
when $\beta_1\gapE\ll1$ and $\beta_2\gapE\ll1$, $\Espst{\jinst_2}$ is proportional to $\beta_1-\beta_2$ with  a coefficient independent of the temperatures. As a consequence,
the partial derivatives 
$\partial \Espst{\jinst_2}^{[\beta_1,\beta_2]}/\partial \beta_1$ and $\partial \Espst{\jinst_2}^{[\beta_1,\beta_2]}/\partial \beta_2$ are opposite to each other, as in the  symmetry property  \eqref{oppositePD} in the very vicinity of the equilibrium limit. Then the difference \eqref{difjinst} is proportional to the difference $(\beta'_1-\beta_1)-(\beta'_2-\beta_2)$,
\be
\label{difjinstBis}
\Espst{\jinst_2}^{[\beta'_1,\beta'_2]}-\Espst{\jinst_2}^{[\beta_1,\beta_2]}
\underset{\substack{(\beta'_1,\beta'_2)\to(\beta_1,\beta_2)\\ \beta_1\gapE\to 0, \, \beta_2\gapE\to 0}}{\sim}
\left[(\beta'_1-\beta_1)-(\beta'_2-\beta_2)\right]\left.\frac{\partial \Espst{\jinst_2}}{\partial \beta_1}\right\vert_{\beta_2}.
\ee
Besides, the thermal conductivity \eqref{expkappathnlin} behaves as
\be
\kappa_\text{th}^\text{nlin}
\underset{\substack{\beta_1\gapE\to 0\\ \beta_2\gapE\to 0}}{\sim}
\frac{1}{T_1T_2}\frac{\nu_1\nu_2}{\nu_1+\nu_2}
\left(\frac{\gapE}{2}\right)^2.
\ee

\subsubsection{Housekeeping entropy flow and mean excess heats}
\label{HousekeepingExcess}

In the long-time limit, whatever the initial configuration probability $\Prob_0$ may be,  the system reaches a stationary state where the Markovian stochastic dynamics enforces that  the cumulated  heats received from each thermostat, namely the random variables $\Heat_1(t)$ and $\Heat_2(t)$, have averages $\Esp{\Heat_1(t)}_{\Prob_0}$  and $\Esp{\Heat_2(t)}_{\Prob_0}$ which both grow linearly in time with opposite coefficients, $-\Esp{\Heat_1(t)}_{\Prob_0}\underset{t\to+\infty}{\sim}\Esp{\Heat_2(t)}_{\Prob_0}\underset{t\to+\infty}{\sim} t\Espst{\jinst_2}^{[\beta_1,\beta_2]}$. Then
\be
\label{limitcomblin}
\lim_{t\to+\infty} \frac{\beta_1 \Esp{\Heat_1(t)}_{\Prob_0}+\beta_2 \Esp{\Heat_2(t)}_{\Prob_0}}{t} =-(\beta_1-\beta_2)\Espst{\jinst_2}^{[\beta_1,\beta_2]}=\left.\frac{\dexch S}{dt}\right\vert_\text{st},
\ee
where  the stationary exchange entropy flow appears by virtue of \eqref{dirrSst}. Meanwhile the  sum $\Heat_1(t)+\Heat_2(t)$ remains bounded at any time and its average tends to   the heat amount corresponding to the mean energy difference between  the final and initial stationary states,
\be
\label{limitsum}
\lim_{t\to+\infty} \Esp{\Heat_1(t)+\Heat_2(t)}_{\Prob_0}=
\Espst{\En}^{[\beta_1,\beta_2]}- \Esp{\En}_{\Prob_0}.
\ee

In the phenomenological framework of steady state thermodynamics \cite{OonoPaniconi1998}, when work is supplied to the system,  the total  heat \textit{given} to the system is usually expressed as the sum of
an ``excess'' heat  $\Heat_\text{exc}$ associated with the energy exchange during transitions between two different steady states 
and a ``housekeeping'' heat $\Heat_\text{hk}$ associated with the energy supplied to maintain the system in the NESS reached in the long-time limit. 
These two heat amounts have been discussed  for a system in contact with only one thermal bath and submitted to a time-dependent external force  which is described by  Langevin dynamics  \cite{Sekimoto1998,HatanoSasa2001,SpeckSeifert2005}.

By analogy, with the standard sign convention, we may introduce a ``housekeeping'' entropy flow supplied to the system
which can be measured as the asymptotic behavior
\be
\label{defDeltaShk}
\sighk[\Probst]\equiv
-\lim_{t\to+\infty} \frac{\beta_1 \Esp{\Heat_1(t)}_{\Prob_0}+\beta_2 \Esp{\Heat_2(t)}_{\Prob_0}}{t},
\ee
and which, by virtue of \eqref{limitcomblin}  coincides with the opposite of  the stationary exchange entropy flow, namely with the stationary entropy production rate (see \eqref{dirrSst})
\be
\sighk[\Probst]=-\left.\frac{\dexch S}{dt}\right\vert_\text{st}=\left.\frac{\dint S}{dt}\right\vert_\text{st}.
\ee
From the explicit expression \eqref{jinstexp} of the mean instantaneaous heat current we obtain the expression for the housekeeping entropy flow \eqref{defDeltaShk}
\be
\label{sighkexp}
\sighk[\Probst]=\frac{\nu_1\nu_2}{\nu_1+\nu_2}
\left(\gamma_1-\gamma_2\right)\left(\beta_1-\beta_2\right)\frac{\gapE}{2}.
\ee

When the system is prepared in a stationary state by thermal contact with heat reservoirs at the inverse temperatures $\beta_1^0$ and $\beta_2^0$ respectively, then 
$\Esp{\En}_{\Prob_0}=\Espst{\En}^{[\beta_1^0,\beta_2^0]}$ and the difference in
\eqref{limitsum} becomes equal to 
$\Espst{\En}^{[\beta_1,\beta_2]}-\Espst{\En}^{[\beta_1^0,\beta_2^0]}$. With the standard convention,  the ``excess'' heats \textit{given} to the system $\left.{\Heatav_{\text{exc}, a}}\right\vert^{[\beta_1,\beta_2]}_{[\beta_1^0,\beta_2^0]}$ with $a=1,2$  can be measured as
\be
\label{defexcessheat}
\left.{\Heatav_{\text{exc}, a}}\right\vert^{[\beta_1,\beta_2]}_{[\beta_1^0,\beta_2^0]}\equiv-\lim_{t\to+\infty} \left[\Esp{\Heat_a(t)}_{\Probst^{[\beta_1^0,\beta_2^0]}}- t\Espst{\jinst_a}^{[\beta_1,\beta_2]}\right].
\ee
Then,   by virtue of the stationary condition $\Espst{\jinst_1}=-\Espst{\jinst_2}$, the equality \eqref{limitsum}  becomes
\be
\label{sumexcess}
-\left.{\Heatav_{\text{exc}, 1}}\right\vert^{[\beta_1,\beta_2]}_{[\beta_1^0,\beta_2^0]}
-\left.{\Heatav_{\text{exc}, 2}}\right\vert^{[\beta_1,\beta_2]}_{[\beta_1^0,\beta_2^0]}
=\Espst{\En}^{[\beta_1,\beta_2]}-\Espst{\En}^{[\beta_1^0,\beta_2^0]}.
\ee
The excess heats $\left.{\Heatav_{\text{exc},a}}\right\vert^{[\beta_1,\beta_2]}_{[\beta_1^0,\beta_2^0]}$'s  defined in
\eqref{defexcessheat} are  explicitly calculated in subsection \eqref{Excessheat} from the expressions of the average heat amounts $\Esp{\Heat_a(t)}_{\Prob_0}$'s at any finite time $t$ (for any initial distribution $\Prob_0$ of the two-spin configuration) with the results given in \eqref{ExpEcxessHeat}.

In the linear response regime where the relative differences $(T_1-T_1^0)/T_1^0$ and $(T_2-T_2^0)/T_2^0$ are infinitesimal, by virtue of the definition \eqref{defheatcapacity} of the generalized heat capacities $C_\text{st}^{[a]}(T_1,T_2)$, with $a=1,2$,
\be
-\left.{\Heatav_{\text{exc}, 1}}\right\vert^{[T_1+dT_1,T_2+dT_2]}_{[T_1,T_2]}
-\left.{\Heatav_{\text{exc}, 2}}\right\vert^{[T_1+dT_1,T_2+dT_2]}_{[T_1,T_2]}\to
C_\text{st}^{[1]}(T_1,T_2) dT_1+C_\text{st}^{[2]}(T_1,T_2)dT_2.
\ee
We notice that the notion of heat capacity has been studied in the case of non equilibrium steady states where the system is submitted to a non-conservative force and is in contact with a single thermostat \cite{BoksenbojmMaesETAL2011}.

\section{Joint probability distribution for heat cumulated exchanges at finite time in the model}

\label{JointProbsection}

Instead of studying the evolution of the probability distribution  $\Prob(\C;t)$ of the spins configuration $\C=\left(\sigma_1,\sigma_2\right)$, we address directly the evolution of the joint probability distribution 
$\Prob\left(\C' \vert \Heat_1,\Heat_2, t\vert  \C\right)$ for the cumulated heats $\Heat_1$ and $\Heat_2$ received from the thermal baths 1 and 2 during a time $t$ when the system is  in configuration $\C=(\sigma_1, \sigma_2)$ at time $t_0=0$ and in configuration $\C'=(\sigma'_1, \sigma'_2)$ at time $t$. In order to obtain results which hold as generally  as possible, the initial probability distribution for configurations is not assumed to have the same symmetry under simultaneous spin flips as the stationary distribution.

Since the two-spin system has only two energy levels separated by the energy gap $\gapE$, the cumulated heats $\Heat_a$  are integer multiples of $\gapE$ and we set
\be
\label{discreteHeat}
\Heat_1=-n_1\gapE \quad\textrm{and}\quad \Heat_2=n_2\gapE.
\ee
The minus sign in the definition of $\Heat_1$ is introduced for the sake of conveniency, because the mean instantaneous heat currents $\Espst{\jinst_1}$ and $\Espst{\jinst_2}$ in the stationary state are 
opposite to each other. In other words, $n_1\gapE$ is the amount of heat dissipated towards  heat bath 1, while $n_2\gapE$ is the amount of heat received from  heat bath 2.
With these notations $\Prob\left(\C_f \vert \Heat_1,\Heat_2, t\vert  \C_0\right)$ can be written as a matrix element of some evolution operator $ \Uev(n_1,n_2 ; t)$
\be
\Prob\left(\C' \vert \Heat_1,\Heat_2, t\vert  \C\right)=\left(\sigma'_1,\sigma'_2\vert \Uev(n_1,n_2 ; t)\vert \sigma_1,\sigma_2\right).
\ee

\subsection{Explicit calculations}

\subsubsection{Constraint from energy conservation}

According to the expression \eqref{energyExp} for the interaction energy between the two spins,
 the energy difference between the final and the initial configurations reads
 \be
\label{EnChange}
\En(\sigma'_1, \sigma'_2)-\En(\sigma_1, \sigma_2)=\frac{\sigma_1\sigma_2-\sigma'_1\sigma'_2}{2}\gapE,
\ee
and it can take only three values $0$, $+\gapE$ and $-\gapE$. On the other hand,  
according to \eqref{discreteHeat}, $\Heat_1 +\Heat_2=(n_2-n_1)\gapE$, namely
\be
\Heat_1 +\Heat_2=\Delta n \times\gapE
\quad\textrm{where}\quad
\Delta n\equiv n_2-n_1.
\ee
Energy conservation entails that the energy variation of the two-spin system is equal to the sum of the heat amounts received from the thermostats:
$\En(\sigma'_1, \sigma'_2)-\En(\sigma_1, \sigma_2)=\Heat_1 +\Heat_2$.
As a consequence the correspondence between the total amount of received heat and the couple of initial and final states reads
\bea
\label{correspondance}
\Delta n=0\qquad &\Leftrightarrow &\qquad\sigma'_1\sigma'_2=\sigma_1\sigma_2
\\
\nonumber
(\Delta n)^2=1\qquad &\Leftrightarrow &\qquad\sigma_1\sigma_2= \Delta n \quad\textrm{and} \quad\sigma'_1\sigma'_2= -\Delta n .
\eea
Therefore it is convenient to introduce the decomposition
\be
\label{decompUev}
\Uev(n_1,n_2 ; t)=\sum_{\Delta n={0,+1,-1}}\delta_{n_2,n_1+\Delta n}\Uev(n_1,n_1+\Delta n ; t).
\ee
In the  basis $\{(+,+), (-,-), (+,-), (-+)\}$ already used in  \eqref{basis} 
the correspondence \eqref{correspondance} enforces that $\Uev(n_1,n_2 ; t)$
 can be decomposed into three $4\times 4$ matrices
\be
\label{matrixdecomposition}
\Uev=\Uev_{[\Delta n=0]}+ \Uev_{[\Delta n=+1]}+\Uev_{[\Delta n=-1]}
\ee
with
\be
\Uev_{[\Delta n=0]}=
\begin{pmatrix}
\mathbb{A}& \mathbb{O}\\
\mathbb{O} & \mathbb{D}
\end{pmatrix}
\quad
\Uev_{[\Delta n=1]}=\begin{pmatrix}
\mathbb{O}& \mathbb{O}\\
\mathbb{C} & \mathbb{O}
\end{pmatrix}
\quad
\Uev_{[\Delta n=-1]}=\begin{pmatrix}
\mathbb{O} & \mathbb{B}\\
\mathbb{O} & \mathbb{O}
\end{pmatrix}.
\ee
The subscript involving $\Delta n$ indicates the unique value of $\Delta n$ which is involved in a history where the initial and final states are $(\sigma_1,\sigma_2)$ and $(\sigma'_1,\sigma'_2)$ respectively.  $\mathbb{O} =\begin{pmatrix} 0 & 0\\ 0& 0 \end{pmatrix}$ and $\mathbb{A}$, $\mathbb{B}$, $\mathbb{C}$ and $\mathbb{D}$ are $2\times 2$ matrices.

\subsubsection{Generating function method}

The evolution equation for $\Uev(n_1,n_2 ; t)$ is easily derived by considering the probability
\be
\label{defUev}
\Prob(\sigma_1,\sigma_2, n_1,n_2; t)=\sum_{\sigma'_1,\sigma'_2}
\left(\sigma_1,\sigma_2\vert \Uev(n_1,n_2 ; t)\vert \sigma'_1,\sigma'_2\right)
\Prob(\sigma'_1,\sigma'_2; t=0).
\ee
 $\Prob(\sigma_1,\sigma_2, n_1,n_2; t)$ is  the probability that the system is in configuration $(\sigma_1,\sigma_2)$ a time $t$ and has received the heat amounts $\Heat_1=-n_1\gapE$ and $\Heat_2=n_2\gapE$ during the time interval $[0,t]$ when the initial probability distribution for the spins is $\Prob(\sigma'_1,\sigma'_2; t=0)$.
The evolution equation for $\Prob(\sigma_1,\sigma_2, n_1,n_2; t)$ is
 a generalization of the master equation \eqref{MarkovEquation} which governs the evolution of $\Prob(\sigma_1,\sigma_2;t)$. By taking into account the explicit expression \eqref{Transexp} for the transition rates we get
\bea
\label{JointProbEvolution}
\frac{2}{\nu_1+\nu_2}\frac{d \Prob(\sigma_1,\sigma_2, n_1,n_2; t)}{dt}&=&
-\left[1-\sigma_1\sigma_2(\nuad_1 \gamma_1+\nuad_2 \gamma_2)\right]\Prob(\sigma_1,\sigma_2, n_1,n_2; t)\\
&&
+\nuad_1\left[ 1+\sigma_1\sigma_2 \gamma_1\right]\Prob(-\sigma_1,\sigma_2, n_1-\sigma_1\sigma_2 ,n_2; t)
\nonumber\\
&&
+\nuad_2\left[ 1+\sigma_1\sigma_2 \gamma_2\right]\Prob(\sigma_1,-\sigma_2, n_1,n_2+\sigma_1\sigma_2 ; t)
\nonumber
\eea
where the dimensionless inverse time scales $\nuad_a$'s are defined in \eqref{defnuad}.

The operator in the r.h.s. of the evolution equation \eqref{JointProbEvolution} is partially diagonalized by considering the generating function
$\Prob(\sigma_1,\sigma_2, z_1,z_2; t)=\sum_{n_1=-\infty}^{+\infty}\sum_{n_2=-\infty}^{+\infty}
z_1^{n_1} z_2^{n_2}\Prob(\sigma_1,\sigma_2, n_1,n_2; t)$ which is absolutely
convergent for $z_1$ and $z_2$ of modulus $1$. Considering the latter generating function is equivalent to introducing
\be
\label{defUevG}
\UevG(z_1,z_2 ; t)\equiv\sum_{n_1=-\infty}^{+\infty}\sum_{n_2=-\infty}^{+\infty}
z_1^{n_1} z_2^{n_2} \Uev(n_1,n_2 ; t).
\ee
Since $\Prob(\sigma_1,\sigma_2, n_1,n_2;
t=0)=\delta_{n_1,0}\delta_{n_2,0}\Prob(\sigma_1,\sigma_2; t=0)$, we infer that
$\UevG(z_1,z_2 ; t=0)=\Id_4$, where $\Id_4$ denotes the identity $4\times 4$ matrix.
The inversion formula which allows to retrieve $\Uev(n_1,n_2 ; t)$ reads
\be
\label{InvUev}
\Uev(n_1,n_2 ; t)
 =\oint_{|z_1|=1}\frac{d z_1}{2 \pi  \iexp}\oint_{|z_2|=1}\frac{d z_2}{2  \pi \iexp}
\frac{\UevG(z_1,z_2 ; t)}{z_1^{n_1+1} z_2^{n_2+1}}.
\ee

The decomposition \eqref{decompUev} of $\Uev(n_1,n_2 ; t)$ leads to a similar decomposition for $\UevG(z_1,z_2;t)$
\be
\label{decompUevG}
\UevG(z_1,z_2;t)=\sum_{\Delta n={0,+1,-1}}\UevG_{[\Delta n]}(z_1,z_2;t).
\ee
The  decomposition \eqref{matrixdecomposition} of  $\Uev(n_1,n_2 ; t)$ into three  $4\times 4$ matrices (enforced by  the constraint \eqref{correspondance}  due to  energy conservation) is also valid for $\UevG(z_1,z_2;t)$.
Moreover
$\UevG_{[\Delta n]}(z_1,z_2;t)$ has necessarily the following dependence upon $z_2$ and $z_1z_2$:
$
\UevG_{[\Delta n]}(z_1,z_2;t)=z_2^{\Delta n}\widehat{\mathbb{V}}_{[\Delta n]}(z_1z_2;t)
$
Therefore, by using the change of variable $z_1\longrightarrow z=z_1z_2$ in \eqref{InvUev}
one gets
\be
\label{relUDnUDn}
\Uev(n_1,n_1+\Delta n ; t)=
\oint_{|z|=1}\frac{d z}{2 \pi  \iexp}
\frac{\UevG_{[\Delta n]}(z_1=z, z_2=1;t)}{z^{n_1+1}}.
\ee

\subsubsection{Diagonalization}

The evolution of $\UevG(z_1,z_2 ; \tad)$ with the dimensionless time variable
\be
\label{deftad}
\tad=\frac{\nu_1+\nu_2}{2}t
\ee
 reads
\be
\label{evolUevG}
\frac{d\UevG(z_1,z_2 ; \tad)}{d\tad}
=\Aev(z_1,z_2)\UevG(z_1,z_2 ; \tad),
\ee
where, from the evolution equation \eqref{JointProbEvolution},
\be
\label{Aev2}
\Aev(z_1,z_2)=-\Id_4+
\begin{pmatrix}
\gammaeff & 0 & b_2 & b_1 \\
0 & \gammaeff &b_1 & b_2 \\
c_2 & c_1 &-\gammaeff  &0 \\
c_1 & c_2 & 0 &-\gammaeff 
\end{pmatrix}
\ee
with the following notations:  $b_1=\nuad_1(1+\gamma_1) z_1$, $b_2= \nuad_2 (1+\gamma_2)\frac{1}{z_2}$, 
$c_1=\nuad_1(1-\gamma_1) \frac{1}{z_1}$ and $c_2=\nuad_2 (1-\gamma_2)z_2$.

Since the transition rates are invariant under the simultaneous  changes of both spin signs, it is convenient to consider the transformed matrix 
\be
\label{defAev}
\Aev'(z_1,z_2)=\Pcb^{-1}\Aev(z_1,z_2)\Pcb
\ee
with
\be
\Pcb^{-1}=\begin{pmatrix}
1 & 1 & 0 & 0\\
0 & 0 & 1 & 1 \\
1 &-1 & 0 & 0 \\
0 & 0 & 1 & -1
\end{pmatrix}.
\ee
The matrix $\Aev'(z_1,z_2)$ corresponds to two sets of decoupled equations,
\be
\Aev'(z_1,z_2)=-\Id_4+\begin{pmatrix}
\B_+ & \mathbb{O}\\
 \mathbb{O} & \B_-
\end{pmatrix}
\ee
where $ \B_\epsilon=\begin{pmatrix}
\gammaeff & \epsilon  b_1+b_2\\ \epsilon c_1+c_2& -\gammaeff 
\end{pmatrix} $ 
for $\epsilon =\pm$. As $\B_\epsilon$ is traceless, $\B^2_\epsilon(z_1,z_2)$ is proportional to $\Id_2$. 
Explicitly 
\be
\B^2_\epsilon(z_1,z_2)=\Deltastar_{\epsilon}(z_1z_2) \Id_2
\ee
with
\be
\label{defDeltastar}
\Deltastar_\epsilon(z)=1-2A+\epsilon\left[(A+B) z +(A-B)\frac{1}{z}\right]
\ee
where
\be
\label{defAB}
A=\nuad_1\nuad_2\left(1-\gamma_1\gamma_2\right)
\quad\textrm{and}\quad
B=\nuad_1\nuad_2\left(\gamma_1-\gamma_2\right).
\ee
We notice that $A\geq | B |$.
As a consequence, 
\be
\label{exponAevp}
e^{\tad  \Aev'(z_1,z_2)}=e^{-\tad} \times\begin{pmatrix}
e^{\tad \B_+(z_1,z_2)} & \mathbb{O}\\
 \mathbb{O} & e^{\tad \B_-(z_1,z_2)} 
\end{pmatrix}
\ee
where
\be
\label{exponB}
e^{\tad \B_\epsilon(z_1,z_2)}=\cosh(\tad \sqrt{\Deltastar_{\epsilon}(z_1z_2)}) \Id_2 +
\frac{\sinh(\tad \sqrt{\Deltastar_{\epsilon}(z_1z_2)})}{ \sqrt{\Deltastar_{\epsilon}(z_1z_2)}}
\B_\epsilon(z_1,z_2).
\ee
Moreover the eigenvalues of the matrix $\frac{1}{2}(\nu_1+\nu_2)\Aev(z_1,z_2)$ are, with the notations
$\epsilon=\pm$ and $\eta=\pm$,
\be
\label{AevEigenvalues}
\mu^{(\epsilon,\eta)}(z)=\frac{\nu_1+\nu_2}{2}\left[-1+\eta \sqrt{\Deltastar_\epsilon(z)}
\right].
\ee

\subsubsection{Results for the generating function}

From \eqref{defAev} we can calculate $\UevG(z_1,z_2;\tad)=e^{\tad \Aev(z_1,z_2)}=\Pcb e^{\tad
  \Aev'(z_1,z_2)}\Pcb^{-1}$.
From the explicit expressions \eqref{exponAevp} and \eqref{exponB} we get the
matrices $\UevG_{[\Delta n]}(z_1,z_2;\tad)$ defined in  \eqref{decompUevG}. The 16 matrix elements can be written in the compact form
\bea
\label{UzResults}
(\sigma_1,\sigma_2\vert \UevG_{[\Delta n=0]}(z_1,z_2;\tad)\vert\sigma_1,\sigma_2)&=&\Ch^{+}(z_1z_2;\tad)+\sigma_1\sigma_2 \gammaeff \Sh^{+}(z_1z_2;\tad)
 \\
\nonumber
(-\sigma_1,-\sigma_2\vert \UevG_{[\Delta n=0]}(z_1,z_2;\tad)\vert\sigma_1,\sigma_2)&=&\Ch^{-}(z_1z_2;\tad)+\sigma_1\sigma_2 \gammaeff \Sh^{-}(z_1z_2;\tad) 
\\
\nonumber
(-\sigma_1,\sigma_2\vert \UevG_{[\Delta n]}(z_1,z_2;\tad)\vert\sigma_1,\sigma_2)&\underset{\Delta n =\pm 1}{=}& \delta_{\sigma_1\sigma_2, \Delta n} \,z_2^{\Delta n} F^+_{\Delta n}(z_1z_2;\tad) 
\\
\nonumber
(\sigma_1,-\sigma_2\vert \UevG_{[\Delta n]}(z_1,z_2;\tad)\vert\sigma_1,\sigma_2)&\underset{\Delta n =\pm 1}{=}&\delta_{\sigma_1\sigma_2, \Delta n} \,z_2^{\Delta n} F^-_{\Delta n}(z_1z_2;\tad)
\eea
with
\be
\Ch^{\pm}(z;\tad) = e^{-\tad}\frac{1}{2}\left[\cosh(\tad\sqrt{\Deltastar_+(z)})\pm\cosh(\tad\sqrt{\Deltastar_-(z)})\right]
\ee
\be
\Sh^{\pm}(z;\tad) =e^{-\tad}\frac{1}{2}\left[
\frac{\sinh(\tad\sqrt{\Deltastar_+(z)})}{\sqrt{\Deltastar_+(z)}}\pm
\frac{\sinh(\tad\sqrt{\Deltastar_-(z)})}{\sqrt{\Deltastar_-(z)}}\right]
\ee
and
\be
F^{\pm}_{\Delta n}(z;\tad)=\frac{1}{z^{\Delta n}}
\nuad_1\left(1-\Delta n \,\gamma_1\right)\Sh^{\pm}(z;\tad)
+\nuad_2\left(1-\Delta n \,\gamma_2\right)\Sh^{\mp}(z;\tad).
\ee

\subsubsection{Results for the joint probability}

$\Uev(n_1,n_1+\Delta n ; t)$ is derived from $\UevG_{[\Delta n]}(z_1,z_2;t)$ through the unit circle integral in \eqref{relUDnUDn}. In fact, because of the parity property of the $\cosh$ and  $\sinh$ functions,  the functions $\Ch^{\pm}(z;\tad)$ and $\Sh^{\pm}(z;\tad)$ are functions not of $\sqrt{\Deltastar_\epsilon(z)}$ but only of $\Deltastar_\epsilon(z)$. According to the definition \eqref{defDeltastar} of $\Deltastar_\epsilon(z)$, the only singular points of $\Deltastar_\epsilon(z)$ are $z=0$  and $z=+\infty$, and the same is true for the integrands  in 
$\oint_{|z|=1} (dz/2\pi\iexp)z^{-(m+1)}\Ch^{\pm}(z;\tad)$ and 
$\oint_{|z|=1}(dz/2\pi\iexp)z^{-(m+1)}\Sh^{\pm}(z;\tad)$. 
Changing $z$ into $-z$ we get that 
\be
\oint_{|z|=1} \frac{dz}{2\pi\iexp}\frac{1}{z^{m+1}}\cosh\left(\tad \sqrt{\Deltastar_-(z)}\right)=
(-1)^m \oint_{|z|=1} \frac{dz}{2\pi\iexp}\frac{1}{z^{m+1}}\cosh\left(\tad \sqrt{\Deltastar_+(z)}\right).
\ee
As a consequence
\be
\oint_{|z|=1} \frac{dz}{2\pi\iexp}\frac{1}{z^{m+1}}\Ch^{\pm}(z;\tad)=\frac{1\pm (-1)^m}{2}
e^{-\tad} \ccstar_m(\tad)
\ee
and
\be
\oint_{|z|=1} \frac{dz}{2\pi\iexp}\frac{1}{z^{m+1}}\Sh^{\pm}(z;\tad)=\frac{1\pm (-1)^m}{2}
e^{-\tad} \csstar_m(\tad),
\ee
where
\be
\label{defccstar}
\ccstar_m(\tad)\equiv \oint_{|z|=1} \frac{dz}{2\pi\iexp} \frac{1}{z^{m+1}}\cosh\left(\tad\sqrt{\Deltastar_+(z)}\right)
\ee
and
\be
\label{defcsstar}
\csstar_m(\tad)\equiv \oint_{|z|=1} \frac{dz}{2\pi\iexp} \frac{1}{z^{m+1}}\frac{\sinh\left(\tad\sqrt{\Deltastar_+(z)}\right)}{\sqrt{\Deltastar_+(z)}}.
\ee

Eventually, the matrix elements of $\Uev(n_1,n_1+\Delta n ; t)$ are derived from the expressions \eqref{UzResults} for the  matrix elements of $\UevG_{[\Delta n]}(z_1,z_2;t)$  with the result
\bea
\label{Uevn1n2Formules0}
(\sigma_1,\sigma_2\vert\Uev(n,n; t)\vert \sigma_1,\sigma_2)
&=&\delta_\text{even}(n)\times U^{(0)}(n,\sigma_1\sigma_2;\tad)
\\
\nonumber
(-\sigma_1,-\sigma_2\vert\Uev(n,n ; t)\vert \sigma_1,\sigma_2)
&=&\delta_\text{odd}(n)\times U^{(0)}(n,\sigma_1\sigma_2;\tad)
\\
\nonumber
(-\sigma_1,\sigma_2\vert\Uev(n,n+\Delta n ; t)\vert \sigma_1,\sigma_2)
&\underset{\Delta n=\pm1}{=}&
 \delta_{\sigma_1\sigma_2,\Delta n}\times
\delta_\text{odd}(n) \times U^{(1)}(n,\Delta n;\tad)
\\
\nonumber
(\sigma_1,-\sigma_2\vert\Uev(n,n +\Delta n; t)\vert \sigma_1,\sigma_2)
&\underset{\Delta n=\pm1}{=}&
\delta_{\sigma_1\sigma_2,\Delta n}\times
\delta_\text{even}(n)\times U^{(1)}(n,\Delta n;\tad),
\eea
where $\delta_\text{even}(n)=\frac{1}{2}\left[1+(-1)^n\right]$ and
$\delta_\text{odd}(n)=\frac{1}{2}\left[1-(-1)^n\right]$
while
\bea
\label{defU0U1Formules0}
U^{(0)}(n,\sigma_1\sigma_2;\tad)&=& 
e^{-\tad}\left[\ccstar_n(\tad)+\gammaeff \sigma_1\sigma_2\csstar_n(\tad)\right]
\\
\nonumber
U^{(1)}(n,\Delta n;\tad)&=&
e^{-\tad}\left[\nuad_2 \left(1-\gamma_2\Delta n\right)\csstar_n(\tad)
+\nuad_1\left(1-\gamma_1 \Delta n\right) \csstar_{n+\Delta n}(\tad)\right].
\eea
We notice that the parity condition factors $\frac{1}{2}\left[1\pm(-1)^n\right]$ have a simple interpretation. During a history such that spin $\sigma_1$ is in the same  state (in flipped states) in the initial and final configurations, thermal bath $1$ has flipped spin $\sigma$ an even (odd) number of times, so that 
the corresponding sum $n \gapE$ of the  successive amounts $\pm \gapE$ dissipated towards  thermal bath $1$ is necessarily an even (odd) multiple of $\gapE$.

\subsection{Various explicit probabilities}
\label{ExplicitProbabilities}

The probability that the system is in configuration $(\sigma_1,\sigma_2)$ at time $t$ when the initial configuration is distributed according to the law $\Prob_0$ can be calculated, by virtue of the definition \eqref{defUevG}, as
\be
\Prob_{\Prob_0}\left(\sigma_1,\sigma_2;t\right)=
\sum_{\Delta n=-1,0,+1}\sum_{\sigma'_1,\sigma'_2}
\left(\sigma_1,\sigma_2\vert\UevG_{[\Delta n]}(z_1=1,z_2=1;t)\vert
  \sigma'_1,\sigma'_2\right)\Prob_0(\sigma'_1,\sigma'_2). 
\ee
The matrix elements of $\UevG_{[\Delta n]}(z_1=1,z_2=1;t)$ are derived from
\eqref{UzResults} where,
according to \eqref{defDeltastar},  $\Deltastar_+(z=1)=1$ while
$\Deltastar_-(z=1)=1-4A$ with $A=\nuad_1\nuad_2(1-\gamma_1\gamma_2)$. Using the identities
$\Prob_0(\sigma_1,\sigma_2)+\Prob_0(-\sigma_1,-\sigma_2)=\frac{1}{2}[1+\sigma_1\sigma_2\gamma_0]$, where $\gamma_0=\Esp{\sigma_1\sigma_2}_{\Prob_0}$, and
$\Prob_0(\sigma_1,\sigma_2)-\Prob_0(-\sigma_1,-\sigma_2)=
\frac{1}{2}\left[\sigma_1\Esp{\sigma'_1}_{\Prob_0}+\sigma_2\Esp{\sigma'_2}_{\Prob_0}\right]$,
where $\Esp{\sigma'_1}_{\Prob_0}$ (resp. $\Esp{\sigma'_2}_{\Prob_0}$) is the
average value of the first (resp. second) spin at time
$0$, 
a straightforward calculation leads to 
\bea
\label{ProbExp}
&&\Prob_{\Prob_0}\left(\sigma_1,\sigma_2;t\right)=
\frac{1}{4}[1+\sigma_1\sigma_2\gammaeff ]+
\frac{1}{4}  \sigma_1\sigma_2[\gamma_0-\gammaeff]e^{-2\tad}
\\
\nonumber
&&
+\frac{1}{8}\left[\sigma_1\Esp{\sigma'_1}_{\Prob_0}+\sigma_2\Esp{\sigma'_2}_{\Prob_0}\right]
\left[\left(1+\sigma_1\sigma_2\frac{\gammaeff}{\alpha}\right)e^{-(1-\alpha)\tad}
+\left(1-\sigma_1\sigma_2\frac{\gammaeff}{\alpha}\right)e^{-(1+\alpha)\tad}\right]
\\
\nonumber
&&
+\frac{1}{8}\left[-\sigma_1\Esp{\sigma'_1}_{\Prob_0}+\sigma_2\Esp{\sigma'_2}_{\Prob_0}\right]
\frac{1}{\alpha}
\left[\nuad_1-\nuad_2+\sigma_1\sigma_2(\nuad_1\gamma_1-\nuad_2\gamma_2)\right]
\left[e^{-(1-\alpha)\tad}-e^{-(1+\alpha)\tad}\right].
\eea
where $\alpha\equiv \sqrt{1-4A}$.
When $\Prob_0$ is invariant under the simultaneous reversal of the spins $\sigma_1$ and $\sigma_2$,
$
\Prob_0(\sigma_1,\sigma_2)=(1/4)\left[1+\sigma_1\sigma_2\gamma_0\right]
$ and only the terms in the first line of \eqref{ProbExp} do contribute. Then the evolution of $\Prob_{\Prob_0}\left(\sigma_1,\sigma_2;t\right)$ towards the stationary distribution 
$\Probst(\sigma_1,\sigma_2)=\frac{1}{4}\left[1+\sigma_1\sigma_2\gammaeff\right]$
involves only one time scale, namely $1/(\nu_1+\nu_2)$ (recall that $\tad=\frac{1}{2}(\nu_1+\nu_2)t$).

For any initial probability distribution $\Prob_0$ of the spins, the probability
$\Prob_{\Prob_0}\left(\sigma_1,\sigma_2,n_1,n_2;t\right)$ that at time $t$ the system is in configuration $(\sigma_1,\sigma_2)$  and  has  received a heat amount $\Heat_1=-n_1\gapE$ from  bath $1$ and $\Heat_2=n_2\gapE$ from  bath $2$ is calculated from  \eqref{Uevn1n2Formules0} with the result
\bea
\label{ProbDetailee}
\Prob_{\Prob_0}\left(\sigma_1,\sigma_2,n,n;t\right)&=&
 U^{(0)}(n,\sigma_1\sigma_2;\tad)
 \left[\delta_\text{even}(n)\Prob_0(\sigma_1,\sigma_2)
+\delta_\text{odd}(n)\Prob_0(-\sigma_1,-\sigma_2)\right]
\\
\nonumber
\Prob_{\Prob_0}\left(\sigma_1,\sigma_2,n,n+\Delta n;t\right)&\underset{\Delta n=\pm1}{=}&
\delta_{\sigma_1\sigma_2,-\Delta n}U^{(1)}(n,\Delta n;\tad)
 \left[\delta_\text{even}(n)\Prob_0(\sigma_1,-\sigma_2)
+\delta_\text{odd}(n)\Prob_0(-\sigma_1,\sigma_2)\right]
\eea
Various joint probabilities can be derived from these expressions.

The joint probability $\Prob_{\Prob_0}\left(n_1, n_2;t\right)$ that at time $t$ the system  has received a heat amount $\Heat_1=-n_1\gapE$ from  bath $1$ and a heat amount $\Heat_2=n_2\gapE$ from  bath $2$ is 
\bea
\label{ProbQ1Q2explicit}
\Prob_{\Prob_0}\left(n, n;\tad\right)
&=&
e^{-\tad}\left[\ccstar_{n}(\tad)+\gammaeff \gamma_0\csstar_{n}(\tad)\right]
\\
\nonumber
\Prob_{\Prob_0}\left(n, n+\Delta n;\tad\right)
&\underset{\Delta n=\pm1}{=}&
e^{-\tad} \frac{1}{2}[1+\Delta n \,\gamma_0]\left[\nuad_2 \left(1-\Delta n\,\gamma_2\right)\csstar_{n}(\tad)
+\nuad_1\left(1-\Delta n \,\gamma_1 \right) \csstar_{n+\Delta n}(\tad)\right],
\eea
where $\gamma_0$ has been defined before \eqref{ProbExp}.

The joint probability $\Prob_{\Prob_0}\left(\sigma_1\sigma_2=\pm1,n_1;t\right)$ that
at time $t$ the system is in a configuration where $\sigma_1\sigma_2$ is equal
to $\pm1$ and has  received a heat amount $\Heat_1=-n_1\gapE$
from  bath $1$ is seen via \eqref{ProbDetailee} to have value
\bea
\label{Probpn1}
&&\Prob_{\Prob_0}\left(\sigma_1\sigma_2=+1, n_1;\tad\right)=
\\
\nonumber
&&
\frac{1}{2}e^{-\tad}\left\{
\left(1+\gamma_0\right) \ccstar_{n_1}(\tad)
+\left[ 2\gammaeff +(\nuad_2-\nuad_1\gamma_1)\left(1-\gamma_0\right)\right]\csstar_{n_1}(\tad)
 +\nuad_1\left(1+\gamma_1\right)\left(1-\gamma_0\right)\csstar_{n_1-1}(\tad)
 \right\}
\eea
and 
\bea
\label{Probmn1}
&&\Prob_{\Prob_0}\left(\sigma_1\sigma_2=-1, n_1;\tad\right)=
\\\nonumber
&&\frac{1}{2}e^{-\tad}\left\{
 \left(1-\gamma_0\right)\ccstar_{n_1}(\tad)
+\left[ -2\gammaeff +(\nuad_2+\nuad_1\gamma_1)\left(1+\gamma_0\right)\right]\csstar_{n_1}(\tad)
 +\nuad_1\left(1-\gamma_1\right)\left(1+\gamma_0\right)\csstar_{n_1+1}(\tad)
 \right\}.
\eea
The expressions for $\Prob_{\Prob_0}\left(\sigma_1\sigma_2=\pm1,
  n_2;\tad\right)$  are obtained from the latter equations by making the
exchanges $\nuad_1 \leftrightarrow \nuad_2$ and  $\gamma_1\leftrightarrow
\gamma_2$ and the replacements $\ccstar_{n_1}\rightarrow \ccstar_{n_2}$,
$\csstar_{n_1}\rightarrow \csstar_{n_2}$, and 
$\csstar_{n_1-1}\rightarrow \csstar_{n_2+1}$ for $\sigma_1\sigma_2=1$, resp.
$\csstar_{n_1+1}\rightarrow \csstar_{n_2-1}$ for $\sigma_1\sigma_2=-1$.

From these expressions 
we get the probability distribution for only one heat amount $\Heat_1$ or $\Heat_2$
\bea
\label{Probn1}
\Prob_{\Prob_0}\left(n_1;\tad\right)&=&
e^{-\tad}\Big\{ \ccstar_{n_1}(\tad)
+\left[\nuad_2+\nuad_1\gamma_1\gamma_0\right]\csstar_{n_1}(\tad)
\\ 
\nonumber 
&&\qquad\qquad
+\frac{1}{2}\nuad_1(1-\gamma_1)(1+\gamma_0)\csstar_{n_1+1}(\tad)
 +\frac{1}{2}\nuad_1(1+\gamma_1)(1-\gamma_0)\csstar_{n_1-1}(\tad)
\Big\}
\eea
and similarly
\bea
\label{Probn2}
\Prob_{\Prob_0}\left(n_2;\tad\right)&=&
e^{-\tad}\Big\{ \ccstar_{n_2}(\tad)
+\left[\nuad_1+\nuad_2\gamma_2\gamma_0\right]\csstar_{n_2}(\tad)
\\ 
\nonumber 
&&\qquad\qquad +\frac{1}{2}\nuad_2(1+\gamma_2)(1-\gamma_0)\csstar_{n_2+1}(\tad)
 +\frac{1}{2}\nuad_2(1-\gamma_2)(1+\gamma_0)\csstar_{n_2-1}(\tad)
\Big\}.
\eea

From the identities 
 $\sum_{n=-\infty}^{+\infty} \ccstar_n(\tad)=\cosh\tad$, and  $\sum_{n=-\infty}^{+\infty} \csstar_n(\tad)=\sinh\tad$, the probability that the total heat amount received from both thermostats is $\Heat_1+\Heat_2= (n_2-n_1)\gapE$ reads
\bea
\label{ProbDeltan}
\Prob_{\Prob_0}\left(n_2-n_1=0 ;\tad\right)&=&
\frac{1}{2}\left[1+\gammaeff \gamma_0\right]
+\frac{1}{2}\left[1-\gammaeff \gamma_0\right]e^{-2\tad}
\\
\nonumber
\Prob_{\Prob_0}\left(n_2-n_1=\Delta n;\tad\right)
&\underset{\Delta n=\pm1}{=}&
\frac{1}{4}\left\{1-\gammaeff\gamma_0
+\Delta n \left[\gamma_0-\gammaeff\right]\right\}
\left[1-e^{-2\tad}\right].
\eea
As a consequence 
\be
\Esp{\left[\Heat_1+\Heat_2\right]_\tad}_{\Prob_0}
=\frac{1}{2}\left[\gamma_0-\gammaeff\right]
\left[1-e^{-2\tad}\right] \gapE,
\ee
and we retrieve property \eqref{limitsum}.

We notice that all  formul\ae\ are still valid in the limit where $T_1$ vanishes, namely where
$\beta_1\gapE$ goes to infinity.

\subsection{Excess heats}
\label{Excessheat}

The excess heats associated with the transition between two different steady
states have been defined in \eqref{defexcessheat}. For the two-spin system they
can be explicitly calculated. Indeed, when the system is initially prepared in
the stationary state with distribution $\Prob_0$ by contact with two thermostats
at the inverse temperatures $\beta_1^0$ and $\beta_2^0$ and then is put at time
$t=0$ in contact with  two heat baths at inverse temperatures $\beta_1$ and
$\beta_2$, the mean heat amount received from the thermostat 1 between time $t=0$ and time $t$ is
 given (with the convention \eqref{discreteHeat}) by 
 \be
 \Esp{\Heat_1(t)}_{\Prob_0}=-\gapE
 \sum_{n_1=-\infty}^{+\infty}n_1\sum_{\sigma_1,\sigma_2}\sum_{\sigma'_1,\sigma'_2}
 \sum_{\Delta n=0,1,-1}
 \left(\sigma'_1,\sigma'_2\vert \Uev (n_1, n_1+\Delta n; t)\vert \sigma_1,\sigma_2\right)\Prob_0(\sigma_1,\sigma_2).
 \ee
 By virtue of the definition \eqref{defUevG} of the relevant generating function and the decomposition \eqref{decompUevG}, the latter expression can be rewritten as
  \be
 \Esp{\Heat_1(t)}_{\Prob_0}=-\gapE \frac{\partial}{\partial z_1}
 \left.\left(\sum_{\sigma_1,\sigma_2}\sum_{\sigma'_1,\sigma'_2} \sum_{\Delta n=0,1,-1}
 \left(\sigma'_1,\sigma'_2\vert \UevG_{[\Delta n]}(z_1,z_2; t)\vert \sigma_1,\sigma_2\right)\Prob_0(\sigma_1,\sigma_2)\right)\right\vert_{z_1=z_2=1}.
 \ee
From the explicit expressions  \eqref{UzResults}  for the matrix elements and 
\eqref {defDeltastar} for $\Deltastar_+(z)$, a straightforward calculation leads to 
\be
 \Esp{\Heat_1(t)}_{\Prob_0}=\Espst{\jinst_1} t +\left[\nuad_1\nuad_2\left(\gamma_1-\gamma_2\right)-\nuad_1\left(\gamma_1-\gammaeff^0\right)\right] \frac{\gapE}{2}\left[1-e^{-(\nu_1+\nu_2)t}\right]
\ee
where $\Espst{\jinst_1}=-\Espst{\jinst_2}$ is given in \eqref{jinstexp} and $\gammaeff^0$ is a function of $\nuad_1$,  $\nuad_2$, $\beta_1^0$ and $\beta_2^0$ written in \eqref{gammaeff}. A similar calculation yields
\be
\label{expExceeHeat}
 \Esp{\Heat_2(t)}_{\Prob_0}=\Espst{\jinst_2} t +\left[-\nuad_1\nuad_2\left(\gamma_1-\gamma_2\right)-\nuad_2\left(\gamma_2-\gammaeff^0\right)\right] \frac{\gapE}{2}\left[1-e^{-(\nu_1+\nu_2)t}\right].
\ee
As a result, with the sign convention of definition \eqref{defexcessheat}, the excess  heat given to the system by heat bath $1$ in the present protocol reads
\be
\label{ExpEcxessHeat}
\left.{\Heatav_{\text{exc}, 1}}\right\vert^{[\beta_1,\beta_2]}_{[\beta_1^0,\beta_2^0]}
=
-\left[\nuad_1\nuad_2\left(\gamma_1-\gamma_2\right)-\nuad_1\left(\gamma_1-\gammaeff^0\right)\right] \frac{\gapE}{2}.
\ee 
The excess  heat given to the system by heat bath $2$, $\left.{\Heatav_{\text{exc}, 2}}\right\vert^{[\beta_1,\beta_2]}_{[\beta_1^0,\beta_2^0]}$,  is obtained from the latter expression by the exchanges $\nuad_1 \leftrightarrow \nuad_2$ and 
$\gamma_1 \leftrightarrow \gamma_2$. These two excess heats  are not opposite to each other, and comparison with the expressions \eqref{EspstEn} for the mean energies in the initial and final stationary states shows that the sum of  the excess heats coming from both thermostats indeed satisfy the identity \eqref{sumexcess}.

\subsection{Symmetry property for reversed heat transfers (when $T_1\not=0$) specific to the model }

\subsubsection{Symmetry arising from modified detailed balance}

The symmetry properties for reversed heat transfers when $T_1\not=0$ are more conveniently exhibited 
 after a change of variable in the complex plane where the integrals involved in $\Prob_{\Prob_0}\left(\Heat_1,\Heat_2, t\right)$
are defined.
The relevant functions $\ccstar_n(\tad)$ and $\csstar_n(\tad)$ are defined in \eqref{defccstar} and \eqref{defcsstar} while $\Deltastar_+(z)$ is given in \eqref{defDeltastar}.
The origin $z=0$ is a singular point in $\Deltastar_+(z)$ and $z\Deltastar_+(z)= (A+B)P(z)$ where the second order polynomial $P(z)=(z-z_{+})(z-z_{-})$ vanishes for two roots
$z_{+}$ and $z_{-}$. The product of the roots is equal to 
\be
\label{defrho}
z_{+}z_{-}=\frac{A-B}{A+B}=\frac{1}{\rho^2}\quad\textrm{with}\quad
\rho\equiv e^{(\beta_1-\beta_2)\gapE/2}
\ee
and the sum of the roots is equal to $-(1-2A)/(A+B)$. 

When $T_1\not=0$, $z_{+}z_{-}$ does not vanish and
by using the variable change $\zeta=z/\sqrt{z_{-}z_{+}}$, namely
\be
\zeta=z \rho,
\ee
the unit circle is changed into a circle with radius $\rho$, while the roots $z_{+}$ and $z_{-}$ are changed into $\zeta_{-}$ and $\zeta_{+}$ with $\zeta_{-}\zeta_{+}=1$. 
Then, for a function such as $\cosh(\sqrt{\Deltastar_+(z))}$ or $\sinh(\tad\sqrt{\Deltastar_+(z)})/\sqrt{\Deltastar_+(z)}$, each of which is in fact  a function of $\Deltastar_+(z)$  denoted by $f(\Deltastar_+(z))$,
\be
\label{RelIntzzeta}
\oint_{|z|=1} \frac{dz}{2\pi\iexp}\frac{1}{z^{n+1}}f(\Deltastar_+(z))=\rho^n 
\oint_{|\zeta|=\rho} \frac{d\zeta}{2\pi\iexp}\frac{1}{\zeta^{n+1}}f(\widetilde{\Delta}_+(\zeta))
\ee
with 
$
\widetilde{\Delta}_+(\zeta)\equiv\Deltastar_+(\zeta/\rho)
$.
$\widetilde{\Delta}_+(\zeta)$ is a symmetric function of $\zeta$ and  $1/\zeta$,
\be
\label{DeltaPlus}
\widetilde{\Delta}_+(\zeta)=b + a \frac{\zeta+\zeta^{-1}}{2}
\ee
with $a=2\sqrt{A^2-B^2}$ and $b=1-2A$, namely
\be
a=2\nuad_1\nuad_2\sqrt{(1-\gamma_1^2)(1-\gamma_2^2)}
\ee
\be
b=1-2\nuad_1\nuad_2\left(1-\gamma_1\gamma_2\right).
\ee
Since the only singular points in the integrand in the r.h.s. of
\eqref{RelIntzzeta}  are $\zeta=0$ and $\zeta=\infty$,  the circle $\vert \zeta\vert=\rho$ can be deformed  into the unit circle and we get the identity
\be
\label{integralDeltastarDelta}
\oint_{|z|=1} \frac{dz}{2\pi\iexp}\frac{1}{z^{n+1}}f(\Deltastar_+(z))=
\rho^n\oint_{|z|=1} \frac{dz}{2\pi\iexp}\frac{1}{z^{n+1}}f(\widetilde{\Delta}_+(z)).
\ee

By inserting the latter identity in the definitions \eqref{defccstar} and \eqref{defcsstar} we get the relations
\bea
\ccstar_n(\tad)&=&\rho^n \cc_n(\tad)
\\
\nonumber
\csstar_n(\tad)&=&\rho^n \cs_n(\tad)
\eea
where
\be
\label{defcc}
\cc_n(\tad)\equiv \oint_{|z|=1} \frac{dz}{2\pi\iexp} \frac{1}{z^{n+1}}\cosh\left(\tad\sqrt{\widetilde{\Delta}_+(z)}\right)
\ee
and
\be
\label{defcs}
\cs_n(\tad)\equiv \oint_{|z|=1} \frac{dz}{2\pi\iexp} \frac{1}{z^{n+1}}\frac{\sinh\left(\tad\sqrt{\widetilde{\Delta}_+(z)}\right)}{\sqrt{\widetilde{\Delta}_+(z)}}.
\ee
Since $\widetilde{\Delta}_+(z)$ is invariant under the exchange of $z$ and $1/z$,
\bea
\cc_{n}(\tad)&=&\cc_{|n|}(\tad)
\\
\nonumber
\cs_{n}(\tad)&=&\cs_{|n|}(\tad).
\eea
Therefore the functions involved in the matrix elements \eqref{Uevn1n2Formules0}  of  $\Uev(n_1,n_2; t)$ can be rewritten as
\bea
\label{defU0U1FormulesBis}
U^{(0)}(n,\sigma_1\sigma_2;\tad)&=& 
e^{-\tad} \rho^{n}\left[\cc_{|n|}(\tad)+\gammaeff \sigma_1\sigma_2\cs_{|n|}(\tad)\right]
\\
\nonumber
U^{(1)}(n,\Delta n;\tad)&=&
e^{-\tad} \rho^{n}\left[\nuad_2 \left(1-\gamma_2\Delta n\right)\cs_{|n|}(\tad)
+\nuad_1\left(1-\gamma_1 \Delta n\right) \rho^{\Delta n}\cs_{|n+\Delta n|}(\tad)\right].
\eea
where $\rho$, defined in \eqref{defrho}, also reads $\rho=\sqrt{(1+\gamma_1)(1-\gamma_2)/(1-\gamma_1)(1+\gamma_2)}$. 

According to paper I, the modified detailed balance 
 implies some time-reversal symmetry property for histories, which itself entails
 some  relation  between probabilities of forward and backward evolutions where the given initial and final configurations are exchanged (and the heat amounts are changed into their opposite values). In the spin model language, with the definitions \eqref{discreteHeat}, the symmetry exhibited in paper I for the probability that the system evolves from an initial configuration $\C_0=(\sigma_1,\sigma_2)$ to a final configuration $\C_f=(\sigma'_1,\sigma'_2)$ while receiving the heat amount $\Heat_1=-n_1\gapE$ and $\Heat_2=n_2\gapE$ reads, for non-vanishing matrix elements,
\be
\label{TimeRevPn1n2}
\frac{(\sigma'_1,\sigma'_2\vert\Uev(n_1,n_2; t)\vert \sigma_1,\sigma_2)}{(\sigma_1,\sigma_2\vert\Uev(-n_1,-n_2; t)\vert \sigma'_1,\sigma'_2)}=e^{(n_1\beta_1-n_2\beta_2)\gapE}.
\ee
Comparison of the latter relation with the expressions \eqref{Uevn1n2Formules0}
implies that
\be
\label{symU0}
\frac{U^{(0)}(n,\sigma_1\sigma_2;\tad)}{U^{(0)}(-n,\sigma_1\sigma_2;\tad)}=e^{[n\beta_1-n\beta_2]\gapE}
\ee
and
\be
\label{symU1}
\frac{U^{(1)}(n,\Delta n;\tad)}{U^{(1)}(-n,-\Delta n;\tad)}=e^{[n\beta_1-(n+\Delta n)\beta_2]\gapE}.
\ee
The latter relations can be checked from the explicit expressions \eqref{defU0U1FormulesBis}.

We  notice that the relation \eqref{TimeRevPn1n2} can also be retrieved by noticing that the modified detailed balance entails that  the matrix $\Aev(z_1,z_2)$, which rules the evolution of $\UevG(z_1,z_2 ; \tad)$ according to \eqref{evolUevG}, obeys the symmetry $\Aev(z_1,z_2)=\Aev^T(e^{-\beta_1\gapE}/z_1,e^{\beta_2\gapE}/z_2)$, where 
$\Aev^T$ denotes the transposed matrix of $\Aev$. Therefore,
after  the variable change $z_1=\zeta_1/\rho_1$ and $z_2=\zeta_2/\rho_2$ with $\rho_1=\exp(\beta_1\gapE /2)$ and $\rho_2=\exp(-\beta_2\gapE /2)$, the matrix $\Aev(z_1,z_2)$ becomes the matrix $\widetilde{\Aev}(\zeta_1,\zeta_2)\equiv\Aev(z_1=\zeta_1/\rho_1,z_2=\zeta_2/\rho_2)$ which  obeys the symmetry $\widetilde{\Aev}^T(\zeta_1,\zeta_2)=\widetilde{\Aev}(1/\zeta_1,1/\zeta_2)$. Then the  derivation of the  symmetry \eqref{TimeRevPn1n2}  is the following. First we make  the
variable change $z_1=\zeta_1/\rho_1$ and $z_2=\zeta_2/\rho_2$
 in the integral representation \eqref{InvUev} for $\Uev(n_1,n_2 ; t)$. Since $\widetilde{\Aev}(\zeta_1,\zeta_2)$ has no non-analyticity, apart from the $1/\zeta_1$ and $1/\zeta_2$ singular terms, the integrals on the circles with radii equal to $\rho_1$ and $\rho_2$ are equal to the integrals with the same integrands on the circles with radii equal to 1. If $\zeta$ is on the unit circle, $1/\zeta$ is also on this circle and we can make the variable change $\zeta_1\to 1/\zeta_1$ and $\zeta_2\to 1/\zeta_2$ ; then the symmetry of $\widetilde{\Aev}$ leads to the
 symmetry  \eqref{TimeRevPn1n2}.
 
As shown in paper I, the consequence \eqref{TimeRevPn1n2} of the modified detailed balance \eqref{ergodicThermo} entails that, if the spin system is in an equilibrium state at inverse temperature $\beta_0$ at time $t=0$ where it is put in contact with the two thermostats, then  the ratio of the
probabilities to measure some given heat amounts $\Heat_1$ and $\Heat_2$ or
their opposite values obeys the fluctuation relation
\be
\label{OppositePQ1PQ2Probcan}
\frac{\Prob_{\Probcan^{\beta_0}}\left(\Heat_1,\Heat_2, ;t\right)}{\Prob_{\Probcan^{\beta_0}}\left(-\Heat_1,-\Heat_2; t\right)}=e^{(\beta_0-\beta_1)\Heat_1+(\beta_0-\beta_2)\Heat_2}.
\ee

\subsubsection{A relation specific to the model }

The present model happens to obey a very specific relation for reversed heat transfers when  the initial state of the system has an arbitrary probability distribution $\Prob_0$. 
According to \eqref{Uevn1n2Formules0}, after summation over the final configuration,
\bea
\label{EvolAllprobbis}
\sum_{\sigma'_1,\sigma'_2}(\sigma'_1,\sigma'_2\vert\Uev(n,n; t)\vert \sigma_1,\sigma_2)
&=&U^{(0)}(n,\sigma_1\sigma_2;\tad)
\\
\nonumber
\sum_{\sigma'_1,\sigma'_2}(\sigma'_1,\sigma'_2\vert\Uev(n,n +\Delta n; t)\vert \sigma_1,\sigma_2)
&\underset{\Delta n=\pm1}{=}&
U^{(1)}(n,\Delta n;\tad)\,\delta_{\sigma_1\sigma_2,\Delta n}.
\eea
Therefore, when the initial configurations are distributed with an arbitrary probability $\Prob_0$
\be
\label{Probnn1}
\sum_{\sigma'_1,\sigma'_2}(\sigma'_1,\sigma'_2\vert\Uev(n,n; t)\vert \Prob_0)
= 
\sum_{\sigma_1,\sigma_2} U^{(0)}(n,\sigma_1\sigma_2;\tad)\Prob_0\left(\sigma_1,\sigma_2\right)
\ee
and
\be
\label{Probnn2}
\sum_{\sigma'_1,\sigma'_2}(\sigma'_1,\sigma'_2\vert\Uev(n,n +\Delta n; t)\vert \Prob_0)
\underset{\Delta n=\pm1}{=}
U^{(1)}(n,\Delta n;\tad)\times  \Prob_0\left(\sigma_1\sigma_2=\Delta n \right)
\ee
where $\Prob_0\left(\sigma_1\sigma_2=\Delta n \right)$ denotes the probability that the product $\sigma_1\sigma_2$ is equal to $\Delta n$ in the initial configuration.

Then from the expressions \eqref{Probnn1} and \eqref{Probnn2} for  $\Prob_{\Prob_0}\left(\Heat_1,\Heat_2, t\right)$ in the cases $\Heat_1=-\Heat_2$ and $\Heat_1\not=-\Heat_2$ respectively, and by virtue of the consequences \eqref{symU0}-\eqref{symU1} of the modified detailed balance, we get the property
\be
\label{OppositePQ1PQ2Prob0}
\frac{\Prob_{\Prob_0}\left(\Heat_1,\Heat_2, ;t\right)}{\Prob_{\Prob_0}\left(-\Heat_1,-\Heat_2; t\right)}=e^{-\beta_1\Heat_1-\beta_2\Heat_2}
\left[\delta_{ \Heat_1+ \Heat_2,0}+
\sum_{\epsilon=\pm1}\delta_{ \Heat_1+ \Heat_2,\epsilon\gapE}
\frac{\Prob(\sigma_1\sigma_2=\epsilon; t=0)}{1-\Prob(\sigma_1\sigma_2=\epsilon; t=0)}.
\right].
\ee
The appearance of the initial
probability for the sign of the spins product  seemingly arises from the fact that, by virtue of
energy conservation, the values of this sign in the final and initial states are
related to the sum $\Heat_2+\Heat_1$ by the constraint \eqref{correspondance}.

We stress that this relation is very specific to the present model. Its interest
lies not in its precise form, but in that the right-hand side involves
the initial distribution : it has sometimes been speculated that the
(experimental) study of the ratio on the left-hand side for general systems
could give some clues about the initial distribution. The above formula
justifies this hope, but shows at the same time that even in the simple case at
hand only partial information can be retrieved, and suggest that for more
general systems even this partial information may be difficult to extract. In
the cases where the initial distribution is equal either to an equilibrium state
distribution at the inverse temperature $\beta_0$ or to the stationary state
distribution which is a canonical distribution at the inverse temperature
$\betaeff$, the relation allows to retrieve the generic relation \eqref{OppositePQ1PQ2Probcan}.

\subsection{Decaying property of joint probabilities for large heat exchanges}
\label{DecayProbn}

All quantities of interest involve the coefficients $\ccstar_m(\tad)$ and/or
$\csstar_m(\tad)$, computed via contour integrals in \eqref{defccstar}
\eqref{defcsstar}. The integrands involve functions which are holomorphic for $z
\in {\mathbb C}^*$, the pointed complex plane: the quantity $\Deltastar_+(z)$
defined in \eqref{defDeltastar} has this property and the functions $\cosh (w)$
and $\sinh (w)/w$ are entire even functions, so that the square roots in
the composed functions $\cosh\left(\tad\sqrt{\Deltastar_+(z)}\right) \text{ and }
\sinh\left(\tad\sqrt{\Deltastar_+(z)}\right)/\sqrt{\Deltastar_+(z)}$
do no harm.

Hence, in the formul\ae\ for $\ccstar_m(\tad)$ and $\csstar_m(\tad)$, contours
can be deformed. For $r\in ]0,+\infty[$, let $C(\tau,r)\equiv \sup_{|z|=r}
|\cosh\left(\tad\sqrt{\Deltastar_+(z)}\right) | < +\infty$ and $S(\tau,r) \equiv
\sup_{|z|=r} \left| \sinh\left(\tad\sqrt{\Deltastar_+(z)}\right)/
 \sqrt{\Deltastar_+(z)}\right|< +\infty$.

Taking $|z|=r$ as integration contour, one gets immediately that, for each $r$,
$\ccstar_m(\tad)\leq C(\tau,r) r^{-m}$ and
$\csstar_m(\tad)\leq S(\tau,r) r^{-m}$. This shows that $\ccstar_m(\tad)$ and
$\csstar_m(\tad)$ are $o(e^{-K|m|})$ at large $|m|$ for any $K$.

With some efforts, we could get some explicit upper bounds for $C(\tau,r)$ and
$S(\tau,r)$. Then we could extremize over $r$ to get a subexponential bound for
$\ccstar_m(\tad)$ and $\csstar_m(\tad)$, but we shall not need this refinement.

In the limit $T_1=0$, $\Deltastar_+(z)$ is in fact holomorphic for $z
\in {\mathbb C}$ so that $\ccstar_m(\tad)$ and $\csstar_m(\tad)$ vanish for
$m=-1,-2,\cdots$.

\clearpage
\section{Heat amount cumulants  for any $T_1$ and $T_2$}

\label{LongTimeCumulantsExp}

\subsection{Generic properties for a system with a finite number of configurations}

\subsubsection{Characteristic function for the heat amount $\Heat_2$}
\label{CharFunction}

The random variable $\Heat_2$ can take only discrete values $n_2\gapE$, where $n_2$ is a positive or negative integer. Therefore its probability density $\Probdist(\Heat;t)$, defined as $\Probdist(\Heat;t)d\Heat=\Prob\left(\Heat_2\in [\Heat, \Heat+d\Heat[;t\right)$, reads
\be
\label{ProdistDiscrete}
\Probdist(\Heat;t)=\frac{1}{\gapE}\sum_{n_2=-\infty}^{+\infty} \delta\left(\frac{\Heat}{\gapE}-n_2\right)
\Prob(n_2;t),
\ee
where $\delta$ stands for the Dirac distribution.
Since $\Prob(n_2;t)$ decays faster than $\exp(- K |n_2|)$ for any  $K>0$ when $|n_2|$ goes to infinity (see subsection \ref{DecayProbn}),
 the Laplace transform $\widetilde{G}(\lambda;t)$ of $\Probdist(\Heat)$, i.e. the characteristic function of the random variable $\Heat_2$, is well defined for any $\lambda$,\be
\widetilde{G}(\lambda;t)\equiv \int_{-\infty}^{+\infty} d\Heat  e^{\lambda \Heat} \Probdist(\Heat;t)
=\Esp{e^{\lambda \Heat_2(t)}}.
\ee
As a consequence, the properties of  the probability density
$\Probdist(\Heat;t)$ can be investigated through those of its Laplace transform,
thanks to the inversion formula
\be
 \label{inverseLT}
\Probdist(\Heat;t)=\int_{-\iexp\infty}^{+\iexp\infty}
 \frac{d \lambda}{2\pi \iexp} e^{-\lambda \Heat} \widetilde{G}(\lambda;t).
\ee
According to the property \eqref{ProdistDiscrete}, $\widetilde{G}(\lambda)$ is a periodic function of $\lambda$ with period equal to $\iexp 2 \pi /\gapE$, and the r. h. s. of the latter formula can be written as 
\be
\frac{1}{\gapE} \int_0^{2\pi} \frac{d \theta}{2\pi}e^{-\iexp \theta \Heat/\gapE}
\widetilde{G}\left(\frac{\iexp \theta}{\gapE};t\right)\sum_{m=-\infty}^{+\infty}e^{-\iexp 2\pi m \Heat/\gapE}.
\ee 
By virtue of Poisson equality $\sum_{m=-\infty}^{+\infty}e^{-\iexp 2\pi m \Heat/\gapE}
=\sum_{n_2=-\infty}^{+\infty} \delta\left(\frac{\Heat}{\gapE}-n_2\right)
$, 
comparison with \eqref{ProdistDiscrete} leads to 
\be
\Prob(n_2;t)=  \int_0^{2\pi} \frac{d\theta}{2\pi} e^{-\iexp \theta n_2}\widetilde{G}\left(\frac{\iexp \theta}{\gapE};t\right).
\ee
The latter equality coincides with the inverse formula  (analogous to \eqref{InvUev}) in terms of the generating function, $G(z;t)=\sum_{n_2=-\infty}^{+\infty} z^{n_2}\Prob(n_2;t)$,
\be
\Prob(n_2;t)= \oint_{|z|=1} \frac{dz}{2\pi \iexp}\frac{1}{z^{n_2+1}}G(z;t)
\ee
 where $G(z;t)\equiv\widetilde{G}((1/\gapE)\ln z;t)$.

\subsubsection{Relation between long-time cumulants per unit time for $\Heat_1$ and $\Heat_2$}

The generic properties of the cumulants of $\Heat_1$ and $\Heat_2$ have been reviewed in paper I. We recall those which will be useful in the following.
For a Markov  process, the long-time behaviors of these cumulants are proportional to the time $t$ elapsed from the beginning of the measurements. The asymptotic behavior of the cumulants per unit time are given by the derivatives of 
\be
\label{defalphaa}
\alpha_a(\lambda)\equiv\lim_{t\to+\infty}\frac{1}{t}\ln \Esp{e^{\lambda\Heat_a(t)}}
\ee
according to
\be
\label{defCumulantpertime} 
\lim_{t\to+\infty}\frac{\kappa^{[p]}_{\Heat_a}}{t}
=\left.\frac{\partial^p \alpha_{a}(\lambda_a)}{\partial \lambda_a^p}\right\vert_{\lambda_a=0}
\quad\textrm{for $a=\{1,2\}$}.
\ee

Moreover, in the case of a system with a finite number of configurations,
$\Heat_1+\Heat_2=\En(\C_f)-\En(\C_0)$ is restricted to some finite interval and
\be
\label{alpha12finiteBis} 
\alpha_1(\lambda)=\alpha_2(-\lambda).
\ee
As a consequence  the long-time cumulants per unit time obey the following relations
\be
\label{longtimekappa}
\lim_{t\to\infty}\frac{\kappa^{[p]}_{\Heat_1}}{t}=(-1)^p\lim_{t\to\infty}\frac{\kappa^{[p]}_{\Heat_2}}{t}.
\ee

\subsection{Explicit formul\ae\ for the cumulants per unit time}
\label{ExplicitCumulants}

According to the relation \eqref{longtimekappa} between the long-time  cumulants per unit time for $\Heat_1$ and $\Heat_2$, we have only to consider the cumulants for the heat amount $\Heat_2$ received from  bath $2$.
For the two-spin system, where $\Heat_2=n_2 \gapE$, it is convenient to introduce the cumulants $\kappa^{[p]}_{n_2}$ for the dimensionless variable $n_2$ and the associated characteristic function $\Esp{e^{\lad n_2}}$, where $\lad$ is a dimensionless variable. According to the relation \eqref{defCumulantpertime} the 
 long-time behavior of the cumulants per unit time  are derived 
 through the relation
\be
\label{kappan1p}
\lim_{t\to+\infty}\frac{1}{t}\kappa_{n_2}^{[p]}
=\left.\frac{\partial^p \alpha_2(\lad)}{\partial \lad^p}\right\vert_{\lad=0}
\ee
with
\be
\label{defalpha2}
\alpha_2(\lad)=\lim_{t\to+\infty}\frac{1}{t}\ln \Esp{e^{\lad n_2}}.
\ee
According to the definition  of  $\Esp{e^{\lad n_2}}$, the relation \eqref{defUev} between the probability $\Prob(\sigma_1,\sigma_2,n_1,n_2;t)$ and the operator $\Uev(n_1,n_2;t)$, together with the definition \eqref{defUevG} of
$\UevG(z_1,z_2 ; t)$ and its evolution equation  \eqref{evolUevG}, the characteristic function may be expressed as
\be
\Esp{e^{\lad n_2}}=\sum_{\sigma_1,\sigma_2}\sum_{\sigma'_1,\sigma'_2}
\left(\sigma_1,\sigma_2\vert e^{\frac{\nu_1+\nu_2}{2}t\,\Aev(z_1= 1, z_2=e^{\lad})} \vert \sigma'_1,\sigma'_2\right)
\Prob(\sigma'_1,\sigma'_2; t=0).
\ee
According to \eqref{Aev2}, $\Aev(z_1= 1, z_2=e^{\lad})+(1+\gammaeff)\Id_4$ is a real positive matrix and the Perron-Frobenius theorem can be applied. Henceforth 
$\alpha_2(\lad)$ coincides with the eigenvalue of the matrix $[(\nu_1+\nu_2)/2]\Aev(z_1= 1, z_2=e^{\lad})$ with the largest modulus (and which is necessarily real). The four eigenvalues to consider are the $\mu^{(\epsilon,\eta)}(z=e^{\lad})$'s which are given by the expression \eqref{AevEigenvalues}, with
$\epsilon=\pm$ and $\eta=\pm$. The  one with the largest modulus corresponds to $(\epsilon ,\eta)=(+,+)$ and reads
\be
\label{alphaAB}
\alpha_2(\lad)=\frac{\nu_1+\nu_2}{2}\left[-1+\sqrt{1-2A+(A+B)e^{\lad}+(A-B)e^{-\lad}}\right]
\ee
where $A$ and $B$ are defined in \eqref{defAB}.

\begin{figure}[h!]
  \centering
 \includegraphics[scale=0.12]{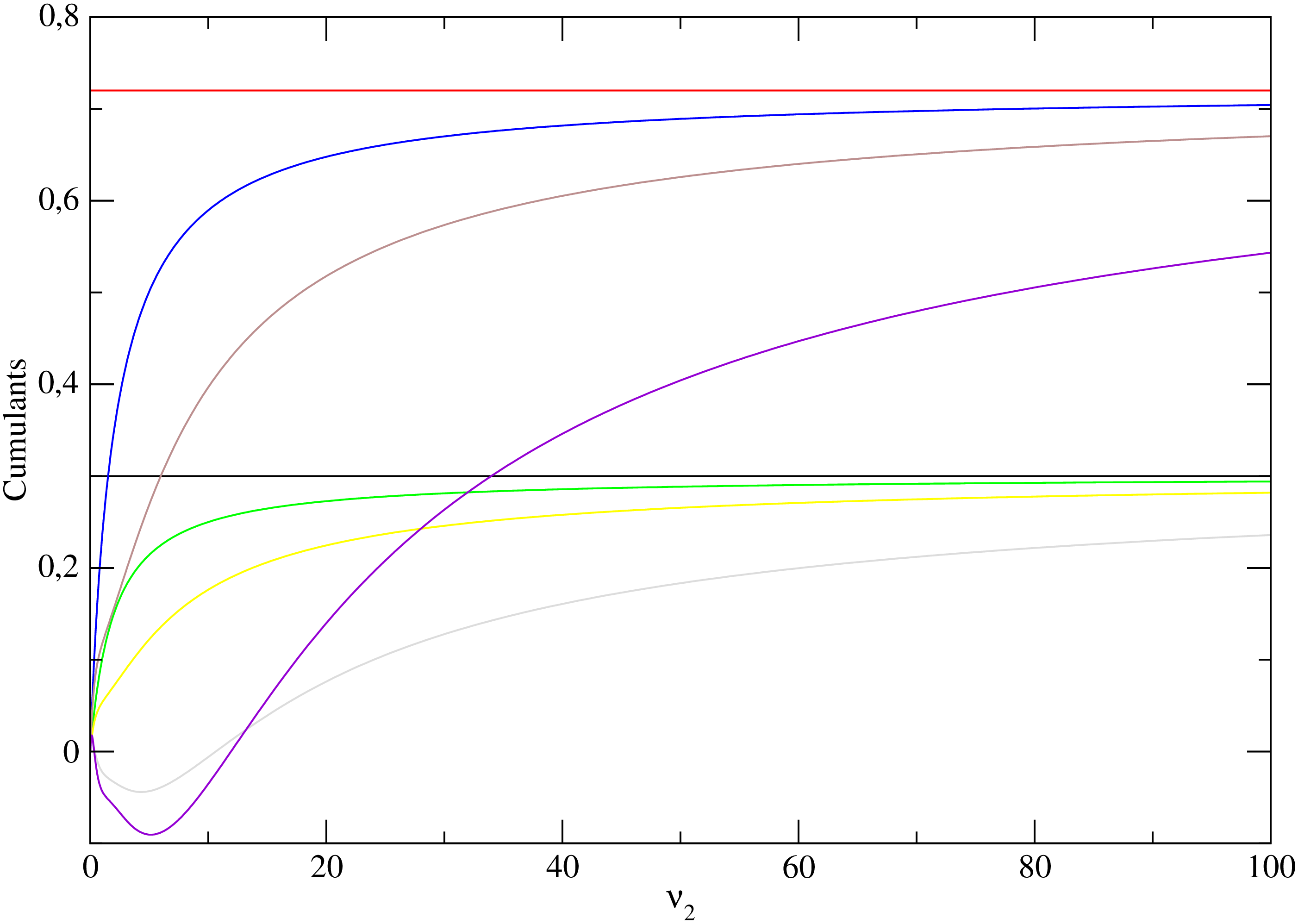}
 \caption{First infinite-time cumulants per unit time as  functions of $\nu_2/\nu_1$. Illustration of the
   large $\nu_2/\nu_1$ behavior with the two asymptotes, $\frac{\nu_1}{2}(\gamma_1-\gamma_2)$ for odd cumulants and 
   $\frac{\nu_1}{2}(1-\gamma_1\gamma_2)$ for even cumulants. The other model parameters are fixed to the
   sample values $\gamma_1=0.7$, $\gamma_2=0.4$, $\nu_1=2$.}\label{LDnuvary1}
\end{figure}

\begin{figure}[h!]
  \centering
 \includegraphics[scale=0.12]{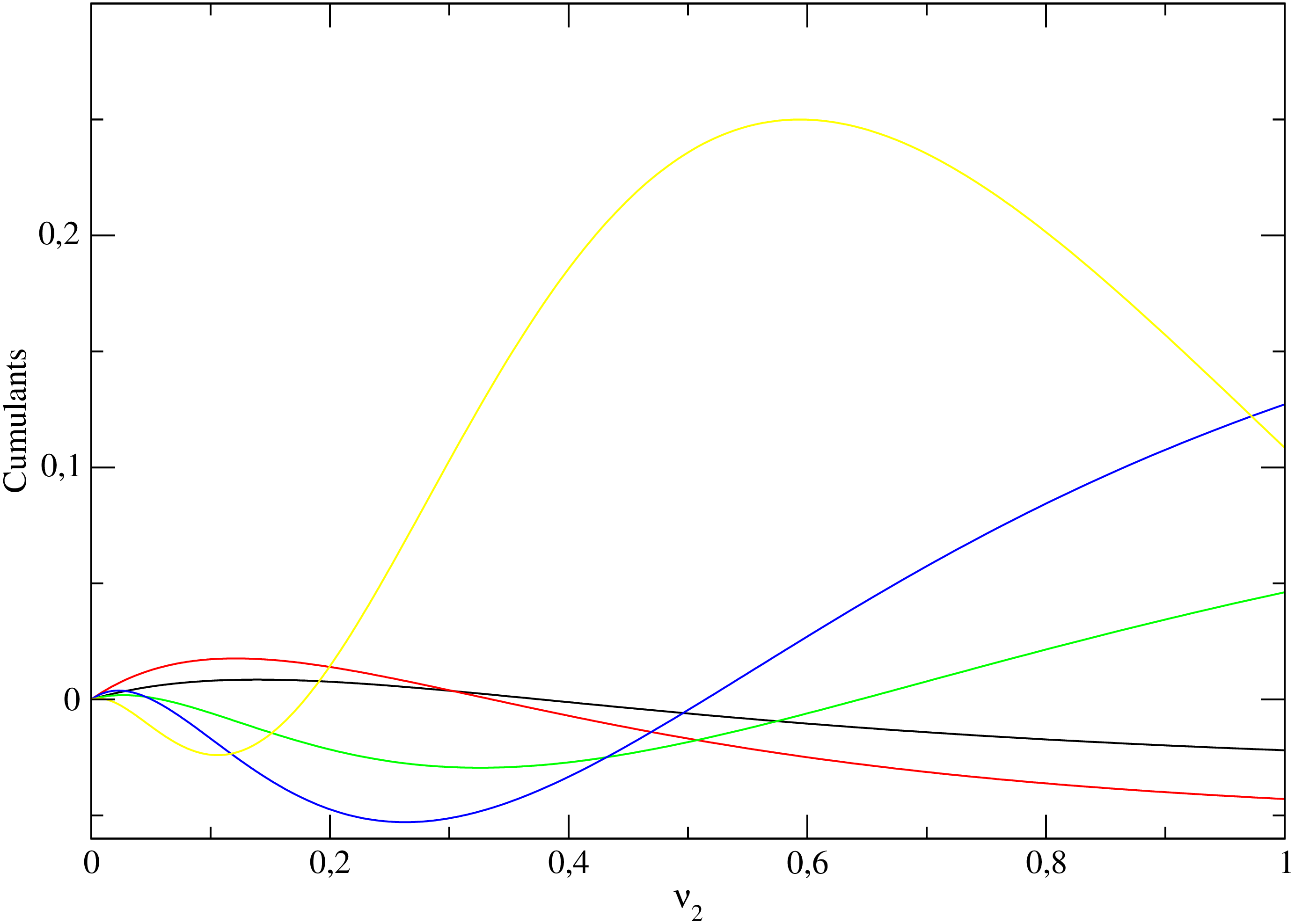}
 \caption{Higher cumulants. Illustration of the oscillations at finite
   $\nu_2/\nu_1$. The other model parameters are fixed to the sample values
   $\gamma_1=0.7$, $\gamma_2=0.4$, $\nu_1=2$.}\label{LDnuvary2}
\end{figure}

It is plain to calculate a number of cumulants per unit time in the infinite-time limit from \eqref{kappan1p}. Their
behavior as a function of the kinetic parameter $\nu_2/\nu_1$ exhibits some
interesting features. For large $\nu_2/\nu_1$, the cumulants go to a limit which
is the same for all odd and for all even cumulants, as will be explained in
section \ref{DependenceTimeScales}. Fig.\ref{LDnuvary1} illustrates this
convergence, which gets slower and slower for higher moments. The first six
cumulants are represented. This figure also shows some oscillations at finite
$\nu_2/\nu_1$.  These oscillations become more and more visible on higher
cumulants.  Fig.\ref{LDnuvary2} illustrates this phenomenon. Cumulants from the
fifth to the ninth are represented. In both figures, the other model parameters
are fixed to the sample values $\gamma_1=0.7$, $\gamma_2=0.4$, $\nu_1=2$.

\clearpage
Only the first cumulants have analytic expressions simple enough to fit on
a line. For the sake of conciseness, the results are first expressed in terms of the dimensionless time $\tad=\frac{1}{2}(\nu_1+\nu_2) t$ as
\bea
\label{longtimekappa1a4}
\lim_{\tad\to+\infty}\frac{1}{\tad}\kappa_{n_2}^{[1]}&=& B
\\
\nonumber
\lim_{\tad\to+\infty}\frac{1}{\tad}\kappa_{n_2}^{[2]}&=& A-B^2
\\
\nonumber
\lim_{\tad\to+\infty}\frac{1}{\tad}\kappa_{n_2}^{[3]}&=& B\left[1-3A+3B^2\right]
\\
\nonumber
\lim_{\tad\to+\infty}\frac{1}{\tad}\kappa_{n_2}^{[4]}&=& A-3A^2
+B^2\left[-4+18A-15B^2\right].
\eea
All odd cumulants are proportional to $B$, because all odd powers of $\lad$ in the expansion of the expression \eqref{alphaAB} for $\alpha_2(\lad)$ are proportional to $B$.
The first three cumulants are rewritten in terms of the model parameters as
\bea
\lim_{t\to+\infty}\frac{\Espst{\Heat_2}}{t}
&=&\nuad_1\nuad_2\left(\gamma_1-\gamma_2\right)\frac{(\nu_1+\nu_2)\gapE}{2} 
=\Espst{\jinst_2}
\nonumber
\\
\label{longtimekappa2}
\lim_{t\to+\infty}\frac{\Espst{\Heat_2^2}-\Espst{\Heat_2}^2}{t}
&=&\nuad_1\nuad_2\left[1-\gamma_1\gamma_2-\nuad_1\nuad_2(\gamma_1-\gamma_2)^2\right]\frac{(\nu_1+\nu_2)(\gapE)^2}{2} 
\\
\nonumber
\lim_{t\to+\infty}\frac{\Espst{\Heat_2^3}^c}{t}
&=&\nuad_1\nuad_2\left(\gamma_1-\gamma_2\right)\left[
1-3\nuad_1\nuad_2(1-\gamma_1\gamma_2)+3\nuad_1^2\nuad_2^2\left(\gamma_1-\gamma_2\right)^2
\right]\frac{(\nu_1+\nu_2)(\gapE)^3}{2}.
\eea
$\Esp{\Heat_2^3}^c$ is the third cumulant, which is   equal to  the third centered moment, namely $\Esp{\Heat_2^3}^c\equiv\Esp{\left[\Heat_2-\Esp{\Heat_2}\right]^3}$.

At equilibrium $\gamma_1=\gamma_2$, so that $B=0$: then, by virtue of the remark
after \eqref{longtimekappa1a4},  the long-time behavior of all odd cumulants of
$\Heat_2$ is subdominant with respect to the elapsed time $t$, and in the
long-time limit $\Prob(\Heat_2;t)$  becomes  an even  function of $\Heat_2$ at
leading order in time $t$. The fourth cumulant of the cumulated heat $\Heat_2$
received from the thermostat $2$ per unit time does not vanish:
$\lim_{t\to+\infty}\frac{1}{t}\ln \Espeq{e^{\lambda \Heat_2}}$ is not quadratic
in $\lambda$, and even in the long time limit the variable $\Heat_2$ has a non-Gaussian distribution, contrarily to the variable $\left[\Heat_2-\Esp{\Heat_2}\right]/\sqrt{t}$ (for which all cumulants of order larger than $3$ vanish in the infinite time limit).
The first two even cumulants per unit time read
\bea
\label{Cumlulant2Eq}
\lim_{t\to+\infty}\frac{\Espeq{\Heat_2^2}-\Espeq{\Heat_2}^2}{t}
&=&\nuad_1\nuad_2\left(1-\gamma^2\right)\frac{(\nu_1+\nu_2)(\gapE)^2}{2} 
\\
\nonumber
\lim_{t\to+\infty}\frac{\Espeq{\Heat_2^4}^c}{t}
&=&\nuad_1\nuad_2\left(1-\gamma^2\right)
\left[1-3\nuad_1\nuad_2(1-\gamma^2)\right]\frac{(\nu_1+\nu_2)(\gapE)^4}{2}.
\eea
$\Esp{\Heat_2^4}^c$ denotes the fourth cumulant, which can be expressed as
$\Esp{\Heat_2^4}^c=\Esp{\left[\Heat_2-\Esp{\Heat_2}\right]^4}-3\Esp{\left[\Heat_2-\Esp{\Heat_2}\right]^2}$.

 For a system weakly out of equilibrium  the Einstein-Green-Kubo relation, namely 
 \be
\label{GreenKuboHeatBis}
\lim_{(\beta_1,\beta_2)\to(\beta,\beta)}\frac{\Espst{\jinst_2}}{\beta_1-\beta_2}
=\frac{1}{2}
\lim_{t\to+\infty} \frac{\Espeq{\Heat_2^2}-\left(\Espeq{\Heat_2}\right)^2}{t},
\ee
is indeed obeyed by the system, as it should be. This can be checked by comparing the expression \eqref{Cumlulant2Eq} with the limit obtained when $(\beta_1,\beta_2)\to(\beta,\beta)$ for the ratio $\Espst{\jinst_2}/(\beta_1-\beta_2)$ which, by  virtue of \eqref{jinstexp}, reads
\be
\label{expLnlinBis}
\frac{ \Espst{\jinst_2}}{\beta_1-\beta_2}=
\nuad_1\nuad_2\frac{\gamma_1-\gamma_2}{(\beta_1-\beta_2)\gapE}\frac{(\nu_1+\nu_2)(\gapE)^2}{2}.
\ee When the system is far from equilibrium, comparison of the latter expression
for $\Espst{\jinst_2}/(\beta_1-\beta_2)$ with the expression
\eqref{longtimekappa2} for the long-time limit of the second cumulant per unit
time shows that ${\Espst{\jinst_2}/(\beta_1-\beta_2)\not=
\lim_{t\to+\infty}[\Espst{\Heat_2^2} -\Espst{\Heat_2}^2]/t}$ , as it should be (see
subsection 5.3 of paper I). Indeed, by virtue of equation \eqref{alphaAB},
$\alpha_2(\lad)$ obeys the symmetry relation
$\alpha_2(\lad)=\alpha_2(-\overline{\Fth} - \lad)$ with $\overline{\Fth}=\ln
(A+B)/(A-B)= (\beta_1-\beta_2)\gapE$, but $\alpha_2(\lad)$ is not a quadratic
function of $\lad$, i.e $\Heat_2$ has a non-Gaussian distribution in the
long-time limit.

\section{Large deviation function for the cumulated heat current $\Heat_2/t$}

\label{LargeDeviationFunction}

In this section, we derive the large deviation function for the cumulative heat
current $\Heat_2/t$ by three methods. The first one is based on the general
theory of large deviations for the definition of large deviation functions and
uses one of its cornerstones, the Gärtner-Ellis theorem. The second and the
third rely on the fact that $\Heat_2$ takes discrete values in a $t$-independent
set, and uses an ad-hoc definition of large deviation functions (see Appendix E
of paper I). Though the general theory of large deviations and the ad-hoc
definition for discrete exchanged quantities do not have to be the same, the
ad-hoc definition is nevertheless a sensible definition of large deviations.
Physically, the general and the ad-hoc definition are expected to yield the same
result in a case as simple as the two-spin system, and our explicit
computations can be seen as a proof of this fact. A natural tool to compute the
ad-hoc large deviation function is via a contour integral representation, but as
we shall see below, this method is surprisingly tricky even for the simple
two-spin system at hand. In contrast with the general theory of large
deviations, the contour integral method is the basis of a systematic expansion
at large times. However, corrections are less universal than the dominant term. 

The cumulative heat current received from  heat bath $2$ during the time interval $t$ takes the values  $\Jcum=\Heat_2/t$, with $\Heat_2=n\gapE$, $n$ integer.
By dimensional analysis,  the large deviation function  $f_{\Heat_2}(\Jcum)$, which has the dimension of an inverse time, must be a function of 
\be
\Jcumad=\frac{\Jcum}{\gapE}=\frac{n}{t},
\ee
and  we shall often consider the expressions of
\be
\ft_{\Heat_2}(\Jcumad)\equiv f_{\Heat_2}(\Jcum)
\ee
rather than those of $f_{\Heat_2}(\Jcum)=\ft_{\Heat_2}(\Jcum/\gapE)$. Moreover, the explicit calculations are more conveniently dealt with if, instead of considering $\Jcumad$, we introduce the dimensionless current $\jad$ associated with the dimensionless time
$\tad=[(\nu_1+\nu_2)/2]t$,
\be
\label{defjad}
\jad=\frac{n}{\tad}=\frac{2}{(\nu_1+\nu_2)}\Jcumad.
\ee
The dimensionless large deviation function $\fad_{\Heat_2}$ of $\jad$ is such that $t\ft_{\Heat_2}(\Jcumad)=\tad \fad_{\Heat_2}(\jad)$, and  the expression of $\ft_{\Heat_2}(\Jcumad)$ can be retrieved  from  that for $\fad_{\Heat_2}(\jad)$ through
\be
\label{relffad}
\ft_{\Heat_2}(\Jcumad)=\frac{\nu_1+\nu_2}{2} \,\fad_{\Heat_2}\left(\frac{2}{(\nu_1+\nu_2)}\Jcumad\right).
\ee

We notice that large deviation functions for other cumulative quantities are related to $f_{\Heat_2}$. Indeed, in  a system with a finite number of configurations $\Heat_1+\Heat_2$ is bounded and, as a consequence of the general theory of large deviations (see e.g.  paper I),
\be
f_{\Heat_1}(\Jcum)=f_{\Heat_2}(-\Jcum).
\ee
In the same vein, as
$\Deltaexch S=\beta_1\Heat_1+\beta_2\Heat_2=-(\beta_1-\beta_2) \Heat_2+\beta_1(\Heat_1+\Heat_2)$, with $\Heat_1+\Heat_2$ bounded, the large deviation function for $\Deltaexch S$ and that for $\Heat_2$ satisfy the simple relation
\be
\label{relLDSLDHea2}
f_{\Deltaexch S}(\Jcum)=f_{\Heat_2}\left(- \frac{\Jcum}{\beta_1-\beta_2}\right).
\ee

\subsection{Derivation from Gärtner-Ellis theorem}
\label{GartnerEllisLD}

\subsubsection{Method}

By analogy with \eqref{defalpha2}, we introduce the dimensionless function
\be
\overline{\alpha}_2(\lad)\equiv\lim_{\tad\to+\infty}\frac{1}{\tad}\ln \Esp{e^{\lad n_2}}.
\ee
A simplified version of the Gärtner–Ellis theorem (see e.g. the review for physicists
 \cite{Touchette2009}  or the mathematical point of view 
 \cite{DemboZeitouni1998}) states that, if $\overline{\alpha}_2(\lad)$  exists and is differentiable for all $\lad$ in $\mathbb{R}$, then the large deviation function of the current $\jad=n_2/\tad$
exists and it can be calculated as the  Legendre-Fenchel  transform of $\overline{\alpha}_2(\lad)$, namely, with the signs chosen in the definitions used in the present paper,
\be
\fad_{\Heat_2}(\jad)=\min_{\lad\in \mathbb{R}}\{\overline{\alpha}_2(\lad)-\lad \jad\}.
\ee
As a consequence, since $\overline{\alpha}_2(\lad)$ obeys the symmetry 
$\overline{\alpha}_2(\lad)=\overline{\alpha}_2(-(\beta_1-\beta_2)\gapE-\lad)$ (as can be checked from \eqref{alphaAB}), 
$\fad(\jad)$ obeys the fluctuation relation $\fad(\jad)-\fad(-\jad)=(\beta_1-\beta_2)\gapE \times \jad$.
Moreover, the cumulant generating function $\ln\Esp{e^{\lad n_2}}$ is necessarily convex (downward). In the present case $\overline{\alpha}_2(\lad)$ is  strictly convex and continuously differentiable for all real $\lad$, so that the minimum in the definition of the Legendre-Fenchel transform can be readily calculated by using the Legendre transform,
\be
\fad_{\Heat_2}(\jad)=\overline{\alpha}_2\left(\lad_c(\jad)\right)-\jad \lad_c(\jad)
\quad\textrm{with}\quad \frac{d\overline{\alpha}_2}{d \lad}(\lad_c)=\jad.
\ee

\subsubsection{Various expressions for $f_{\Heat_2}$ and its properties}

From the  relation $\alpha_2(\lad)=[(\nu_1+\nu_2)/2] \overline{\alpha}_2(\lad)$ and the expression \eqref{alphaAB} for $\alpha_2(\lad)$, when $T_1\not=0$ ($\gamma_1\not=1$),  $A\not=B$ and we get
\be
\label{fjadAB}
\fad_{\Heat_2}(\jad)=\jad\ln \sqrt{\frac{A+B}{A-B}}
-|\jad| \cosh^{-1}\left[\frac{Y(\jad)}{\sqrt{A^2-B^2}} \right]
-1+\sqrt{1-2A+2Y(\jad)}.
\ee
$\cosh^{-1}x$ denotes the positive real whose hyperbolic cosine is equal to $x$, namely
$ \cosh^{-1}x=\ln \left[x+\sqrt{x^2-1}\right]$, and
\be
\label{defYjad}
Y(\jad)= \jad^2 +\sqrt{\jad^4+(1-2A) \jad^2+A^2-B^2}.
\ee
The expression for $\fad_{\Heat_2}(\jad)$ involves the combinations of the model parameters
\be
\ln \sqrt{\frac{A+B}{A-B}}=(\beta_1-\beta_2)\frac{\gapE}{2},
\quad
A^2-B^2=\nuad_1^2\nuad_2^2 \left(1-\gamma_1^2\right) \left(1-\gamma_2^2\right),
\quad
A=\nuad_1\nuad_2(1-\gamma_1\gamma_2).
\ee

The expression \eqref{fjadAB} for $\fad_{\Heat_2}(\jad)$ can be
rewritten in two different forms according to the sign of $\jad$.
By using the identity
$ \cosh^{-1}x=\ln \left[x+\sqrt{x^2-1}\right]$, the $\cosh^{-1}$ term in \eqref{fjadAB} can be split into two contributions and, according to the sign of $\jad$, we get 
\be
\label{fjneg}
\fad_{\Heat_2}(\jad)\underset{\jad<0}{=} -\jad \ln (A-B)
+\jad \ln\left[Y(\jad) + \sqrt{Y^2(\jad)-\left(A^2-B^2\right)}\right] 
-1+\sqrt{1-2A+2Y(\jad)},
\ee
while
\be
\label{fjpos}
\fad_{\Heat_2}(\jad)\underset{\jad>0}{=}+\jad \ln (A+B)
-\jad \ln\left[Y(\jad) + \sqrt{Y^2(\jad)-\left(A^2-B^2\right)}\right] 
-1+\sqrt{1-2A+2Y(\jad)}.
\ee
In the limit where $T_1$ vanishes ($A\to B$), the latter expressions yield the results discussed in section \ref{T1zero}.

The thermodynamical and kinetic parameters of the heat baths are disentangled if, in place of  $A$ and $B$, we consider the parameters
\be
\label{defp+p-}
p_+=\frac{1}{2}(1+\gamma_1)(1-\gamma_2)
\quad\textrm{and}\quad
p_-=\frac{1}{2}(1-\gamma_1)(1+\gamma_2).
\ee
The relations with $A$ and $B$ are $A=\nuad_1\nuad_2 \left(p_++p_-\right)$ and $B=\nuad_1\nuad_2 \left(p_+-p_-\right)$. Therefore, 
${(A+B)/(A-B)=p_+/p_-}$, $\sqrt{A^2-B^2}=2\nuad_1\nuad_2 \sqrt{p_+p_-}$. Then, by virtue of the relation \eqref{relffad} and the expression \eqref{fjadAB} for $\fad_{\Heat_2}(\jad)$,  
$\ft_{\Heat_2}(\Jcumad)$ reads
\be
\label{fjadpppm}
\ft_{\Heat_2}(\Jcumad)=\Jcumad\ln \sqrt{\frac{p_+}{p_-}}
-|\Jcumad| \cosh^{-1}\left[\frac{Z(\Jcumad)}{\sqrt{p_+p_-}} \right]
+\frac{\nu_1+\nu_2}{2}\left[-1+\sqrt{1-2\nuad_1\nuad_2\left[p_++p_- -2Z(\Jcumad)\right]}\right].
\ee
where, with the definition $Y(\jad)/\sqrt{A^2-B^2}\equiv Z(\Jcumad)/\sqrt{p_+p_-}$,
\be
\label{defZj}
Z(\Jcumad)= \frac{1}{\nu_1\nu_2}\left[2\Jcumad^2 +\sqrt{4\Jcumad^4+\left[(\nu_1+\nu_2)^2-2\nu_1\nu_2(p_++p_-)\right]\Jcumad^2+(\nu_1\nu_2)^2p_+p_-}\right].
\ee
By virtue of the definitions \eqref{defp+p-} of $p_+$ and $p_-$,  the thermodynamic parameters of the thermal baths appear in $\ft_{\Heat_2}(j)$ through  the following combinations
\be
\label{pppmparameters}
\sqrt{\frac{p_+}{p_-}}=e^{(\beta_1-\beta_2)\frac{\gapE}{2}},
\qquad \quad\sqrt{p_+p_-}=\frac{1}{2}\sqrt{\left(1-\gamma_1^2\right)\left(1-\gamma_2^2\right)},
\qquad \quad p_++p_-=1-\gamma_1\gamma_2.
\ee

\begin{figure}[h!]
  \centering
  \includegraphics[width=\textwidth]{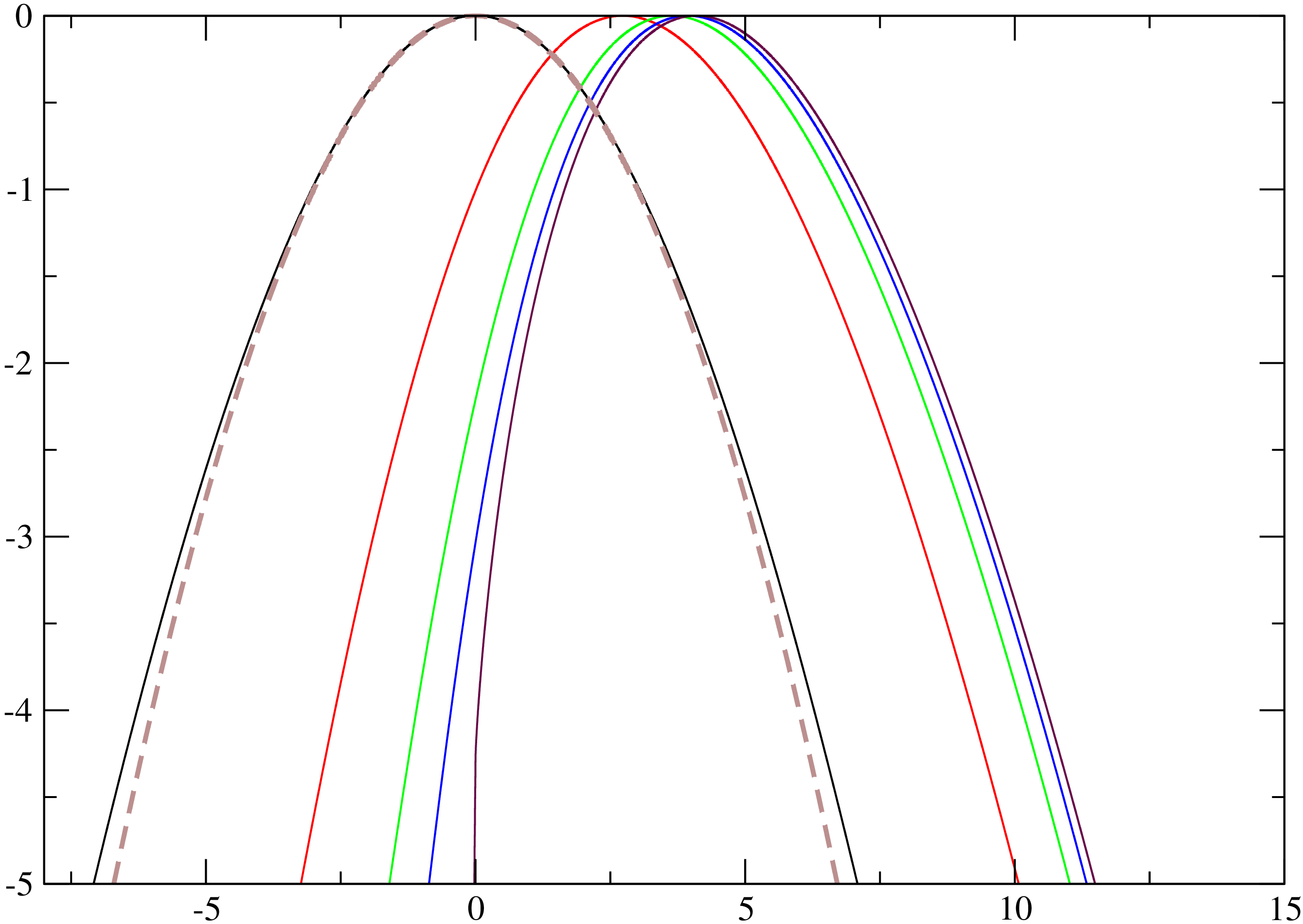} 
  \caption{Effect of increasing the non-equilibrium driving parameter
    $\gamma_1-\gamma_2$. A few large deviation functions are represented  for various values of $\gamma_1$, when
    the other parameters are fixed to the sample values $\gamma_2=0.1$, $\nu_1=100$,
    $\nu_2=10$.  The
    leftmost curve (for $\gamma_1=0.1$) is the equilibrium large deviation function and the dashed
    curve is the quadratic with the same curvature at the origin. The rightmost
    curve is for $\gamma_1=1$ i.e. heat bath 1 at zero temperature. For the intermediate curves, from left to right, $\gamma_1$ takes the values  $0.7$, $0.9$, $0.9666667$.}\label{LDnoneqvary}
\end{figure}

At equilibrium the large deviation function is even. As $\gamma_1-\gamma_2$
increases, the large deviation function becomes more and more asymmetric. In the
zero temperature limit $\gamma_1=1$, the large deviation function becomes
infinite for $\Jcumad < 0$. Fig.\ref{LDnoneqvary} illustrates the changes in the shape
of the large deviation function, with increasing departure from equilibrium.

Some generic properties of a large deviation function can be checked in the case of the above explicit formulae.
By virtue of \eqref{longtimekappa1a4} $\Espst{\jad}=B$ and one checks that $\fad_{\Heat_2}(\Espst{\jad})=0$, $\fad_{\Heat_2}'(\Espst{\jad})=0$ and $\fad_{\Heat_2}''(\Espst{\jad})=-1/(A-B^2)$, namely
\be
{\fad}''_{\Heat_2}(\Espst{\jad})=-\left[\lim_{\tad\to+\infty} \frac{\kappa_{n_2}^{[2]}}{\tad} \right]^{-1}
\ee
The expression \eqref{fjadAB} for $\fad_{\Heat_2}(\jad)$ is the sum of a term
$\jad (\beta_1-\beta_2)\frac{\gapE}{2}$ and an even function of $\jad$. As a
consequence, we check again that
$
\fad_{\Heat_2}(\jad)-\fad_{\Heat_2}(-\jad)= \jad (\beta_1-\beta_2)\gapE
$
namely, by virtue of \eqref{relffad}, $f_{\Heat_2}(\Jcum)$ obeys the fluctuation relation
$f_{\Heat_2}(\Jcum)-f_{\Heat_2}(-\Jcum)=(\beta_1-\beta_2) \Jcum$.
Moreover the absolute value of $\jad$ in the expression \eqref{fjadAB} for
$\fad_{\Heat_2}(\jad)$ is responsible for a (rather mild) singularity in the curve 
$f_{\Heat_2}(\Jcum)$ at $\Jcum =0$: a jump in the third derivative.

We notice that the large current behavior of $\fad_{\Heat_2}(\jad)$ in the present model reads
\be
\fad_{\Heat_2}(\jad)\underset{|\jad|\to +\infty}{\sim}-2 |\jad| \ln |\jad|.
\ee

\clearpage

\subsection{Derivation from a saddle-point method}
\label{DetailsCol}

Before embarking on the derivation, let us explain why the saddle point method
is not straightforward for this model. 

The saddle point approximation or expansion is well-suited for the asymptotic
study of integrals of the form $\int_{\Gamma} dz \, \Psi(z) e^{\tad \Phi(z)}$
where $\tad$ is some large real parameter and the functions $ \Psi(z)$,
$\Phi(z)$ are holomorphic in a domain large enough that the initial integration
contour $\Gamma$ can be deformed to a steepest descent path while remaining
within the holomorphicity domain during the deformation. One may also have to
encircle some singularities when deforming the contour, and then one must keep
track of their contributions, which may or may not dominate the saddle point
contribution. This can of course be generalized to a finite sum $\sum_i
\int_{\Gamma_i} dz \, \Psi_i(z) e^{\tad \Phi_i(z)}$ when each individual term
satisfies the hypotheses above. Note however that to get the leading behavior
one may have to take into account possible destructive interferences between
different pieces, for instance if the real parts of saddle point
values are the same for  several $\Psi_i$'s, or if the saddle points for
certain terms compete with encircled singularities for other terms.

In our case, we deal with an integral of the type $\int_{\Gamma} dz \, \sum_i
\Psi_i(z) e^{\tad \Phi_i(z)}$ where there is a single integration contour, and
the sum\footnote{Which in our case consists of only two terms.}  $\sum_i
\Psi_i(z) e^{\tad \Phi_i(z)}$ has nice holomorphicity properties that allow to
deform contours (almost) freely, but each term in itself has singularities and
cuts. So we have to face a kind of dilemma: either we want to keep
holomorphicity, then the large parameter does not appear in an exponential
-- and to our knowledge no straightforward constant phase technique applies -- or
we look at each pure exponential piece individually, and then some branch cuts
may prevent from deforming the contour purely as a constant phase steepest
descent path: the steepest descent path is not closed, some parts of the
original path are deformed along the cuts and they may dominate the saddle
point. But also, the contribution of the pure exponential pieces may interfere.
In our case, we have managed to show that in fact the interferences between
contributions of one pure exponential and cut contributions from another pure
exponential are destructive (with reminder terms controlled explicitly),
leaving the contribution of only a single saddle point (not one saddle point for
each pure exponential). But our argument relies on some tricks and features that
appear to us at this stage as coincidences: we have not been able to identify a
general framework avoiding our tedious analysis. And indeed, examples are known
\cite{Farago2002,vanZonCohen2004PRE}  where (depending possibly on parameters) the cut contributions
do or do not dominate the saddle point. 

To conclude these comments, let us mention one general direction that seems
worth pursuing, though we have not been able to use it to simplify significantly
our argument even in our simple case. In physical problems, the functions
$\Phi_i(z)$ will often be closely related to the different branches of a single
algebraic function.  For instance, the functions $\Phi_i(z)$ are often closely
related to the eigenvalues of some $z$-dependent matrix. So a natural route
would be to regard the integrals not in the $z$ plane, but on the appropriate
uniformizing Riemann surface, in our case an elliptic curve.

We now turn to the detailed analysis.

\subsubsection{Method}

 The current probability density $\Probdist(\jad;\tad)$ is related to the probability $\Prob\left(\Heat_2/(\tad \gapE) \in [\jad, \jad+d \jad[ ; \tad \right)$ 
   by the definition
$\Prob\left(\Heat_2/(\tad \gapE) \in [\jad, \jad+d \jad[ ; \tad \right)\equiv  \Probdist(\jad; \tad) d \jad$. 
Since  $\Heat_2/\gapE$ 
can take only integer values, the density distribution $\Probdist(\jad; \tad)$ is a sum of Dirac distributions
\be
\Probdist(\jad; \tad)= \sum_{n=-\infty}^{+\infty} \delta\left( \jad-\frac{n}{\tad} \right)
\tad \Prob\left(\frac{\Heat_2}{\gapE}=n;\tad\right).
\ee
In the long-time limit
\be
\label{defProbdistas}
\Probdist(\jad; \tad) \underset{\tad\to+\infty}{\sim} \sum_{n=-\infty}^{+\infty}
\delta\left(\jad- \frac{n}{\tad}\right) \tad \Prob^\text{as}(\jad ;\tad) \ee
where $\Prob^\text{as}(\jad ;\tad)$ is a function of the continuous parameter
$\jad$ that we shall compute below, and which is such that the
following asymptotic behavior holds: 
\be
\label{defProbas}
 \left. \Prob\left(\frac{\Heat_2}{\gapE}=\tad \jad;\tad\right)\right\vert_{\text{$\tau\jad$ integer}}
 \underset{\tad\to+\infty}{\sim}\Prob^\text{as}(\jad  ;\tad).
\ee
The notation $\left. g(\jad, \tad)\right\vert_{\text{$\tad\jad$ integer}}$ is a reminder of the rule that if the  function $g(\jad, \tad)$ is given by an integral representation, the latter must be calculated in the case where $\tad\jad$ is an integer. 
By using one of the ad-hoc definitions of the large deviation function introduced in Appendix E.2 of paper I, the function $\Prob^\text{as}(\jad  ;\tad)$ can be rewritten as
\be
\label{defLDfunctionTer}
\Prob^\text{as}(\jad  ;\tad)=A(\jad,\tad) e^{\tad \fad_{\Heat_2}(\jad)}
 \quad\textrm{with}\quad
\lim_{\tad\to+\infty} \frac{1}{\tad}\ln A(\jad,\tad)=0.
\ee
When one is interested only in the large deviation function, the only information to be retained from the latter equation is merely
\be
\label{LFforHeat2}
\fad_{\Heat_2}(\jad)=\lim_{\tad\to+\infty}\frac{1}{\tad}\ln\left.\Prob(\Heat_2=\tau \jad \gapE ;\tad)\right\vert_{\text{$\tau\jad$ integer}}.
\ee

Consequently, $\fad_{\Heat_2}(\jad)$ can be investigated by means of a saddle-point method applied to the representation of $\left.\Prob(\Heat_2=\tau \jad \gapE ;\tad)\right\vert_{\text{$\tau\jad$ integer}}$ in the complex $z$ plane given by \eqref{Probn2}. In the latter expression   $\Prob(\Heat_2=n \gapE ;\tad)$ is equal 
to $e^{-\tad}$ times  a linear combination of $\ccstar_n(\tad)$, $\csstar_n(\tad)$, $\csstar_{n+1}(\tad)$ and $\csstar_{n-1}(\tad)$. When $T_1=0$ the expressions 
\eqref{defccstar} and \eqref{defcsstar} of the latter functions are convenient for studying the large $\tad$ behavior of $\ccstar_{n=\tad\jad}(\tad)$, $\csstar_{n=\tad\jad}(\tad)$. When $T_1\not=0$ the study is slightly  more complicated and it is more conveniently performed by considering the related coefficients
defined by $\ccstar_n(\tad)=\rho^n\cc_n(\tad)$ and $\csstar_n(\tad)=\rho^n\cs_n(\tad)$ where the  expression \eqref{defrho}  of $\rho$ is  finite when $T_1\not=0$.
We present the details in the case where $T_1\not=0$.

When $T_1\not=0$,
in the long-time limit, we have to consider the behaviors of the functions
\be
\label{defKc}
\Kc(\jad;\tad)\equiv \cc_{n=\tad \jad}\left(\tad\right)
=\oint_{|z|=1} \frac{dz}{2\pi\iexp}\frac{1}{z^{\tad \jad+1}}\cosh\left(\tad\sqrt{\widetilde{\Delta}_+(z)}\right)
\ee
and
\be
\label{defKs}
\Ks^{(\Delta n)}(\jad;\tad)\equiv  \cs_{\tad \jad+\Delta n}\left(\tad \right)
=\oint_{|z|=1} \frac{dz}{2\pi\iexp} \frac{1}{z^{\Delta n}}\frac{1}{z^{\tad\jad+1}}\frac{\sinh\left(\tad\sqrt{\widetilde{\Delta}_+(z)}\right)}{\sqrt{\widetilde{\Delta}_+(z)}}
\ee
with $\Delta n=0,1,-1$.
 It is sufficient to exhibit the derivation of the long-time behavior of $\Kc(\jad;\tad)$, because the calculation of the  long-time behavior of $\Ks^{(\Delta n)}(\jad;\tad)$ follows the same lines.
 Moreover, according to the property $\cc_{n}(\tad)=\cc_{|n|}(\tad)$, we have to consider only the case where $\jad>0$.

For the study of the large $\tad$ limit,  the $\cosh$ function in  the integrand of $\Kc(\jad;\tad)$ is split into two exponentials, and $\Kc(\jad;\tad)$ appears as the sum of two integrals
\be
\label{decompKc}
2\Kc(\jad;\tad)=\oint_{|z|=1} \frac{dz}{2\pi\iexp z} e^{\tad\Phi^{(+)}(z)}
+\oint_{|z|=1} \frac{dz}{2\pi\iexp z}e^{\tad\Phi^{(-)}(z)}
\ee
where
\be
\label{defPhipm}
 \Phi^{(\pm)}(z)=-\jad \ln z\pm\sqrt{b+\frac{a}{2}\left(z+\frac{1}{z}\right)}.
\ee
We notice that, since $\tad\jad$ is in fact an integer, $\exp(-\tad\jad \ln z)$ is single valued and there is no cut  in the complex plane $z$ associated with the logarithmic function. However, since the $\cosh$ function has been split into two exponentials, we have to consider the two cuts associated with $\sqrt{\widetilde{\Delta}_+(z)}$. These cuts are
\be
]-\infty, -x_>]\quad\textrm{and}\quad [-x_<,0]
\ee
where $-x_>$ and $-x_<$ are the two negative real roots of the second-order polynomial
$z\widetilde{\Delta}_+(z)$ where $\widetilde{\Delta}_+(z)$  is given in \eqref{DeltaPlus}. The    roots  are such that $0\leq x_< < 1<x_>$.

\subsubsection{Deformation of contours}

\begin{figure}[h!]
  \centering
  \includegraphics[width=0.5 \textwidth]{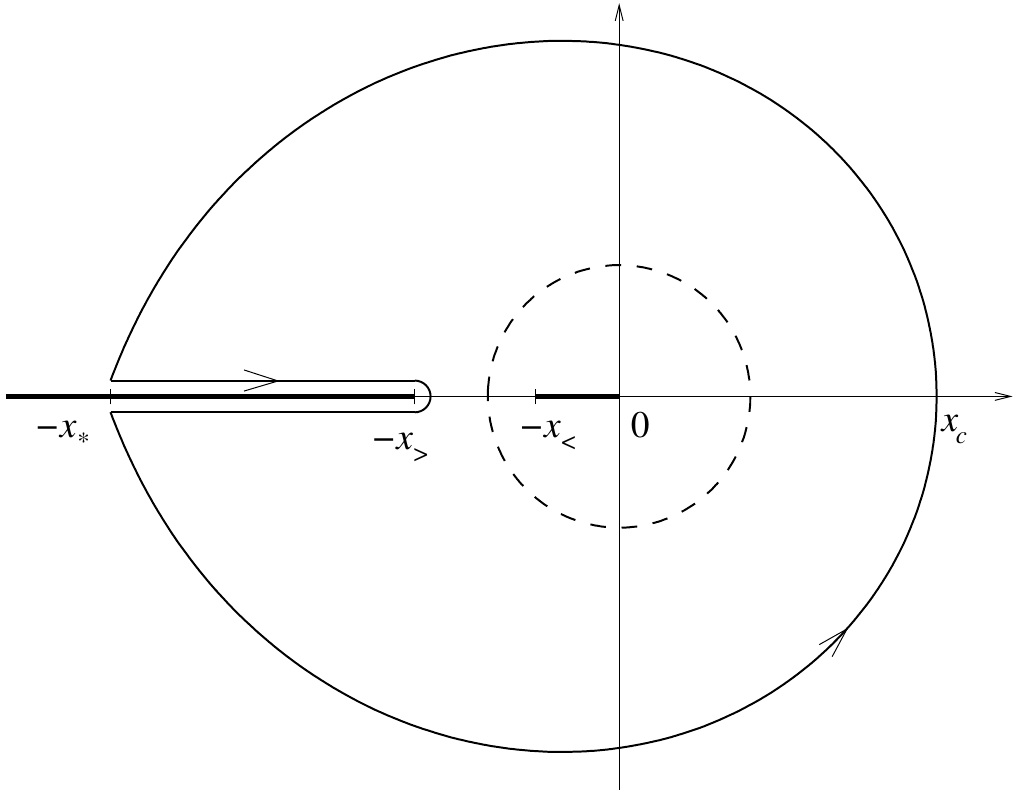}
    \caption{Deformed contour for $e^{\tad\Phi^{(+)}(z)}$ which decomposes as a steepest-descent contour  ${\Gamma^\star}$ which goes through the saddle-point $\xcp$ in the anti clockwise sense and a piece  which circumvents part of the cut $]-\infty, -x_>]$ in the clockwise sense. The other cut $[-x_<, 0]$ lies inside the unit circle represented by a dotted line.}\label{ContourPhip}
\end{figure}

The large $\tad$ behavior of  $\Kc(\jad;\tad)$ can be investigated  by applying the saddle-point method to the contribution from  the  integral involving $\Phi^{(+)}(z)$.
For that purpose we have to find a way to deform the unit circle into a contour that goes through a saddle point along a constant phase path  where $\Phi^{(+)}(z)$ is maximum at the saddle point. 
It can be easily found that the function $\Phi^{(+)}(z)$ has two real  saddle points where $\Phi^{(+)}(z)$ as well as its second derivative are real,  but only one of them  corresponds to a maximum of  $\Phi^{(+)}(z)$ when the real axis is crossed perpendicularly. The latter saddle point is $\xcp =\exp\left[\cosh^{-1}\left(\xp(\jad)\right)\right]$, namely by using $\cosh^{-1}\left(x\right)=\ln [x+\sqrt{x^2-1}]$,
\be
\label{valuexc}
\xcp =\xp(\jad)+\sqrt{\xp^2(\jad)-1}
\ee
with
\be
\label{defxp}
\xp(\jad)\equiv\frac{2}{a}\left[\jad^2+\sqrt{\jad^4+b\jad^2+\frac{a^2}{4}} \right].
\ee
The constant phase contour ${\Gamma^\star}$ which crosses the real axis perpendicularly at $\xcp$ can be looked for in the form $z_\star(\theta)=e^{\lambda_\star(\theta)+\iexp \theta}$. It proves to be  
\be
z_\star(\theta)=e^{\cosh^{-1}\left( y_\star(\theta;\jad)\right)+\iexp \theta}
\ee
where $\theta\in ]-\pi,\pi[$ and
\be
y_\star(\theta;\jad)= u(\theta;\jad)\cos\theta+\sqrt{1+\frac{2b}{a}u(\theta;\jad)+u^2(\theta;\jad)} 
\ee
with
\be
 u(\theta;\jad)=\frac{\theta^2}{\sin^2\theta} \frac{2\jad^2}{a}.
\ee
The contour ${\Gamma^\star}$ crosses the negative real axis at the point $z_\star(\theta=\pi)\equiv -x_\star (\jad)$ with
\be
\label{defxstar}
-x_\star (\jad)=-\exp\left[\cosh^{-1}\left(\frac{b}{a}+\pi^2 \frac{\jad^2}{a}\right)\right]
\ee
which lies on the cut $]-\infty, -x_>]$, because $-x_>=-\exp\left[-\cosh^{-1}\left(\frac{b}{a}\right)\right]$.
As a consequence, the unit circle can be deformed into the contour ${\Gamma^\star}$ 
and a contour ${\cal C}^{(+)}_{[-x_\star , -x_>]}$ that goes around the cut $]-\infty, -x_>]$ between the points $-x_\star (\jad)$ and $-x_>$ in the clockwise sense 
(see Fig.\ref{ContourPhip})
 \be
\label{Kcplus}
\oint_{|z|=1} \frac{dz}{2\pi\iexp z} e^{\tad\Phi^{(+)}(z)}=
\oint_{{\Gamma^\star}} \frac{dz}{2\pi\iexp z} e^{\tad\Phi^{(+)}(z)}
+\oint_{{\cal C}^{(+)}_{[- x_\star, -x_>]}} \frac{dz}{2\pi\iexp z} e^{\tad\Phi^{(+)}(z)}.
\ee

\begin{figure}[h!]
  \centering
  \includegraphics[width= 0.5\textwidth]{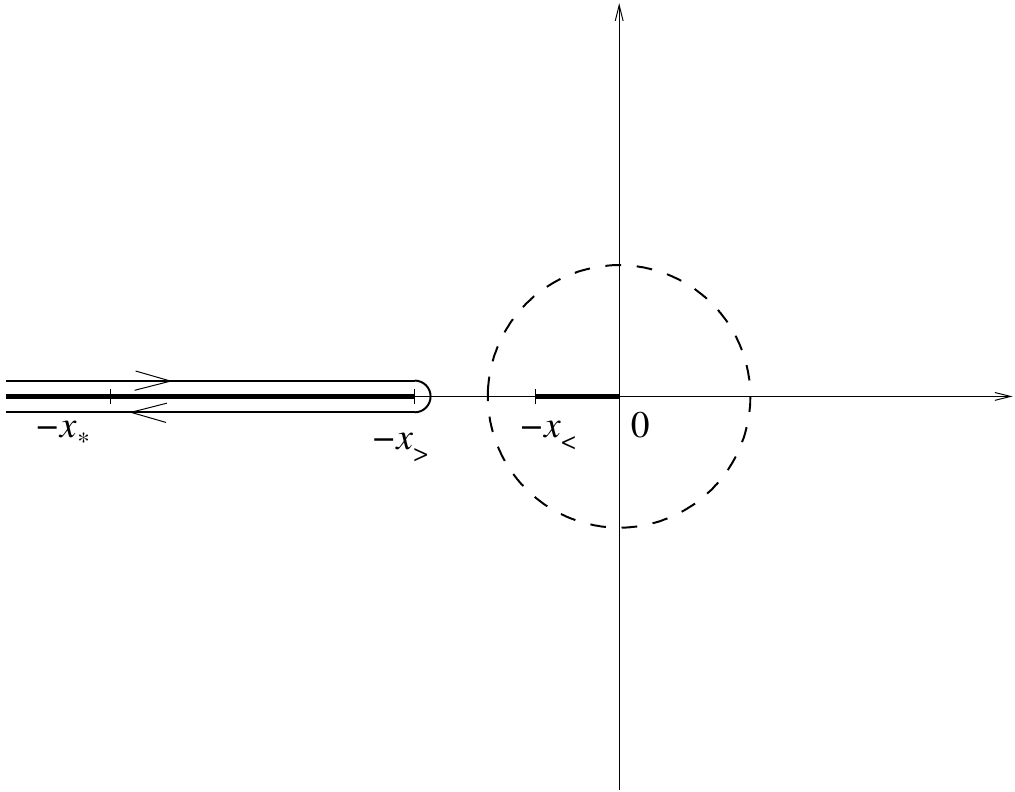}
  \caption{Deformed contour for $e^{\tad\Phi^{(-)}(z)}$ consisting of the circle at infinity, not  represented here, and a path which circumvents the whole cut $]-\infty, -x_>]$ in the clockwise sense.}\label{ContourPhim}
\end{figure}

On the other hand, in the integral involving $\Phi^{(-)}(z)$ the unit circle can be deformed into a circle, minus the point on the negative real axis,  with radius $R$ that goes to infinity and a path around  the cut 
$]-\infty, -x_>]$.
By using the parametrization $z=Re^{\iexp \theta}$, with $\theta\not=\pi$, we get the following large $|z |$ behavior: $
\vert \exp\left[\tad\Phi^{(-)}(z)\right]\vert \underset{|z|\to +\infty}{\sim}
\exp\left[-\tad\sqrt{a R}\cos(\theta/2)\right]$, so that the contribution of the integral along a circle of radius $R$ vanishes in the limit where $R$ goes to infinity. Consequently, (see Fig.\ref{ContourPhim}),
\be
\label{Kcmoins0}
\oint_{|z|=1} \frac{dz}{2\pi\iexp z} e^{\tad\Phi^{(-)}(z)}=\oint_{{\cal C}^{(+)}_{]-\infty, -x_>]}} \frac{dz}{2\pi\iexp z} e^{\tad\Phi^{(-)}(z)}.
\ee

\clearpage
\noindent The crucial point is then to notice that the sum $e^{\tad\Phi^{(+)}(z)}+e^{\tad\Phi^{(-)}(z)}$ is an analytic function of $z$, which has no cut along the interval $]-\infty, -x_>]$. As a consequence, the integral along the contour ${\cal C}^{(+)}_{]-\infty, -x_>]}$ with $\Phi^{(-)}(z)$ can be replaced by the opposite of the the same integral with $\Phi^{(+)}(z)$ in place of $\Phi^{(-)}(z)$, and the equality \eqref{Kcmoins0} becomes 
\be
\label{Kcmoins}
\oint_{|z|=1} \frac{dz}{2\pi\iexp z} e^{\tad\Phi^{(-)}(z)}
=\oint_{{\cal C}^{(-)}_{]-\infty, -x_>]}} \frac{dz}{2\pi\iexp z} e^{\tad\Phi^{(+)}(z)},
\ee
where ${\cal C}^{(-)}_{]-\infty, -x_>]}$ is the contour which goes around the cut $]-\infty, -x_>]$  in the anti clockwise sense.

\begin{figure}[h!]
  \centering
  \includegraphics[width= 0.5\textwidth]{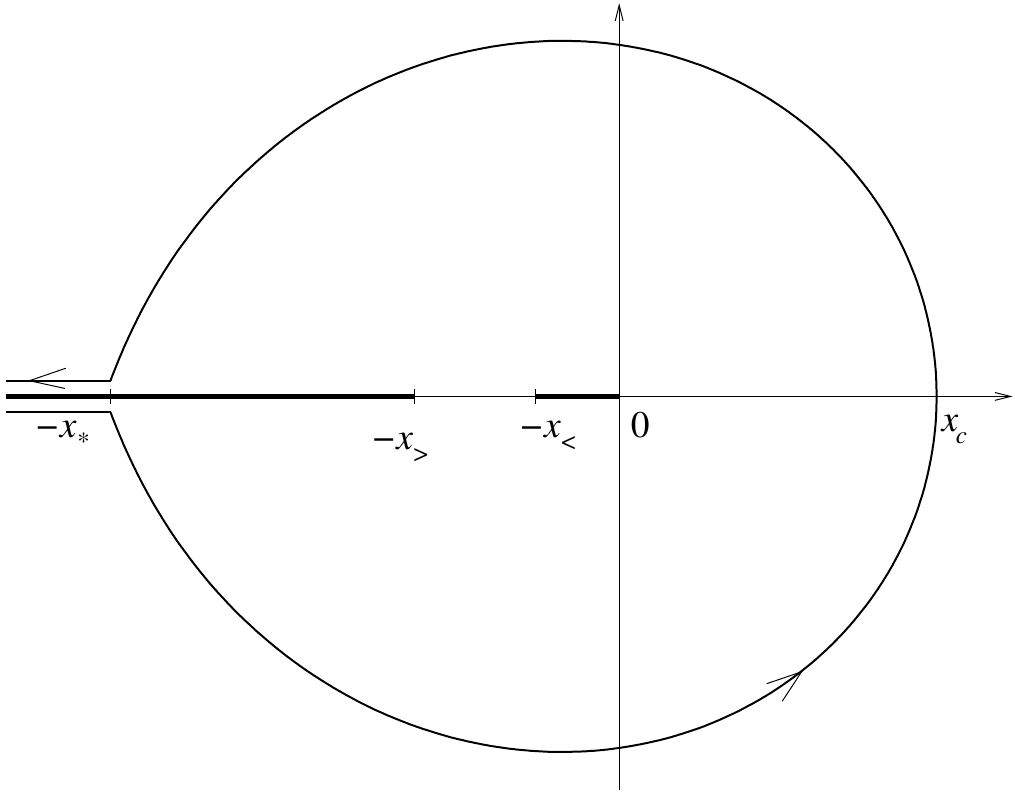}
  \caption{Contour of integration for $e^{\tad\Phi^{(+)}(z)}$ in the integral representation \eqref{decompbisKc} of $\Kc(\jad;\tad)$.}
  \label{ContourPhipm}
\end{figure}

When the contributions \eqref{Kcplus} et \eqref{Kcmoins} from the integrals involving respectively $\Phi^{(+)}$ and $\Phi^{(-)}$ are summed according to the definition \eqref{decompKc} of $\Kc(\jad;\tad)$, we get 
\be
\label{decompbisKc}
2\Kc(\jad;\tad)=\oint_{{\Gamma^\star}} \frac{dz}{2\pi\iexp z} e^{\tad\Phi^{(+)}(z)}
+\oint_{{\cal C}^{(-)}_{]-\infty,-x_\star]}} \frac{dz}{2\pi\iexp z} e^{\tad\Phi^{(+)}(z)}
\ee
We stress that $\exp\left[\tad\Phi^{(+)}(z)\right]$ diverges when $|z|$ goes to infinity, except on the negative real axis, so that the contour integral along the cut $]-\infty,-x_\star]$ cannot be closed at the point  $z=-\infty$. The expression \eqref{decompbisKc} corresponds to integrate
$e^{\tad\Phi^{(+)}(z)}/(2\pi\iexp z)$ along the contour in Fig.\ref{ContourPhipm}.

On the contour ${\cal C}_{]-\infty,- x_\star]}$, $z=e^{\lambda +\iexp \sigma \pi}$
where $\sigma=+1$ if $z$ is above the cut and  $\sigma=-1$ otherwise.
Since  $\tad\jad=n$ where $n$ is an integer
\be
\oint_{{\cal C}_{]-\infty,-x_\star]}^{(-)}} \frac{dz}{2\pi\iexp z} e^{\tad \Phi^{(+)}(z)}=(-1)^n\int_{\ln x_\star} ^{+\infty}\frac{d \lambda}{\pi}e^{-\tad \jad \lambda} \sin\left(\tad\sqrt{a\cosh\lambda-b}\right).
\ee
The sign of this contribution changes for two consecutive values of $\jad$, but its absolute value is bounded,
\be
\label{borneOscil}
\left\vert\oint_{{\cal C}_{]-\infty,-x_\star]}^{(-)}} \frac{dz}{2\pi\iexp z} e^{\tad\Phi^{(+)}(z)}\right\vert
\leq \frac{1}{\pi \tad \jad}e^{-\tad \jad\ln x_\star}.
\ee

\clearpage
\subsubsection{Large $\tad$ behavior}

According to the saddle-point formula
\be
\label{AsKcGammap}
\oint_{{\Gamma^\star}} \frac{dz}{2\pi\iexp z} e^{\tad\Phi^{(+)}(z)}\underset{\tad\to+\infty}{\sim}
\frac{1}{\sqrt{\tad}}\frac{1}{\xcp\sqrt{2\pi \times d^2\Phi^{(+)}/dz^2\vert_{z=\xcp}}}e^{\tad\Phi^{(+)}(\xcp)}.
\ee
By using the inequalities, $-\ln x_\star<-\ln \xcp$, derived from the expression \eqref{valuexc}-\eqref{defxp} and \eqref{defxstar}, and $- \jad \ln \xcp \leq \Phi^{(+)}(\xcp)$, derived from \eqref{defPhipm},
the bound exhibited in \eqref{borneOscil} implies that
\be
\label{bornesup}
\oint_{{\cal C}_{]-\infty,-x_\star]}^{(-)}}\frac{dz}{2\pi\iexp z} e^{\tad \Phi^{(+)}(z)}=o\left(e^{\tad\Phi^{(+)}(\xcp)}\right)
\ee
where $o\left(e^{\tad g(j)}\right)
$ denotes a function which decays faster than $e^{\tad g(j)}$ when $\tad$ goes to $+\infty$.

Eventually, the definition \eqref{defKc} of $\Kc(\jad;\tad)$ and the decomposition \eqref{decompbisKc} together with the behaviors  \eqref{AsKcGammap} et \eqref{bornesup} lead to
\be
\left. \cc_{\tad \jad}\left(\tad\right)\right\vert_{\text{$\tau\jad$ integer}}\underset{\tad\to+\infty}{\sim}
A_{\cc} \times e^{\tad\Phi^{(+)}(\xcp)}
\ee
where
\be
\Phi^{(+)}(\xcp)=-\jad \ln [\xp(\jad)+\sqrt{\xp^2(\jad)-1}]+\sqrt{b+a\xp(\jad)}
\ee
and
\be
A_{\cc}=\frac{1}{\sqrt{\tad}}\frac{1}{2\sqrt{2\pi}}\frac{1}{\xcp\sqrt{ d^2\Phi^{(+)}/dz^2\vert_{z=\xcp}}}
\ee
with  $\xcp =\xp(\jad)+\sqrt{\xp^2(\jad)-1}$ and
\be 
d^2\Phi^{(+)}/dz^2\vert_{z=\xcp}=\frac{2}{a}\jad \frac{ \sqrt{\jad^4+b\jad^2 +
    \frac{a^2}{4}}}{\left[\xp(\jad)+\sqrt{\xp^2(\jad)-1}\; \right]^2\sqrt{\xp^2(\jad)-1}}.
\ee
The same argument can be performed for $\Ks^{(\Delta n)}(\jad;\tad)$ defined in  \eqref{defKs}, with the result
\be
\label{Ksasymptot}
\left. \cs_{\tad \jad+\Delta n}\right\vert_{\text{$\tau\jad$ integer}}\left(\tad \right)\underset{\tad\to+\infty}{\sim}
A_{\cs}\times e^{\tad\Phi^{(+)}(\xcp)}
\quad\textrm{with}\quad A_{\cs}=\left[\xcp^{\Delta n} \sqrt{\widetilde{\Delta}_+(\xcp)}\right]^{-1}A_{\cc}.
\ee
As a consequence, 
\be
\label{LDfromccn}
\lim_{\tad\to+\infty} \frac{1}{\tad} \ln \left[e^{-\tad}\rho^{\tad\jad} \cc_{\tad \jad}\left(\tad\right)\right]=\fad(\jad)
\ee
and
\be
\label{LDfromcsn}
\lim_{\tad\to+\infty} \frac{1}{\tad} \ln  \left[e^{-\tad}\rho^{\tad\jad}\cs_{\tad \jad+\Delta n}\left(\tad \right)\right]=\fad(\jad)
\ee
with $\fad(\jad)=-1+\jad \ln\rho+\Phi^{(+)}(\xcp)$. By virtue of \eqref{defrho} 
$\ln\rho=\ln\sqrt{\frac{A+B}{A-B}}$ and, according to the definitions in \eqref{DeltaPlus}, $\cosh^{-1}\left(\frac{1-b}{a}\right)=\cosh^{-1}\left(\frac{A}{\sqrt{A^2-B^2}}\right)=\ln\sqrt{\frac{A+B}{A-B}}$. Therefore $\fad(\jad)$ reads
\be
\label{fjadab}
\fad(\jad)=-1+\jad \cosh^{-1}\left(\frac{1-b}{a}\right)- |\jad| \ln\left[\xp(\jad)+\sqrt{\xp^2(\jad)
-1}\right]+\sqrt{b +a\xp(\jad)}
\ee
where $\xp(\jad)$ is defined in \eqref{defxp}.

Eventually, according to \eqref{Probn2}, $\Prob\left(\frac{\Heat_2}{\gapE}=n;\tad\right)$ is a finite linear combination of functions of $n$ plus a finite increment $\Delta n$, which can be rewritten as
\be
\Prob\left(\frac{\Heat_2}{\gapE}=n;\tad\right)
\underset{n=\tad\jad}{=}\sum_{\Delta n =0,1,-1}
b_{\Delta n} \,\,g_{\Delta n} (\tad\jad+\Delta n ; \tad),
\ee
and all functions $g_{\Delta n}$ prove to have the same ``large deviation function'' $\fad(\jad)$ in the sense of definition \eqref{LFforHeat2}, 
\be
g_{\Delta n} (\tad\jad+\Delta n ; \tad)
 \underset{\substack{\tad\to+\infty\\ \tad\jad \,\text{integer}}}{\sim}
 A_{\Delta n} (\jad, \tad) e^{\tad \fad(\jad)}.
\ee
Therefore, by comparison with \eqref{defProbas} and \eqref{defLDfunctionTer} we get
\be
\label{LDHeat2fad}
\fad_{\Heat_2}(\jad)=\fad(\jad)
 \quad\textrm{and}\quad
A(\jad,\tad)=\tad \sum_{\Delta n =0,1,-1}
b_{\Delta n} A_{\Delta n} (\jad, \tad) 
\ee
where the expression \eqref{fjadab} of $\fad(\jad)$ indeed coincides with the result \eqref{fjadAB}.

\subsection{Derivation by Laplace's method on a discrete sum}
\label{LaplacesMethod}

As the reader may have noticed, the computation of the large deviation function via contour integrals is a bit tricky and clumsy due to the cuts, and relies on some compensations which are not totally obvious to foresee.

In the case at hand it is possible to derive the large deviation function via Laplace's method applied to a discrete sum of non-negative contributions. We illustrate this briefly in the case of 
$\Ks^{(\Delta n)}$.

The key is an explicit formula for $\sin\left( \tad \sqrt{\widetilde{\Delta}_+(z)}\right)/\sqrt{\widetilde{\Delta}_+(z)}$ as a Laurent series in $z$. From the symmetry $z\leftrightarrow 1/z$ we can concentrate on positive powers of $z$. We start with 
\be
\frac{\sin\left(\tad\sqrt{\widetilde{\Delta}_+(z)}\right)}{\sqrt{\widetilde{\Delta}_+(z)}}=\sum_{k=0}^{+\infty} \frac{\tad^{2k+1}}{(2k+1)!}[\widetilde{\Delta}_+(z)]^k
\ee
and expand $[\widetilde{\Delta}_+(z)]^k$ as a Laurent polynomial in $z$,
\be
\left(b+\frac{a}{2}\left(z+\frac{1}{z}\right)\right)^k=\sum_{\substack{l,m\geq0\\l+m\leq k}}
z^{l-m}\left(\frac{a}{2}\right)^{l+m}b^{k-l-m} \frac{k!}{l!m!(k-l-m)!}.
\ee
So for $n\geq0$ one gets (take $l=m+n$ above) 
\be
\oint_{|z|=1}\frac{dz}{2\pi\iexp}\frac{1}{z^{n+1}}\frac{\sin\left(\tad\sqrt{\widetilde{\Delta}_+(z)}\right)}{\sqrt{\widetilde{\Delta}_+(z)}}
=\sum_{m\geq0}\,\,\sum_{k\geq2m+n} \frac{\tad^{2k+1}}{(2k+1)!}\left(\frac{a}{2}\right)^{2m+n}
b^{k-2m-n}\frac{k!}{m!(m+n)!(k-2m-n)!}.
\ee
As $\tad$, $a$ and $b$ are $>0$, this is a (double) sum of positive terms, and we are interested in the limit 
\be
\tad\to+\infty\qquad n=\tad\jad+\Delta n \quad\textrm{with}\quad \Delta n \in\{0,-1,1\}.
\ee
It is straightforward to check that in this limit the maximal term in the (double) sum is in the bulk (i.e. not for $m=0$ or $k=2m+n$) and such that $k$ and $m$ scale linearly with $\tad$. One can use the Stirling approximation for all factorials and obtain the large deviation function straightforwardly, the most painful part of the computation being the location of the maximal term. We omit all details.

\clearpage
\section{Dependence upon typical time scales of the thermostats}

\label{DependenceTimeScales}

In this section we are interested in the influence of the typical time scales in the limit where the heat exchanges with one of the two thermostats become infinitely fast. 
We only explicitly consider the limit when $\nu_2/\nu_1$ goes to infinity,
namely when  heat bath 2 exchanges heat with the two-spin system far faster
than heat bath 1 does. Indeed, the opposite limit when $\nu_2/\nu_1$
vanishes involves similar calculations (where the roles of $\Heat^d_1\equiv -\Heat_1$ and
$\Heat_2$ are interchanged), and the results in both limits are essentially the same when they are stated in terms of quantities pertaining either to the ``fast'' heat bath or to the ``slow'' heat bath. The main results are summarized with the latter terminology in subsection \ref{ContentsSec}.

\subsection{Stationary mean values in the infinite  $\nu_2/\nu_1$ limit}

The evolution of the probability distribution $\Prob(\sigma_1,\sigma_2;t)$ is given in \eqref{ProbExp} where the time scales involve the parameter $\alpha=\sqrt{1-4A}$ with $A=\nuad_1\nuad_2\left(1-\gamma_1\gamma_2\right)$.
When the ratio of inverse time scales $\nu_2/\nu_1$ goes to $+\infty$, $\lim_{\nu_2/\nu_1\to+\infty}\frac{1}{2}(\nu_1+\nu_2)\left[1+\alpha\right]=\nu_2$ and 
$\lim_{\nu_2/\nu_1\to+\infty}\frac{1}{2}(\nu_1+\nu_2)\left[1-\alpha\right]=\nu_1(1-\gamma_1\gamma_2)$, so that
 the probability distribution $\Prob(\sigma_1,\sigma_2;t)$  goes exponentially fast to its stationary value over the time scale $1/[\nu_1(1-\gamma_1\gamma_2)]$. Moreover,
  since $\gammaeff\equiv \nuad_1\gamma_1+\nuad_2\gamma_2$,
$\lim_{\nu_2/\nu_1\to+\infty}\gammaeff=\gamma_2$, and, according to \eqref{structPst}, the  stationary probability distribution $\Probst(\sigma_1,\sigma_2)$ coincides with the canonical distribution at the inverse temperature $\beta_2$ of the fast thermostat, namely
\be
\label{structPstas}
\lim_{\nu_2/\nu_1\to+\infty}\Probst(\sigma_1,\sigma_2)=\frac{1}{4}(1+\gamma_2\sigma_1\sigma_2).
\ee
As a consequence the heat capacities $C_\text{st}^{[a]}(T_1,T_2)$'s  corresponding to a variation of the  temperature $T_a$ of bath $a$  (for $a=1,2$) and given in \eqref{expCa}, become
$\lim_{\nu_2/\nu_1\to+\infty}C_\text{st}^{[1]}(T_1,T_2)=0$ and
$\lim_{\nu_2/\nu_1\to+\infty}C_\text{st}^{[2]}(T_1,T_2)= C_\text{eq}(T_2)$ respectively,
while the heat capacity $C_\text{st}(T_1,T_2)$  corresponding to  equal variations of both bath  temperatures, and given at the end of \eqref{expC}, becomes
$
\lim_{\nu_2/\nu_1\to+\infty}C_\text{st}(T_1,T_2)=C_\text{eq}(T_2)
$.

On the other hand the time scale of the mean currents of exchanged quantities  is that of the slow thermostat. Indeed, according to \eqref{jinstexp},
\be
\lim_{\nu_2/\nu_1\to+\infty}\Espst{\jinst_2}=\left(\gamma_1-\gamma_2\right)\frac{\nu_1\gapE}{2},
\ee
and the thermal conductivity is determined by the typical  time scale $1/\nu_1$ of the slow thermostat. 
Similarly   the housekeeping entropy flow \eqref{sighkexp}, which is equal to the opposite of the mean exchange entropy flow in the stationary state $d_\text{exch} S/dt\vert_\text{st}$,  becomes
\be
\lim_{\nu_2/\nu_1\to+\infty}\sighk[\Probst]=\nu_1\left(\gamma_1-\gamma_2\right)\left(\beta_1-\beta_2\right)\frac{\gapE}{2}.
\ee

\subsection{Various probabilities in the infinite  $\nu_2/\nu_1$ limit}
\label{RandomWalk}

The stationary probability that the heat received from bath 1 is equal to $\Heat_1=-n_1\gapE$ is given by \eqref{Probn1} where $\gamma_0$ is to be replaced by 
$\lim_{\nu_2/\nu_1\to+\infty}\gammaeff=\gamma_2$, with the result
\be
\label{limProbst}
\lim_{\nu_2/\nu_1\to +\infty}\Probst(n_1;t)=\lim_{\nu_2/\nu_1\to +\infty}e^{-\tad} \left[\ccstar_{n_1}(\tad)+\csstar_{n_1}(\tad)\right].
\ee
According to the definitions \eqref{defccstar} and \eqref{defcsstar},  the integral representations of $\ccstar_n(\tad)$ and $\csstar_n(\tad)$ in the complex plane involve the discriminant 
$\Deltastar_+(z)$. 
In order to  discuss the dependence upon the inverse time scales $\nu_1$ and $\nu_2$, it is convenient to  rewrite the expression \eqref{defDeltastar} of  $\Deltastar_+(z)$ 
 in terms of the  parameters  $p_+$ and $p_-$ defined in \eqref{defp+p-}.
The correspondence with $A$ and $B$ reads $A=\nuad_1\nuad_2 \left(p_++p_-\right)$ and $B=\nuad_1\nuad_2 \left(p_+-p_-\right)$, and
\be
\label{DeltsstarBis}
\Deltastar_+(z)=1+2\nuad_1\nuad_2\left[-(p_++p_-)+p_+z+ p_-\frac{1}{z}\right].
\ee
Therefore $\lim_{\nu_2/\nu_1\to +\infty}\Deltastar_+(z)=1$ and the leading order of
$\frac{1}{2}(\nu_1+\nu_2)[-1+\sqrt{\Deltastar_+(z)}]$ is merely \\
$\frac{1}{2}\nu_1\left[-(p_++p_-)+p_+z+ p_-(1/z)\right]$.
Consequently $e^{-\tad}\ccstar_{n_1}(\tad)$ and $e^{-\tad}\csstar_{n_1}(\tad)$ have the same asymptotic behavior and
\be
\lim_{\nu_2/\nu_1\to +\infty}e^{-\tad} \left[\ccstar_{n}(\tad)+\csstar_{n}(\tad)\right]
= \oint_{|z|=1} \frac{dz}{2\pi\iexp} \frac{1}{z^{n+1}}G_{\scriptscriptstyle RW}(z;\nu_1 t),
\ee
where
\be
\label{GRW}
G_{\scriptscriptstyle RW}(z;\nu_1t)= \exp\left\{-\frac{1}{2} (p_++p_-)\nu_1 t+\frac{1}{2}\left[p_+  z+ p_-\frac{1}{z}\right]\nu_1 t\right\}.
\ee
This expression can be interpreted as $G_{\scriptscriptstyle RW}(z;\nu_1 t)=\sum_{n=-\infty}^{+\infty}z^n\Prob_{\scriptscriptstyle RW}(n; \nu_1t)$, namely  $G_{\scriptscriptstyle RW}(z;\nu_1t)$ is the generating function of   the probability $\Prob_{\scriptscriptstyle RW}(n;\nu_1t)$ for  the  continuous-time random walk, also referred to as the ``randomized'' random walk  (see for instance page 59 of Ref.\cite{Feller1971}), which is determined by  the Markov evolution equation 
\be
\label{RWequation}
\frac{d \Prob_{\scriptscriptstyle RW}(n; \nu_1t)}{dt}=\frac{\nu_1}{2}\left[-\left(p_++p_-\right)\Prob_{\scriptscriptstyle RW}(n; \nu_1t)+ p_
+\Prob_{\scriptscriptstyle RW}(n-1; \nu_1t) +p_-\Prob_{\scriptscriptstyle RW}(n+1; \nu_1t)\right],
\ee 
and the initial condition $\Prob_{\scriptscriptstyle RW}(n;t=0)=\delta_{n,0}$.  By virtue of the identity which defines the generating function of modified Bessel functions $I_{n}(x)$,
\be
e^{\frac{1}{2}\left[p_+  z+ p_-\frac{1}{z}\right] \nu_1 t}=\sum_{n=-\infty}^{+\infty}
\left(z\sqrt{\frac{p_+}{p_-}}\right)^n I_{n}\left(\nu_1 t\sqrt{p_+p_-}\right),
\ee
where
\be
\label{defBessel}
I_n(x)\equiv\int_0^{2\pi}\frac{d\theta}{2\pi} e^{-\iexp n \theta} e^{x \cos\theta}.
\ee
As can be derived from the latter integral representation, the modified Bessel function $I_{n}(x)$ is an even function of $n$, $I_{-n}=I_{n}$. 
Therefore the series representation of  the expression   \eqref{GRW} for the generating function $G_{\scriptscriptstyle RW}(z; \nu_1t)$  yields
\be
\label{ProbRW}
\Prob_{\scriptscriptstyle RW}(n; \nu_1t)=\left(\sqrt{\frac{p_+}{p_-}}\right)^n I_{|n|}\left(\sqrt{p_+p_-}\times \nu_1 t\right)
e^{-\frac{1}{2} (p_++p_-)\nu_1t}.
\ee

Eventually the limit in \eqref{limProbst} reads
\be
\label{Probn1as}
\lim_{\nu_2/\nu_1\to +\infty}\Probst(n_1;t)=\Prob_{\scriptscriptstyle RW}(n_1;\nu_1t),
\ee
where the thermodynamic parameters of the thermal baths appear through  the combinations
of $p_+$ and $p_-$ explicitly given in \eqref{pppmparameters}.
On the other hand the probability that at time $t$ the system has received a heat amount $\Heat_2=n_2\gapE$ from bath $2$ is given by \eqref{Probn2}. Since $\lim_{\nu_2/\nu_1\to+\infty}e^{-\tad}\ccstar_{n}(\tad)=\frac{1}{2}\Prob_{\scriptscriptstyle RW}(n;\nu_1t)$ and 
$\lim_{\nu_2/\nu_1\to+\infty}e^{-\tad}\csstar_{n}(\tad)=\frac{1}{2}\Prob_{\scriptscriptstyle RW}(n; \nu_1t)$,
\be
\label{Probn2as}
\lim_{\nu_2/\nu_1\to +\infty}\Probst\left(n_2;t\right) =
\frac{1+\gamma_2^2}{2}\Prob_{\scriptscriptstyle RW}(n_2; \nu_1t)
+\frac{1-\gamma_2^2}{4}\left[\Prob_{\scriptscriptstyle RW}(n_2+1; \nu_1t)+\Prob_{\scriptscriptstyle RW}(n_2-1; \nu_1t)\right].
\ee

We notice that at equilibrium, namely in the stationary  state where $\beta_1=\beta_2=\beta$, $p_+=p_-$ according to \eqref{pppmparameters}, and the limit \eqref{Probn1as} reads
\be
\lim_{\nu_2/\nu_1\to +\infty}\Probeq\left(n_1;t\right) 
= I_{|n_1|}\left(\frac{1}{2}(1-\gamma^2)\nu_1 t\right)
e^{-\frac{1}{2}(1-\gamma^2)\nu_1 t}.
\ee
Similar formul\ae\ hold for the three contributions in $\lim_{\nu_2/\nu_1\to +\infty}\Probeq\left(n_2;t\right)$, which is derived from \eqref{Probn2as}.

\subsection{Interpretation: mean-field regime}

When the initial state is distributed according to the stationary measure, the probability that at time $t$ the system is in a configuration where $\sigma_1\sigma_2$ is equal to $\pm1$ and that the system has received a heat amount $\Heat_1=-n_1\gapE$ from  bath $1$ is given by \eqref{Probpn1} and \eqref{Probmn1} where 
$\gammaeff$ is to be replaced by $\lim_{\nu_2/\nu_1\to+\infty}\gammaeff=\gamma_2$, namely
\be
\lim_{\nu_2/\nu_1\to +\infty}\Probst\left(\sigma_1\sigma_2=\pm1,n_1;t\right) =
\frac{1}{2}\left[1\pm \gamma_2\right]
\lim_{\nu_2/\nu_1\to +\infty}e^{-\tad} \left[\ccstar_{n_1}(\tad)+\csstar_{n_1}(\tad)\right].
\ee
Comparison with \eqref{structPstas} and \eqref{limProbst} shows that 
the latter equation can be interpreted as
\be
\label{factorization}
\Probst\left(\sigma_1\sigma_2=\pm1, n_1;t\right) 
\underset{\nu_2/\nu_1\to +\infty}{\sim}\Probst\left(\sigma_1\sigma_2=\pm1\right)
\times \Probst(n_1;t).
\ee
This is a mean-field property: between two flips of  spin $\sigma_1$,  spin $\sigma_2$ is flipped so many times by  thermostat 2 that, when  spin $\sigma_1$ is flipped again, the sign of  $\sigma_1\sigma_2$ is no longer correlated to its value when the previous flip of $\sigma_1$ occurred. Therefore  the variation of $n_1$, which is generated by  the flip of $\sigma_1$ and the value of which is  determined by the sign of $\sigma_1\sigma_2$, is no longer correlated to the sign  which  $\sigma_1\sigma_2$  had when the previous variation of $n_1$ occurred: the probability distributions of $\sigma_1\sigma_2$ and $n_1$ are independent from each other.

On the other hand, the probability that at time $t$ the system is in a configuration where $\sigma_1\sigma_2$ is equal to $1$ and that the system has received a heat amount 
$\Heat_2=n_2\gapE$ from  bath $2$ has an expression given by the remark after \eqref{Probpn1} and \eqref{Probmn1}. We get
\be
\label{Probsigmapn2as}
\lim_{\nu_2/\nu_1\to +\infty}\Probst\left(\sigma_1\sigma_2=+1,n_2;t\right) =
\frac{(1+\gamma_2)^2}{4}\Prob_{\scriptscriptstyle RW}(n_2; \nu_1t)
+\frac{1-\gamma_2^2}{4}\Prob_{\scriptscriptstyle RW}(n_2+1; \nu_1t),
\ee
while
\be
\label{Probsigmapm2as}
\lim_{\nu_2/\nu_1\to +\infty}\Probst\left(\sigma_1\sigma_2=-1,n_2;t\right) =
\frac{(1-\gamma_2)^2}{4}\Prob_{\scriptscriptstyle RW}(n_2; \nu_1t)
+\frac{1-\gamma_2^2}{4}\Prob_{\scriptscriptstyle RW}(n_2-1; \nu_1t).
\ee
 Comparison with \eqref{Probn2as} shows that there  is no factorization similar to \eqref{factorization}. In other words the  variables $n_2$ and $\sigma_1\sigma_2$ are still correlated.

\subsection{Symmetry property specific to the probability of $\Heat_1$}

The probability $\Prob_{\scriptscriptstyle RW}(n; \nu_1t)$ for the continuous-time random walk, recalled in \eqref{ProbRW}, obeys the symmetry
\be
\ln\frac{\Prob_{\scriptscriptstyle RW}(n; \nu_1t)}{\Prob_{\scriptscriptstyle RW}(-n; \nu_1t)}=n\ln \frac{p_+}{p_-}\quad\textrm{at any time $t$}.\quad
\ee
 As a consequence, according to \eqref{Probn1as} and the relations
\eqref{pppmparameters}, the probability distribution for the heat amount dissipated towards the slow bath, $\Heatd_1=-n_1\gapE$, obeys the finite-time symmetry
\be
\label{FRHeat1}
\ln\frac{\Probst(\Heatd_1;t)}{\Probst(-\Heatd_1;t)}
\underset{\nu_2/\nu_1\to +\infty}{=}(\beta_1-\beta_2)\Heatd_1.
\ee
However, by virtue of \eqref{Probn2as}, there is no similar finite-time symmetry property for
$\Probst(\Heat_2;t)$.

\subsection{Cumulants per unit time for $\Heatd_1$ and $\Heat_2$}

At any time, according to  \eqref{GRW}, the characteristic function $G_{\scriptscriptstyle RW}(e^{\lad}; \nu_1t)$ for the continuous-time random walk takes the very simple form
\be
G_{\scriptscriptstyle RW}(e^{\lad}; \nu_1t)=\exp[t\alpha_{\scriptscriptstyle RW}(\lad; \nu_1)]
\ee
where
\be 
\label{defalphaRW}
\alpha_{\scriptscriptstyle RW}(\lad; \nu_1)=\frac{\nu_1}{2}\left[-\left(p_++p_-\right)+\ p_+ e^{\lad} + p_-e^{-\lad}\right].
\ee
On the other hand, according to \eqref{Probn1as},
$\sum_{n_1=-\infty}^{+\infty}e^{\lad n_1}\lim_{\nu_2/\nu_1\to
  +\infty}\Probst(n_1;t)=G_{\scriptscriptstyle RW}(e^{\lad}; \nu_1t)$.   Therefore,
the cumulants for $\Heatd_1$ are given  at any finite time by the
 formul\ae\ $\kappa_{n_1}^{[q]}=\partial^q \ln G_{\scriptscriptstyle RW}(e^{\lad};\nu_1 t)/\partial \lambda^q\vert_{\lambda=0}$  and  the cumulants per unit time read for $p\geq 0$
\be
\frac{1}{t}\kappa_{n_1}^{[2p+1]}
\underset{\nu_2/\nu_1\to +\infty}{=}
\frac{\nu_1}{2}\left(p_+-p_-\right)=\frac{\nu_1}{2}(\gamma_1-\gamma_2)
\ee
and for $p\geq 1$
\be
\frac{1}{t}\kappa_{n_1}^{[2p]}
\underset{\nu_2/\nu_1\to +\infty}{=}
\frac{\nu_1}{2} \left(p_++p_-\right)
=
\frac{\nu_1}{2}(1-\gamma_1\gamma_2).
\ee
In the case of $\Heat_2$, by virtue of \eqref{Probn2as}
\be
\sum_{n_2=-\infty}^{+\infty}e^{\lad n_2}\Probst(n_2;t)
\underset{\nu_2/\nu_1\to +\infty}{=}
\frac{1}{2}\left[1+\gamma_2^2+ (1-\gamma_2^2)\cosh\lad\right]
G_{\scriptscriptstyle RW}(e^{\lad};\nu_1 t).
\ee
The cumulants per unit time of $\Heat_2$ coincide with the cumulants per unit time of $\Heatd_1$  only  in the long-time limit, in agreement with \eqref{longtimekappa} and because $\Heatd_1=-\Heat_1=n_1\gapE$,
\be
\lim_{t\to+\infty}\frac{1}{t}\kappa_{n_2}^{[p]}
=\lim_{t\to+\infty}\frac{1}{t}\kappa_{n_1}^{[p]}.
\ee
When the system is at equilibrium $\gamma_1=\gamma_2$ and the long-time behavior of all odd cumulants per unit time vanish, as already noticed in subsection \ref{ExplicitCumulants}.

We notice that the previous results can also be retrieved directly from the expression for
the generating function $\alpha_2(\lad)$ of the long-time cumulants per unit time. (The limits $\nu_2/\nu_1\to +\infty$ and $t\to+\infty$ do commute with each other.)
The expression of $\alpha_2(\lad)$ is given in \eqref{alphaAB}, and,  according to the relations
after  \eqref{defp+p-}, it reads
\be
\label{alphap+p-}
\alpha_2(\lad)=\frac{1}{2}\left\{-(\nu_1+\nu_2)+\sqrt{(\nu_1+\nu_2)^2+2\nu_1\nu_2\left[-(p_++p_-)+p_+e^{\lad}+ p_-e^{-\lad}\right]}\right\}.
\ee
$\alpha_2(\lad)$ is a symmetric function of $\nu_1$ and $\nu_2$. In the limit $\nu_2/\nu_1\to +\infty$ the generating function of the long-time cumulants per unit time  becomes
\be
\label{alphalimitnu2infini}
\lim_{\nu_2/\nu_1\to +\infty}\alpha_2(\lad)=\alpha_{\scriptscriptstyle RW}(\lad;\nu_1),
\ee
where $\alpha_{\scriptscriptstyle RW}(\lad;\nu_1)$ is given in \eqref{defalphaRW}.

\subsection{Long-time current distribution in the infinite $\nu_2/\nu_1$ limit }

According to the definition \eqref{defBessel} of the modified Bessel function
\be
I_{n=t\Jcumad}(\alpha t)=\int_0^{2\pi}\frac{d\theta}{2\pi}e^{t g(\theta;\Jcumad)}
\quad\textrm{with}\quad
g(\theta;\Jcumad)=-\iexp \Jcumad\theta + \alpha \cos\theta,
\ee
where $\alpha$ denotes some parameter.
In the complex plane where the affix reads $z=\theta +\iexp \theta'$, $g(z;\Jcumad)$, the analytic continuation of $g(\theta;\Jcumad)$, is a periodic function of $z$ with period $2\pi$ when $t \Jcumad$ is equal to an integer. Therefore 
$I_{n=t \Jcumad}$ can be rewritten as $I_{n=t \Jcumad}=\int_{[-\pi,\pi]}(dz/2\pi)\exp[t g(z;\Jcumad)]$ and, by applying a saddle-point method to the latter integral, with a deformation of the initial contour in order to exhibit the constant phase path which goes through the saddle-point in the direction where it is indeed a maximum, as done in subsection \ref{DetailsCol}, one obtains (with the relevant saddle point $z_c=-\iexp \ln[(|\Jcumad|/\alpha)+\sqrt{(\Jcumad/\alpha)^2+1}]$) that
\be
I_{n=t\Jcumad}(\alpha t)\underset{t\to +\infty}{\sim}\frac{1}{\sqrt{2\pi \alpha t \sqrt{(\Jcumad/\alpha)^2+1}}}
\exp\left[t\left(\alpha\sqrt{\left(\frac{\Jcumad}{\alpha}\right)^2+1}-|\Jcumad|\ln\left[\frac{|\Jcumad|}{\alpha}+\sqrt{\left(\frac{\Jcumad}{\alpha}\right)^2+1}\right] \right)\right].
\ee
(The latter asymptotic behavior can  also be directly read at page 378 of Ref.\cite{AbramowitzStegun1972}).
The  long-time behavior of a current density $\Probdist(\Jcumad;t)$ is given in terms of $\Prob(n=t\Jcumad;t)$ by \eqref{defProbdistas}-\eqref{defProbas}. From the expression \eqref{ProbRW} for 
$\Prob_{\scriptscriptstyle RW}(n=t\Jcumad;\nu_1t)$  with $\alpha=\nu_1\sqrt{p_+p_-}$, we get
\be
\label{ProbdistRW}
\Prob^\text{as}_{\scriptscriptstyle RW}(\Jcumad,t;\nu_1)
=A_{\scriptscriptstyle RW}(\Jcumad,t;\nu_1)e^{t f_{\scriptscriptstyle RW}(\Jcumad;\nu_1)}
\ee
with $A_{\scriptscriptstyle RW}(\Jcumad,t;\nu_1)=1/\sqrt{2\pi \nu_1 t\sqrt{p_-p_-+(\Jcumad/\nu_1)^2}}$ and
\be
\label{fRW}
f_{\scriptscriptstyle RW}(\Jcumad;\nu_1)
=-\frac{\nu_1}{2}(p_++p_-)+\Jcumad\ln\sqrt{\frac{p_+}{p_-}}+|\Jcumad| \ln \sqrt{p_+p_-}
 +\nu_1\sqrt{\frac{\Jcumad^2}{\nu_1^2}+p_+p_-}
-|\Jcumad| \ln\left[\frac{|\Jcumad|}{\nu_1}+\sqrt{\frac{\Jcumad^2}{\nu_1^2}+p_+p_-}\right]
\ee
$f_{\scriptscriptstyle RW}(\Jcumad;\nu_1)$ 
is the large deviation function for the randomized random walk described by the Markov equation \eqref{RWequation}. We notice that, as predicted by large deviation theory and in particular the Gärtner-Ellis theorem, $f_{\scriptscriptstyle RW}(\Jcumad;\nu_1)$ can also be retrieved as the inverse Legendre transform of the generating function of the long-time cumulants per unit time 
$\alpha_{\scriptscriptstyle RW}(\lad;\nu_1)$ written in \eqref{defalphaRW}.

The probability density of the cumulative heat current $\Heatd_1/t$ in the long-time limit is given by \eqref{defProbdistas}, \eqref{Probn1as} and \eqref{ProbdistRW}, with the result
\be
\Probdist^\text{as}\left(\frac{\Heatd_1}{t\gapE}=\Jcumad;t\right)
\underset{\nu_2/\nu_1\to+\infty}{=} \sum_{n=-\infty}^{+\infty}
\delta\left(\Jcumad- \frac{n}{t}\right)
 t A_{\scriptscriptstyle RW}(\Jcumad,t;\nu_1)e^{t f_{\scriptscriptstyle RW}(\Jcumad;\nu_1)}
\ee
Similarly, according to \eqref{Probn2as}, by an argument similar to that leading to \eqref{LDHeat2fad}, 
 the probability density of the cumulative heat current $\Heat_2/t$ in the long-time limit is shown to read 
 \be
\label{Probdistasheat0}
 \Probdist^\text{as}\left(\frac{\Heat_2}{t\gapE}=\Jcumad;t\right)
\underset{\nu_2/\nu_1\to+\infty}{=} \sum_{n=-\infty}^{+\infty}
\delta\left(\Jcumad- \frac{n}{t}\right)
t A_2(\Jcumad,t;\nu_1)e^{t f_{\scriptscriptstyle RW}(\Jcumad;\nu_1)},
\ee
with $A_2(\Jcumad,t;\nu_1)\not=A_{\scriptscriptstyle RW}(\Jcumad,t;\nu_1)$.
We retrieve that the cumulative heat $\Heatd_1=n_1 \gapE$ and $\Heat_2=n_2\gapE$  have the same large deviation function, and, more precisely,
\be
\label{expfnu2infty}
\lim_{\nu_2/\nu_1\to +\infty} f_{\Heat_2}(\Jcum)=f_{\scriptscriptstyle RW}\left(\frac{\Jcum}{\gapE};\nu_1\right).
\ee

We notice that the expression \eqref{fRW} for the large deviation function 
$f_{\scriptscriptstyle RW}(\Jcumad;\nu_1)$ agrees with the limit of the expression \eqref{fjadpppm} for $\ft_{\Heat_2}(\Jcumad)$, when $\nu_2/\nu_1$ goes to infinity and $\Jcumad/\nu_1$ is fixed. Indeed, in the  expression \eqref{fjadpppm}, which is valid when $T_1\not=0$, the function $Z(\Jcumad)$ given in  \eqref{defZj} is such that
\be
\lim_{\substack{\nu_2/\nu_1\to +\infty \\ \Jcumad/\nu_1 \,\text{fixed}}} Z(\Jcumad)
=\sqrt{\frac{\Jcumad^2}{\nu_1^2}+p_+p_-},
\ee
while $(\nu_1+\nu_2)\left(-1+\sqrt{1-2\nuad_1\nuad_2[p_++p_- -2Z(\Jcumad)]}\right)
\sim - \nu_1\left[p_++p_- -2Z(\Jcumad)\right].$

\section{Case where $T_1=0$: pure energy dissipation towards thermal bath $1$}

\label{T1zero}

\subsection{Microscopic irreversibility}

When the temperature $T_1$ of the colder bath vanishes, in the sense that $\beta_1\gapE$ goes to infinity, the microscopic reversibility \eqref{MicroRevCond} is broken,
\be
(-\sigma,\sigma\vert\Trans\vert \sigma,\sigma)=0
\quad\textrm{whereas}\quad 
(\sigma,\sigma\vert\Trans\vert -\sigma,\sigma)\not=0
\ee
by virtue of the expression \eqref{Transexp} for the transition rates when $\gamma_1=1$.
In other words the thermal bath at zero temperature cannot provide energy to the system, i.e., it cannot flip  spin $\sigma_1$ if the flip corresponds to an increase of the two-spin system energy. There is only energy dissipation towards the zero-temperature bath.

When $\beta_1\gapE=+\infty$, the ratio in the modified detailed balance \eqref{ergodicThermo} vanishes or is infinite when two configurations differs from each other by the sign of $\sigma_1$, and when  spin $\sigma_1$ is flipped by  thermostat 1  the 
corresponding variation of the thermostat  entropy,  $\delta S_1^{\scriptscriptstyle TH}
(\C'\leftarrow \C)\equiv - \beta_1 \delta\Heatm_1(\C'\leftarrow\, \C)$ with definition \eqref{defdeltaq},
is infinite. All direct consequences of the modified detailed balance \eqref{ergodicThermo} are no longer valid.

However the  Markov matrix  \eqref{defMarokovM} of the configurations evolution is still irreducible (see the definition after \eqref{defMarokovM}) because histories such as 
\be
(\sigma,\sigma)\to (\sigma,-\sigma)\to (-\sigma,-\sigma) \to (-\sigma,\sigma)
\ee
correspond to a succession of flips with non-vanishing  transition rates. Therefore, according to the Perron-Frobenius theorem, there still exists a single stationary distribution and in the latter distribution every configuration has a non-vanishing weight.
The stationary probability given by \eqref{structPst} is still a canonical
distribution with an effective inverse temperature $\beta_*^0=(2/\gapE)\tanh^{-1}\gammaeff^0$,
with $\gammaeff^0=\nuad_1+\nuad_2\gamma_2$,
\be
\lim_{\beta_1\gapE\to +\infty}\Probst(\sigma_1,\sigma_2)=\frac{1}{4}[1+\gammaeff^0 \sigma_1\sigma_2].
\ee
According to \eqref{jinstexp}, the mean current is finite,
\be
\lim_{\beta_1\gapE\to +\infty}\Espst{\jinst_2}=\frac{\nu_1\nu_2}{\nu_1+\nu_2}\left(1-\gamma_2\right)\frac{\gapE}{2}.
\ee
Since $\beta_1\gapE=+\infty$, the  stationary exchange entropy flow \eqref{dirrSst}
 is infinitely negative in the stationary state
\be
\lim_{\beta_1\gapE\to +\infty}\left.\frac{d_\text{exch} S}{dt}\right\vert_\text{st}=-\infty,
\ee
while the rate of entropy production, which has the opposite value in the stationary state,  is infinitely positive, $\lim_{\beta_1\gapE\to +\infty}\dint \SG/dt\vert_\text{st}=+\infty$.
We also notice that the heat capacity with respect to a variation of the temperature $T_1$ from the zero value, $C_\text{st}^{[1]}(T_1=0,T_2)$,  defined in \eqref{defheatcapacity}
vanishes according to \eqref{expCa}.

As a consequence of the fact that the thermal bath  at zero temperature cannot give energy to the system, $\Heat_1=-n_1 \gapE$ is necessarily negative and
\be
\Prob(\Heat_1,\Heat_2;t)=0
\quad\textrm{if $\Heat_1>0$}.
\ee
This can be checked on the explicit expressions of subsection \ref{ExplicitProbabilities} as follows. The probability that the system is in configuration $(\sigma_1,\sigma_2)$
at time $t_0=0$, in configuration $(\sigma'_1,\sigma'_2)$ at time $t$ and receives $\Heat_1=-n \gapE$ and $\Heat_2=(n+\Delta n) \gapE$ during the time interval $[0,t]$  is
$(\sigma'_1,\sigma'_2\vert\Uev(n,n+\Delta n; t)\vert \sigma_1,\sigma_2)$.
According to \eqref{Uevn1n2Formules0}-\eqref{defU0U1Formules0} where $\gamma_1$ is to be set equal to 1, the latter matrix elements  involve the functions
$\ccstar_n(\tad)$ and $\csstar_n(\tad)$  defined in \eqref{defccstar} and \eqref{defcsstar}. When $T_1$ vanishes,  
$\gamma_1$ tends to $1$, $A-B=\nuad_1\nuad_2(1-\gamma_1)(1+\gamma_2)$ goes to zero and, by virtue of \eqref{defDeltastar},
$
\lim_{\beta_1\gapE\to +\infty}\Deltastar_+(z)= 1-2A +2A z.
$
As a consequence,  $\cosh(\tad \sqrt{\Deltastar_+(z)})=\sum_{p=0} [1/(2p)!] \tad^{2p}(1-2A +2A z)^p$ contains only positive powers of $z$, and so does 
$\sinh(\tad \sqrt{\Deltastar_+(z)})/\sqrt{\Deltastar_+(z)}$. Consequently $\ccstar_n(\tad)$ and $\csstar_n(\tad)$ vanish for $n<0$ and
$
\lim_{\beta_1\gapE\to +\infty}(\sigma'_1,\sigma'_2\vert\Uev(n,n+\Delta n; t)\vert \sigma_1,\sigma_2)=0\quad\textrm{for any $n<0$}.
$

\subsection{Long-time behavior}

The explicit values of the infinite-time limit for the cumulants per unit time of the heats $\Heatd_1$ and $\Heat_2$ can be calculated as in subsection \ref{ExplicitCumulants}. When $T_1$ vanishes, $\gamma_1=1$, $A$ tends to $A^0=\nuad_1\nuad_2(1-\gamma_2)$, and $A-B$ vanishes. Then the expression \eqref{alphaAB} of $\alpha_2(\lad)$ is reduced to
\be
\label{alphaABT1zero}
\lim_{\beta_1\gapE\to +\infty}\alpha_2(\lad)=\frac{1}{2}\left\{-(\nu_1+\nu_2)+\sqrt{(\nu_1+\nu_2)^2+2\nu_1\nu_2 (1-\gamma_2)\left[-1+e^{\lad}\right]}\right\}.
\ee
The expressions of the first three cumulants can be retrieved by setting $\gamma_1=1$ in the expressions \eqref{longtimekappa2}.

As in section \ref{LargeDeviationFunction},
the large deviation function can be derived either as the Legendre transform of 
$\alpha_2(\lad)$ or by a saddle-point method similar to that performed in subsection \ref{DetailsCol} for $\Prob_{\Prob_0}(n_2;t)$ given by \eqref{Probn2}, which also provides the amplitude of the probability, or it can be retrieved directly by taking the limit $\gamma_1\to 1$ in  the expressions 
\eqref{fjneg}-\eqref{fjpos} for the large deviation function $\fad_{\Heat_2}(\jad)$, as follows. In the limit where $T_1$ vanishes, so does $A-B$, and $Y(\jad)$  tends to 
$Y^0(\jad)=\jad\left[ \jad +\sqrt{\jad^2+1-2A^0}\right]$ where $A^0=\nuad_1\nuad_2(1-\gamma_2)$, while, according to \eqref{jinstexp} and \eqref{defjad},
\be
\Espst{\jad}^0=\nuad_1\nuad_2(1-\gamma_2).
\ee
As a result,
\bea
\label{fjnegposT1zero}
\lim_{\beta_1\gapE\to+\infty} f_{\Heat_2}(\Jcum)&\underset{\Jcum<0}{=}&-\infty
\\
\nonumber
\lim_{\beta_1\gapE\to+\infty}f_{\Heat_2}(\Jcum)&\underset{\Jcum>0}{=}&\frac{\nu_1+\nu_2}{2}
 \left(-1 + \left[1+\ln \Espst{\jad}^0\right] \jad+\sqrt{\jad^2+1-2\Espst{\jad}^0} 
-\jad\ln\jad\left[\jad+\sqrt{\jad^2+1-2\Espst{\jad}^0}\right]\right)
\eea
where $\jad=[2/(\nu_1+\nu_2)] \Jcum/\gapE$. 

\subsection{Limit where $\nu_2/\nu_1$ becomes infinite}

\subsubsection{Finite-time behaviors}

The discussion can be performed along the same lines as in section \ref{DependenceTimeScales}. We only point out the features which are qualitatively different when $\gamma_1=1$. The discrepancies are due to the fact that,
according to \eqref{defp+p-}, when $\gamma_1=1$, though $p_+$ remains finite, 
\be
\label{p+p-T1zero}
p_-=0,
\ee
and the random walk process associated to the variation of the heat amounts $\Heatd_1$ or
$\Heat_2$ can have only positive increments. As shown below, a Poisson process shows off as a randomized random walk which can proceed only in the sense of increasing positive $n_2$.

Since $p_-$  vanishes, while $p_+=1-\gamma_2$, 
 the expression \eqref{DeltsstarBis} for $\Deltastar_+(z)$  becomes 
$\lim_{\beta_1\gapE\to+\infty}\Deltastar_+(z)=1+2\nuad_1\nuad_2(1-\gamma_2)\left[-1+z\right]$ and the generating function
$G_{\scriptscriptstyle RW}(z;\nu_1 t)$ which appears  in subsection \ref{RandomWalk}
(see \eqref{GRW})  is to be replaced by 
\be
G_\text{Pois}(z;\nu_1 t)= e^{[-1+z]p_+\nu_1t}
\ee
with 
\be \label{eq:j0star}
p_+\nu_1=\frac{1}{2}(1-\gamma_2)\nu_1= \lim_{\nu_2/\nu_1\to+\infty}
\Espst{\Jcumad}^0\equiv \Espst{\Jcumad}^{0\, \star}.
\ee
$G_\text{Pois}(z; \nu_1 t)$ is the generating function 
$G_\text{Pois}(z; \nu_1 t)=\sum_{n=0}^{+\infty}z^n\Prob_\text{Pois}(n;\nu_1 t)$ for the Poisson process ruled by the Markov evolution equation 
\be
\label{Poisequation}
\frac{d \Prob_\text{Pois}(n;\nu_1 t)}{dt}=p_+\nu_1\left[-\Prob_\text{Pois}(n;\nu_1 t)
+\Prob_\text{Pois}(n-1;\nu_1 t)\right],
\ee 
and the initial condition $\Prob_\text{Pois}(n;t=0)=\delta_{n,0}$.  The solution reads
\be
\label{ProbPois}
\Prob_\text{Pois}(n;\nu_1 t)= \frac{\left(p_+\nu_1 t\right)^n}{n!}e^{-p_+\nu_1t}.
\ee

Eventually the probability that at time $t$ the system has dissipated a heat amount $\Heatd_1=n_1\gapE$ towards  bath $1$ at zero temperature reads
\be
\label{Probn1asT1nulle}
\lim_{\nu_2/\nu_1\to +\infty}\lim_{\beta_1\gapE\to
  +\infty}\Probst(n_1;t)=\Prob_\text{Pois}(n_1;\nu_1 t).  \ee 
A formula similar to \eqref{Probn2as}, where $\Prob_{\scriptscriptstyle
  RW}(n_2;\nu_1 t)$ is to be replaced by $\Prob_\text{Pois}(n_2;\nu_1 t)$, holds
for the probability that at time $t$ the system has received a heat amount
$\Heat_2=n_2\gapE$ from  bath $2$.

\subsubsection{Long-time behavior}

The infinite-time cumulants per unit time for the heat $\Heat_2$ can be obtained by noticing that, when $\nu_1/\nu_2$ vanishes, the expression \eqref{alphaABT1zero} for 
 $\lim_{\beta_1\gapE\to+\infty}\alpha_2(\lad)$ becomes
\be
\lim_{\nu_2/\nu_1\to+\infty}\lim_{\beta_1\gapE\to+\infty}\alpha_2(\lad)
=\alpha_\text{Pois}\left(\lad;\nu_1\right),
\ee
where 
\be
\alpha_\text{Pois}\left(\lad;\nu_1\right)
=\frac{\nu_1}{2} (1-\gamma_2)\left[-1+e^{\lad}\right]
\ee
is the cumulant generating  function for a Poisson process with average $\nu_1 p_+=\frac{\nu_1}{2}(1-\gamma_2)$.
We notice that the two limits can be taken in the reverse order: by virtue of \eqref{alphalimitnu2infini}
$\lim_{\beta_1\gapE\to+\infty}\lim_{\nu_2/\nu_1\to+\infty}\alpha_2(\lad)=
\lim_{\beta_1\gapE\to+\infty} \alpha_{\scriptscriptstyle RW}(\lad;\nu_1)=\alpha_\text{Pois}(\lad;\nu_1)$.  
The long-time cumulants per unit time are all equal
\be
\lim_{t\to+\infty}\frac{1}{t}
\left(\lim_{\nu_2/\nu_1\to +\infty}\lim_{\beta_1\gapE\to +\infty}\kappa_{n_2}^{[p]}\right)
=
\frac{\nu_1}{2} (1-\gamma_2).
\ee

The large deviation function can be retrieved 

- either as the Legendre transform of 
$\alpha_2(\lad)=\alpha_\text{Pois}(\lad;\nu_1)$, 

- or by a saddle-point method  applied to the expression \eqref{ProbPois} of $\Prob_\text{Pois}(n;\nu_1 t)$ and
similar to that performed for $\Prob_{\scriptscriptstyle RW}(n;\nu_1t)$, with the result, 
\be
\label{ProbasPoisson}
\left.\Prob_\text{Pois}(t \Jcumad;\nu_1 t)\right\vert_\text{$t \Jcumad$ integer}\underset{t\to+\infty}{\sim}
\frac{1}{\sqrt{2\pi t\Jcumad}}e^{tf_\text{Pois}(\Jcumad;\nu_1)}
\ee
with 
\be
f_\text{Pois}(\Jcumad;\nu_1)=-\Espst{\Jcumad}^{0\, \star} +\Jcumad-\Jcumad\ln\frac{\Jcumad}{\Espst{\Jcumad}^{0\, \star}},
\ee
where $\Espst{\Jcumad}^{0\, \star}$ is defined in \eqref{eq:j0star},

- or directly by taking the limit $\nu_2/\nu_1\to+\infty$ with $\Jcumad/\nu_1$ fixed
in  the expression for the large deviation function $\lim_{\beta_1\gapE\to+\infty}f_{\Heat_2}(\Jcum)$ given in  \eqref{fjnegposT1zero} (and by noticing, that $\Espst{\jad}^0$ is of order $\frac{\nu_1}{\nu_2}$ while $\jad^2$ is of order $( \frac{\Jcumad}{\nu_1})^2\times \left(\frac{\nu_1}{\nu_2}\right)^2$),
 
- or by  taking  first the expression for $\Jcum>0$ of
$\lim_{\nu_2/\nu_1\to+\infty}f_{\Heat_2}(\Jcum)$ given by \eqref{expfnu2infty}
and \eqref{fRW} and then taking the limit  $\beta_1\gapE\to+\infty$, namely $p_-
\to 0$.

Eventually, the large deviation function takes the simple form
\bea
\label{LDPoisson}
\lim_{\nu_2/\nu_1\to +\infty}\lim_{\beta_1\gapE\to +\infty}f_{\Heat_2}(\Jcum)&\underset{\Jcum<0}{=}&-\infty
\\
\nonumber
\lim_{\nu_2/\nu_1\to +\infty}\lim_{\beta_1\gapE\to +\infty}f_{\Heat_2}(\Jcum)&\underset{\Jcum>0}{=}&
f_\text{Pois}\left(\frac{\Jcum}{\gapE}; \nu_1\right)=\frac{1}{\gapE}\left[
 -\Espst{\Jcum_2}^{0\, \star} +\Jcum
-\Jcum \ln\frac{\Jcum}{\Espst{\Jcum_2}^{0\, \star}}\right]
\eea
where $\Espst{\Jcum_2}^{0\, \star}=(\nu_1/2) (1-\gamma_2)\gapE$.

The expression of $\Espst{\Jcum_2}^{0\, \star}$ can be interpreted as follows. Since  bath $1$ is at zero temperature,  spin $\sigma_1$ may be flipped only when it is opposite to  spin 
$\sigma_2$. Moreover, since $\nu_2\gg \nu_1$, once  spin $1$ has been flipped so that $\sigma_1\sigma_2=1$, on average  spin $2$  is flipped a great odd number of  times with a net energy  transfer $\gapE$ from  heat bath $2$ until spin $\sigma_1$ is again flipped with an energy transfer $\gapE$ to  heat bath $1$ so that $\sigma_1\sigma_2=1$ again. As a consequence the mean  energy current through the spins system is equal to $\gapE$ times the typical inverse time $\nu_1$ between two possible flips induced by  thermal bath $1$ times the probability that $\sigma_2$ is opposite to $\sigma_1$, namely $(1-\gamma_2)/2$.

\subsection{Limit where $\nu_2/\nu_1$ tends to zero}

In the reverse limit where $\nu_2\ll \nu_1$, the roles of the two heat baths in
the discussion of section \ref{DependenceTimeScales} are interchanged (see comment after \eqref{Probpn1} and \eqref{Probmn1}, as well as comparison of \eqref{Probn1} and \eqref{Probn2}). The slow
thermostat is  heat bath $2$ and the evolution of $\Heat_2$ is a Poisson
process with the kinetic parameter $\nu_2$.

The stationary mean heat current received by the
system is now $\Espst{\Jcum_2}^{0\, \star}=(\nu_2/2) (1-\gamma_2)\gapE$. The interpretation of the latter expression is the following. Since $\nu_2\ll \nu_1$, as soon as  spin $\sigma_2$ is flipped to a value opposite to  spin $\sigma_1$ with an energy transfer $\gapE$ from  heat bath $2$,  bath $1$ flips  spin $\sigma_1$ so that $\sigma_1\sigma_2=1$ and an energy $\gapE$ is transferred to  heat bath $1$. The next flip can be only a flip of  spin  $\sigma_2$ and  its probability per time unit is the value of the transition rate of spin $2$ when $\sigma_1\sigma_2=1$, namely $(1/2)\nu_2(1-\gamma_2)$. As a consequence the mean energy current through the spins system is equal to $\gapE$ times $(\nu_2/2)(1-\gamma_2)$.

\section{Thermal cycles}

As recalled in the introduction, part of the physical relevance of the two-spin
system is as an idealized mesoscopic thermal machine, with heat flowing from the
high temperature reservoir to the low temperature reservoir in average. This
flow of heat results from thermal cycles made by the system. After each thermal
cycle, the two spins have returned to their original state, but an amount of
heat $2\gapE$ has been transferred from heat bath $2$ (the hot bath) to heat
bath $1$ (the cold bath). Thermal fluctuations do occur however, and with this
interpretation two questions come naturally. What is the distribution of
the time needed to make a thermal cycle ? What is the probability that the
machine will perform a thermal cycle in the wrong direction ? 

The graph showing the possible transitions in the two-spin system looks as follows 
\[\begin{array}{ccc} (+,+) & \leftrightarrow & (+,-) \\ \updownarrow  & &
  \updownarrow \\ (-,+) & \leftrightarrow & (-,-) \end{array}.\]
This graph looks like a square, i.e. a cycle with $4$ edges. Most of what we
shall have to say applies equally well to a Markov process with a finite number
of states and whose associated transition graph is a general cycle. As the
probabilistic reasoning is more transparent in this more general framework, we
shall devote a separate section to it for completeness. It is likely that the
forthcoming analysis has already been performed (more than once) in the
literature, but we have not found it. 

\subsection{Markov processes with a cyclic transition graph}

We label the $n \geq 3$ configurations as $1,\cdots,n$, and identify
configuration $m$ with configuration $m+n$ in all subsequent formul\ae . We view
the process as the motion of a particle along the cycle, jumping from time to
time from a site to one of its two neighbors. We choose arbitrarily an
orientation of the cycle. A jump from $m$ to $m+1$ (resp. $m-1$) is said to be
clockwise (resp. anti clockwise). We let $c_m$ be the transition rate from
configuration $m$ to configuration $m+1$ and $a_m$ be the transition rate from
configuration $m$ to configuration $m-1$. All other transition rates vanish. We
set $b_m\equiv a_m+c_m$. If the particle sits at $m$ at time $t$, the
probability that the next jump will be clockwise (resp. anti clockwise) is
$c_m/b_m$ (resp. $a_m/b_m$). By saying that the graph associated to the Markov
process is a cycle we mean that all $a_m$'s and $c_m$'s are $>0$. To be totally
explicit, with the conventions of this article, the generator of the Markov process looks like
\be \label{eq:genmark}
\left(\begin{array}{ccccccc}
-b_1 & a_2 & 0 & \cdots & \cdots & 0  & c_n \\
c_1 & -b_2 & a_3 & 0 & \cdots & \cdots & 0 \\
0 & c_2 & -b_3 & a_4 & 0 & \cdots & 0 \\
\vdots & \vdots & \vdots & \vdots & \vdots &\vdots &\vdots \\
0 & \cdots & 0 & c_{n-3} & -b_{n-2} & a_{n-1} & 0 \\
0 & \cdots & \cdots & 0 & c_{n-2} & -b_{n-1} & a_n \\
a_1 & 0 & \cdots & \cdots & 0 & c_{n-1} & -b_n\\
\end{array}\right)
\ee

Starting at some arbitrary reference configuration, each later visit to this
configuration defines an integer, namely the algebraic number of times the cycle
has been traversed, i.e. the homotopy class of the trajectory between the two
passages at the reference configuration, or the winding number. The variation of
winding number between two successive visits to the reference configuration
belongs to $\{-1,0,1\}$. We can refine the definition of winding to have it
defined at all times, i.e. as a process $W_t$, via the following trick: $nW_t$
is just the number of clockwise jumps minus the number of anti clockwise jumps
that have occurred up to time $t$ (included). Note that $W_t$ is an integer if
and only if the positions of the particle on the cycle are the same at time $0$
and $t$, and then $W_t$ is simply the previously defined winding number.

\subsubsection{Forward and backward thermal cycles}

The easiest question to answer is whether or not the winding number will ever
reach $\pm 1$ ? The (strong) Markov property is the crucial ingredient. 

Let
$\pi_m^-$ be the probability that the winding number of a trajectory started at
$m$ ever reaches the value $-1/n$. Then by the Markov property 
\be
\label{relrecurrpi} 
\pi_m^-=\frac{a_m}{b_m} + \frac{c_m}{b_m} \pi_{m+1}^-\pi_m^-.  
\ee 
The meaning of this equation is clear : either the first jump is anti clockwise
(probability $a_m/b_m$) and the winding number reaches its target $-1/n$ or the
first jump is clockwise (probability $c_m/b_m$), and then the particle has
``lost'' a winding $1/n$ so it has to go from $m+1$ to $m$ with winding number
$-1/n$ to compensate (probability $\pi_{m+1}^-$), and take a new chance.

Luckily, we do not need to solve the full system: the probability that starting
from $m$ the winding number reaches $-1$ is, by the Markov property again,
$\prod_{l=0}^{n-1}\pi_{m-l}^-$, which is independent of $m$. So we denote this
probability simply by $\Pi^-\equiv \prod_{l=0}^{n-1}\pi_{m-l}^-$. We rewrite
\eqref{relrecurrpi}  as 
\be 
\label{eq:prodwind} a_m(1-\pi_m^-) =c_m \pi_m^- (1-\pi_{m+1}^-).  
\ee 
A first consequence is that if $\pi_m^- =1$ for some $m$ then also
$\pi_{m+1}^-=1$ and so on, so that all $\pi_m^-$'s are equal to one, and the
probability to reach winding number $-1$ is unity. On the other hand, in the
case when no $\pi_m^-$ equals one,
\be 
\Pi^-=\prod_{m=1}^n\frac{a_m}{c_m}. 
\ee 
Indeed, in \eqref{eq:prodwind} take the product over all $m$'s in the cycle and simplify both sides by
$\prod_{m=1}^n (1-\pi_m^-) \neq 0$ to get $\prod_{m=1}^n a_m =\prod_{m=1}^n c_m
\prod_{m=1}^n \pi_m^-$, i.e. $\Pi^-= \prod_{m=1}^n a_m/c_m$. As $\Pi^-$ is a
probability, this is possible only if $\prod_{m=1}^n a_m/c_m \leq 1$. 

We have already proved the following : if $\prod_{m=1}^n a_m/c_m \geq 1$ then
$\Pi^- =1$, i.e. winding number $-1$ is reached with probability one. We could
reproduce the above argument with $\pi_m^+$, the probability that the winding of
a trajectory started at $m$ ever reaches the value $1/n$, and $\Pi^+$. We would
get : if $\prod_{m=1}^n a_m/c_m \leq 1$ then $\Pi^+ =1$, i.e. winding number $1$
is reached with probability one. In particular, if $\prod_{m=1}^n a_m=
\prod_{m=1}^nc_m$ then the probability to reach winding numbers $-1$ and $1$ is
unity, and by the Markov property, the probability to reach any winding number
an infinite number of times is also unity: the winding number $W_t$ will
oscillate and take arbitrarily large positive and negative values as
$t\rightarrow +\infty$.

To deal with the case $\prod_{m=1}^n a_m \neq \prod_{m=1}^nc_m$, we need a
deeper result which we shall not prove here: the ergodic theorem for a finite
state Markov process, which, stated informally and adapted to the case at hand,
says that $W_t/t$ will be close to $\Espst{W_t}/t$ with probability close to $1$
at large times. We
shall ``prove'' below a formula for $\Espst{W_t}$ : 
\be 
\label{eq:espstwt}
\Espst{W_t}=t\frac{\prod_{m=1}^n c_m - \prod_{m=1}^n a_m}{P_{n-1}(a_.,c_.)}, 
\ee 
where $P_{n-1}(a_.,c_.)$ is the principal minor of the generator
\eqref{eq:genmark} of the Markov process (note that by construction the
determinant of the generator is $0$). It is a homogeneous polynomial of degree
$n-1$ in the $a_m$'s and $c_m$'s given explicitly by
\be \label{eq:winding} 
P_{n-1}(a_.,c_.)\equiv\sum_{l=1}^{n}\sum_{m=1}^{n} \prod_{1\leq j < l }
a_{m+j}\prod_{l\leq k< n} c_{m+k}, 
\ee
where as usual an empty product stands for $1$.  We observe simply that the
denominator in \eqref{eq:espstwt} is a sum of positive terms so that the sign of $\Espst{W_t}/t$ is
that of the numerator. So if $\prod_{m=1}^n c_m > \prod_{m=1}^na_m$, $W_t\to+\infty$ with probability 1, 
implying that $\Pi^+$, the probability to make one clockwise cycle, is 1 as well.
The above discussion implies that 
\be \Pi^- = \min \{1, \prod_{m=1}^n \frac{a_m}{c_m}\} , \quad \Pi^+ = \min \{1,
\prod_{m=1}^n \frac{c_m}{a_m}\}.\ee

We observe that the quantity $\Aff \equiv - \ln \min \{\Pi^-,\Pi^+\}$ plays an
important role in the context of chemical reactions, where, after multiplication by $\kB T$, it is called the
affinity of the cycle properly oriented. If $\Aff = - \log \Pi^-$ (resp.  $\Aff
= - \log \Pi^+$) the reaction has a tendency to evolve clockwise (resp.
anticlockwise).

Coming back to the physical interpretation, we may view the cycle as a
mesoscopic thermal machine, and imagine that when the winding number changes by
one, some reference heat bath has collected one unit of energy. This is what
happens in the simple two-spin system, the ``unit'' being $2\gapE$. For
arbitrary $n$, a microscopic implementation of this behavior is not so obvious,
but is not needed for the discussion either. By the ergodic theorem for a finite
state Markov process as recalled above, $W_t/t$ will be close to $\Espst{W_t}/t$
with probability close to $1$ at large times. So the behavior of the thermal
machine is deterministic at large times.  But fluctuations may occur, and
$e^{-\Aff}$ is the probability that a time exists at which the net heat transfer
is opposite to that of an expected working cycle, i.e. the thermal machine has
performed the equivalent of a working cycle in the wrong direction.  By the
Markov property, $e^{-k\Aff}$, $k=1,2,\cdots$ is the probability that a time
exists at which the net heat transfer is $-k$ times that of an expected working
cycle.

\clearpage
\subsubsection{Fluctuations in the time it takes to make a thermal cycle} 

We can now come to the second question, namely what are the fluctuations of the
time it takes to make one cycle in the direction the machine is built for? 

As usual, the object that satisfies simple equations is a Laplace transform. We
assume that $\prod_{m=1}^n c_m > \prod_{m=1}^n a_m$, i.e. that winding number
grows in average. We denote by $t_m$ the random time it takes, starting from
$m$ to reach winding number $1/n$.  Our real interest is the random time $T$ it
takes to reach winding number $1$, and by the (strong) Markov property, $T$ is
distributed as a sum of $n$ independent random variables each of which is
distributed like a $t_m$. We write $f_m(\lambda)\equiv\Esp{e^{-\lambda t_m}}_m$
where $\Esp{\cdots}_m$ is expectation with respect to trajectories starting at
$m$. This is a bit redundant here, but we want to stress that $T$ is a cyclic
invariant so that we may write $F(\lambda)\equiv\Esp{e^{-\lambda T}}$ because the
expectation with respect to any initial probability distribution gives the same
result. 

The Markov property says that $F=\prod_{m=1}^n f_m$. The $f_m$'s
satisfy 
\be \label{eq:Fm}
f_m=\frac{c_m}{b_m+\lambda}+\frac{a_m}{b_m+\lambda}f_{m-1}f_m.
\ee  
The interpretation of this equation is analogous to that for $\pi_m^\pm$. If the
particle is at $m$ at some time, it waits an exponential time with parameter
$b_m$ and then jumps to $m+1$ (resp. $m-1$) with probability $c_m/b_m$ (resp.
$a_m/b_m$).  The computation of the Laplace transform of the waiting time gives
\be \int_0^{+\infty} dt b_m e^{-b_m t} e^{-\lambda t} =\frac{b_m}{b_m+\lambda},
\ee 
which multiplied by the jump probability $c_m/b_m$ (resp. $a_m/b_m$) yields the
prefactors above. Then writing $t_m$ as a sum of the exponential waiting time to
leave $m$ and some other (independent of the waiting time at $m$ by the Markov
property) random time, we note that if the jump is to $m+1$ this random time is
$0$, whereas if the jump is to $m-1$ this random time is, by the (strong) Markov
property, distributed like the sum of two independent random times, one
distributed like $t_{m-1}$ and the other like $t_m$.

A standard trick to deal with the quadratic equations for the $f_m$'s is by a
Riccatti transform, to linearize them. Then $F(\lambda)$ acquires an
interpretation as an holonomy. So we write $f_m\equiv g_{m-1}/g_m$ for
$m=1,\cdots,n$ and observe that $F=g_0/g_n$: whereas the sequence $f_m$ is
periodic by construction, the sequence $g_m$ is not, and $F$ is the holonomy
along the cycle. With this in mind, we set $g_{-1}\equiv Fg_{n-1}$.  Then we
define 
\be 
\mathbb{F}_m\equiv\left(\begin{smallmatrix} (b_m+\lambda)/a_m & -c_m/a_m \\ 1 &
    0\end{smallmatrix}\right) 
\ee 
 and check that \eqref{eq:Fm} turns into the linear equation 
\be \left(\begin{smallmatrix} g_{m-2} \\ g_{m-1}\end{smallmatrix} \right)=
\mathbb{F}_m \left(\begin{smallmatrix} g_{m-1} \\ g_{m}\end{smallmatrix} \right)
\ee 
valid for $m=1,\cdots,n$. Setting $\mathbb{F}\equiv\mathbb{F}_1 \cdots
\mathbb{F}_n$, one finds by iterating the above formula that 
\be
 \label{eq:Fn} F\left(\begin{smallmatrix} g_{n-1} \\ g_{n}\end{smallmatrix}
\right)= \left(\begin{smallmatrix} g_{-1} \\ g_{0}\end{smallmatrix} \right)
=\mathbb{F} \left(\begin{smallmatrix} g_{n-1} \\ g_{n}\end{smallmatrix}
\right)
\ee 
so that $F$ is an eigenvalue of the transfer (or Bloch-Floquet or $\cdots$
depending on the community) matrix $\mathbb{F}$, i.e. a solution of 
\be
 \label{eq:F}
F^2-F \, \text{Tr} \, \mathbb{F}+\text{Det} \, \mathbb{F}=0.
\ee
This formula shows clearly that $F$ is a cyclic invariant, because $\text{Tr} \,
\mathbb{F}$ and $\text{Det} \, \mathbb{F}$ are. By continuity, $F(0)=1$, and we
shall see shortly that this allows to choose the right branch. 

Equation \eqref{eq:Fn} allows to express $f_n=g_{n-1}/g_n$
in terms of $F$ and the matrix elements of $\mathbb{F}$. A moment thinking
shows that the same formula expresses
any other $f_m$ in terms of the same $F$ and the matrix elements of the matrix
obtained by applying a cyclic permutation of order $m$ to the factors defining
$\mathbb{F}$. 

The determinant of $\mathbb{F}$ is easily seen by multiplicativity to be
$e^{\Aff}=\prod_{m=1}^n c_m/a_m$ (which by the way does not depend on
$\lambda$), so the affinity has also something to say on $F$.

The trace of $\mathbb{F}$ can be computed in any specific case, but it is
complicated even for our simple two-spin system. We conclude this short
digression in the world of general cycles by computing $\mathbb{F}$ to first
order in $\lambda$. This will allow us to give a ``proof'' of formula
\eqref{eq:espstwt} for $\Espst{W_t}$. We observe that
$\mathbb{F}_m(\lambda)=\mathbb{F}_m(0)+\lambda/a_m \left(\begin{smallmatrix} 1 &
    0\\ 0 & 0\end{smallmatrix}\right)$, so that
\begin{eqnarray} 
\text{Tr} \, \mathbb{F}'(0) & = & \text{Tr}\left(\sum_{m=1}^n  \mathbb{F}_{1}(0)
  \cdots 
\mathbb{F}_{m-1}(0) \frac{1}{a_m} \left(\begin{array}{cc} 1 & 0\\ 0 & 0
\end{array}\right) \mathbb{F}_{m+1}(0) \cdots \mathbb{F}_{n}(0)\right) \\
& = & \sum_{m=1}^n \frac{1}{a_m} \left[\mathbb{F}_{m+1}(0) \cdots \mathbb{
    F}_{n}(0) \mathbb{F}_{1}(0) \cdots \mathbb{F}_{m-1}(0)\right]_{11} 
\end{eqnarray}
where $[\cdots]_{11}$ stands for the top left corner matrix element of the matrix
product. 
Expanding $\eqref{eq:F}$ to first order in $\lambda$
we get $1-\text{Tr} \, \mathbb{F}(0) +\text{Det} \, \mathbb{F}=0$ and
$\left[1-\text{Det} \, \mathbb{F}\right] F'(0)=\text{Tr} \, \mathbb{F}'(0)$. The
first equation implies that the right branch for $\eqref{eq:F}$ is (under our
assumption that $A>0$)
\be \label{eq:FF}
F=\frac{\text{Tr} \, \mathbb{F} - \sqrt{(\text{Tr} \, \mathbb{F})^2-4\text{Det}
    \, \mathbb{F}}}{2},
\ee
and the second then gives  
\be \Esp{T}=\frac{\text{Tr} \, \mathbb{F}'(0)}{\text{Det} \,
  \mathbb{F}-1}.
\ee 

By computing $\text{Tr} \, \mathbb{F}'(0)$ we shall now show that
\be
\Esp{T}=\frac{P_{n-1}(a_.,c_.)}{\prod_{m=1}^n c_m -\prod_{m=1}^n a_m}.\ee
 This is
nothing but the inverse of the value announced in \eqref{eq:espstwt} for
$\Espst{W_t}/t$. That the relation $\Esp{T}\Espst{W_t}/t=1$ should hold is
intuitively clear: $\Espst{W_t}/t$ is the average of the growth of the winding
number per unit time, and $\Esp{T}$ is the average time it takes to increase the
winding number by one unit. Intuition is not proof, but modulo that, we shall
have given a proof of the formula for $\Espst{W_t}/t$. Just note that in
$\Esp{T}$, the expectation is with respect to any initial distribution, while
the average of $W_t$ is exactly proportional to $t$ only if expectation is taken
with respect to the stationary measure.

We sketch the derivation that $(\prod_{l=1}^n a_l) \text{Tr} \,
\mathbb{F}'(0)=P_{n-1}(a_.,c_.)$, which gives immediately the announced formula
for $\Esp{T}$. We first note that $ \mathbb{F}_m(0)$ has
$\left(\begin{smallmatrix} 1 \\ 1 \end{smallmatrix} \right)$ as right
eigenvector with eigenvalue $1$ whatever the value of $m$, so that the same is
true for any product of $ \mathbb{F}_m(0)$'s. The generic matrix with this
property can be parameterised by $\mathbb{K}(x,y)\equiv\left(\begin{smallmatrix}
    1+x & -x \\ 1+y & -y \end{smallmatrix}\right)$.  We observe that
$\mathbb{F}_m(0)=\mathbb{K}(c_m/a_m,0)$ and a simple recursive computation shows
that $\mathbb{K}(x_1,0) \cdots \mathbb{K}(x_k,0)=\mathbb{K}(x_k+x_kx_{k-1}+
\cdots +x_kx_{k-1}\cdots x_1,x_k+x_kx_{k-1}+\cdots +x_kx_{k-1}\cdots x_2)$. In
particular, the top left matrix element of $ \mathbb{F}_{m+1}(0) \cdots
\mathbb{F}_{n}(0) \mathbb{F}_{1}(0) \cdots \mathbb{F}_{m-1}(0)$ is \be
1+\frac{c_{m-1}}{a_{m-1}}+\cdots \frac{c_{m-1}\cdots c_1}{a_{m-1}\cdots
  a_1}+\frac{c_{m-1}\cdots c_1c_{m+1}}{a_{m-1}\cdots
  a_1a_{m+1}}+\cdots+\frac{c_{m-1}\cdots c_1c_{m+1}\cdots c_n}{a_{m-1}\cdots
  a_1a_{m+1}\cdots a_n}.  \ee Multiplying by $(\prod_{l=1}^n a_l)/a_m$ and
summing over $m$ one recovers the formula \eqref{eq:winding} for
$P_{n-1}(a_.,c_.)$.

\subsubsection{Degeneration to random walks}
\label{subsubsec:drw}

There is a degeneration of the above family of Markov processes for which
computations are totally explicit. It corresponds to the case $n=1$ in which each
single step is interpreted as a cycle. So, suppressing indices, $c$ is the rate
for making a cycle in the clockwise direction, and $a$ is the rate for making a
cycle in the anticlockwise direction. We assume $c \geq a$. The evolution of the
winding number in that case is is that of a simple continuous-time random walk.
Equation \eqref{eq:Fm} for the distribution of first passage times degenerates to a
simple quadratic equation
\be \label{eq:Fmn=1}
F=\frac{c}{b+\lambda}+\frac{a}{b+\lambda}F^2,
\ee
which, as was to be expected, is the same as \eqref{eq:F} when $n=1$. As $a\leq
c$, the solution taking value $1$ at $\lambda=0$ is
$F=\frac{b+\lambda-\sqrt{(b+\lambda)^2-4ac}}{2a}$. When $a=c$, $F$ is
non-analytic in $\lambda$, a behavior consistent with the well-known divergence
of the first passage time in that case. Setting $\bar{T}\equiv (c-a)^2 T$ it is
easily checked that
\be 
\label{eq:RWdegen}
\lim_{a,c \rightarrow g}
  \frac{1}{c-a}\ln \Esp{e^{-\lambda \bar{T}}}=\frac{1-\sqrt{1+4g\lambda}}{2g},
   \ee
which is essentially the generating function for Catalan numbers.

\subsection{The case of the two-spin system}

We apply the formul\ae\ of the previous subsection to our concrete model. We
choose the cyclic order 
\be \begin{array}{ccc} (+,+) & \rightarrow & (+,-) \\ \uparrow  & &
  \downarrow \\ (-,+) & \leftarrow & (-,-) \end{array},\ee
with corresponding $a;c$ transition coefficients (remember $c$ is for the clockwise
transition and $a$ for the anticlockwise one)
\be \begin{array}{ccc} \nu_1(1-\gamma_1)/2;\nu_2(1-\gamma_2)/2 & \leftrightarrow &
  \nu_2(1+\gamma_2)/2;\nu_1(1+\gamma_1)/2 \\ \updownarrow  & & 
  \updownarrow \\  \nu_2(1+\gamma_2)/2;\nu_1(1+\gamma_1)/2 & \leftrightarrow &
  \nu_1(1-\gamma_1)/2;\nu_2(1-\gamma_2)/2 \end{array}.\ee 
The meaning of this diagram is that, starting at the upper-left corner $(+,+)$ for instance, the rate from $(+,+)$ to
$(+,-)$ is the clockwise coefficient $\nu_2(1-\gamma_2)/2$, while the rate from  
$ (+,+)$ to $(-,+)$ is the anticlockwise coefficient  $\nu_1(1-\gamma_1)/2$. 

The product $\prod_{m=1}^n a_m/c_m$ is readily evaluated to be
$e^{-\Aff}=e^{-2(\beta_1-\beta_2)\gapE}$. Note that this quantity has a purely
thermodynamic interpretation : this is due to the modified detailed balance.
Under our standard assumption $\beta_1 \geq \beta_2$, we see that in average the
thermal machine works in the clockwise direction, and the probability that at
some time the net heat transfer is $-2k \gapE$ ($k=0,1,\cdots$) is
$e^{-2k(\beta_1-\beta_2)\gapE}$.

Using the formul\ae\ of the previous subsection, a closed formula for
$\Esp{e^{-\lambda T}}$ is not difficult to write down, but it is very
complicated and not particularly illuminating. We content to give the first two
cumulants:
\be \Esp{T}=\left( \frac{1}{\nu_1}+\frac{1}{\nu_2}\right)\frac{4}{\gamma_1
  -\gamma_2}, \ee 
\be \Esp{T^2}-\Esp{T}^2 = \left[(1-\gamma_1\gamma_2) \left(
    \frac{1}{\nu_1^2}+\frac{1}{\nu_2^2} \right) +
  (2-\gamma_1^2-\gamma_2^2)\frac{1}{\nu_1\nu_2} \right] \frac{8}{(\gamma_1
  -\gamma_2)^3}.\ee
As expected, the cumulants of  $T$ diverge when $\gamma_1-\gamma_2\searrow 0$ :
for $\gamma_1=\gamma_2$ there is no net heat current and the winding number is
just a simple symmetric random walk. 

We conclude by observing that the formul\ae\  for the normalized  random variable
\be \bar{T}\equiv T(\gamma_1 -\gamma_2)^2 \nu_1\nu_2/2\ee
are slightly simpler: writing 
\be \frac{1}{2(\gamma_1 -\gamma_2)}\log \Esp{e^{-\lambda \bar{T}}} \equiv
\sum_{k=1}^{+\infty} \frac{p_k}{k!}\lambda^k,\ee
the $p_k$'s are polynomials in $\gamma_1,\gamma_2,\nu_1,\nu_2$, with integral
coefficients, homogeneous of
degree $k$ in $\nu_1,\nu_2$ by dimensional analysis, non-homogeneous but of
degree $2k-2$ in $\gamma_1,\gamma_2$ and symmetric both under the exchange of
$\nu_1,\nu_2$ and under the exchange of $\gamma_1,\gamma_2$. It is easy to check
that in the limit when $\gamma_1=\gamma_2\equiv \gamma$ the formul\ae\ simplify
dramatically. In fact, setting $\nu\equiv \nu_1+\nu_2$, one has
\be 
\lim_{\gamma_1,\gamma_2 \rightarrow \gamma} \frac{1}{2(\gamma_1
  -\gamma_2)}\ln \Esp{e^{-\lambda \bar{T}}}= \frac{1-\sqrt{1+2\lambda \nu
    (1-\gamma^2)}}{(1-\gamma^2)},
    \ee 
which is again essentially the generating function for Catalan numbers, a result reminiscent of the case of the simple random walk recalled in subsection \ref{subsubsec:drw}, see in particular \eqref{eq:RWdegen}.

\par\medskip
\par\medskip
\textbf{ACKNOWLEDGMENTS} 

\par\medskip
F. Cornu is indebted to C. Van den Broeck for bringing her interest to thermal contact modelization a few years ago.

\bibliographystyle{unsrt}

\end{document}